\pdfoutput=1

\documentclass[12pt]{article}
\usepackage{a4wide}  % Ignore l2tabu.
\usepackage{axodraw2}
\usepackage{url}
\usepackage{hyperref}
\usepackage{graphicx}
\usepackage{bm}
\usepackage[tbtags]{amsmath}
\usepackage{amssymb,lmodern,mathtools}
\usepackage{algorithm2e}
\usepackage{array}
\usepackage{longtable,booktabs}

\newcommand{\FORM}{{\sc Form}}

\newcommand{\MINCER}{{\sc Mincer}}
\newcommand{\FORCER}{{\sc Forcer}}

\newcommand{\MSbar}{{\ensuremath{\overline{\text{MS}}}}}

\usepackage{color}

\definecolor{TodoColorB}{rgb}{0.902,0.624,0}
\definecolor{TodoColorT}{rgb}{0.337,0.706,0.914}
\definecolor{TodoColorJ}{rgb}{0,0.62,0.451}
\definecolor{TodoColor4}{rgb}{0,0.447,0.698}
\definecolor{TodoColor5}{rgb}{0.835,0.369,0}
\definecolor{TodoColor6}{rgb}{0.8,0.475,0.655}
\definecolor{TodoColor7}{rgb}{0.941,0.894,0.259}
\definecolor{TodoColor}{rgb}{1,0,0}
\newcommand{\reportnumber}[1]{%
  \begin{picture}(0,0)%
    \unitlength=1cm%
    \put(0.5,2.0){%
      \parbox{\textwidth}{%
        \begin{flushright}%
          \normalsize #1%
        \end{flushright}%
      }%
    }%
  \end{picture}%
}

\begin{document}
\setcounter{page}{0}
\thispagestyle{empty}

\reportnumber{Nikhef 2017-019}

\begin{center}
{\LARGE \FORCER{}, a \FORM{} program for the parametric reduction of four-loop 
massless propagator diagrams}
\end{center}
\vspace{5mm}
\begin{center}
{\large B. Ruijl$^{\, a,b}$, T. Ueda$^{\, a}$ and J.A.M. Vermaseren$^{\, a}$} 
\vspace{1cm}\\
{\it $^a$Nikhef Theory Group \\
\vspace{0.1cm}
Science Park 105, 1098 XG Amsterdam, The Netherlands} \\
{\it $^b$Leiden University \\
\vspace{0.1cm}
Science Niels Bohrweg 1, 2333 CA Leiden, The Netherlands} \\
\vspace{2.0cm}
\end{center}
\vspace{5mm}

\begin{abstract}
We explain the construction of \FORCER{}, a \FORM{} program for the reduction of four-loop 
massless propagator-type integrals to master integrals. The resulting program 
performs parametric IBP reductions similar to the three-loop \MINCER{} program. We 
show how one can solve many systems of IBP identities parametrically in a
computer-assisted manner. Next, we discuss the structure
of the \FORCER{} program, which involves recognizing reduction actions for each topology,
applying symmetries, and
transitioning between topologies after edges have been removed.
This part is entirely precomputed and automatically generated. 
We give examples of recent applications of \FORCER{}, and 
study the performance of the program. Finally we demonstrate how to use the 
\FORCER{} package and sketch how to prepare physical diagrams
for evaluation by \FORCER{}.
\end{abstract}

\newpage
%--#] Startup : 
%--#[ Introduction :

\section{Introduction}
 
Over the years particle physics experiments have become more and more 
precise. This creates the need for more accurate calculations of the 
underlying processes. In particular for QCD with its large coupling 
constant, three-loop calculations for processes such as 
the production of Higgs particles prove to be important for high precision
predictions~\cite{Anastasiou:2015ema,
Anastasiou:2016cez}. This in turn necessitates the evaluation of the 
four-loop splitting functions to determine the parton distributions inside 
the proton. For a variety of reasons a complete calculation at the 
four-loop level is currently out of the question. The next best solution is 
to evaluate a number of Mellin moments as was done at the three-loop level 
over the past 25 years~\cite{Larin:1991fx,Larin:1993vu,Larin:1996wd,
Retey:2000nq,Blumlein:2004xt}. One way to obtain such moments is by 
converting them to massless propagator integrals by expanding
in terms of the parton momentum. The computer program that could deal with 
the resulting three-loop integrals is called \MINCER{}~\cite{Gorishnii:1989gt,
Larin:1991fz} and its algorithms are based on Integration-By-Parts (IBP) 
identities~\cite{Chetyrkin:1981qh}. To obtain higher moments, \MINCER{} 
has been heavily optimized and was essential in an $N=29$ moment 
calculation for polarized scattering~\cite{Moch:2014sna}.

The construction of a similar program for four-loop propagator integrals is 
a far more formidable task. This has led to the exploration of different 
techniques, such as the $1/D$ expansions of Baikov~\cite{Baikov:1996rk,
Baikov:1996iu,Baikov:2005nv}. Instead of solving the systems of IBP 
equations parametrically as was done in \MINCER{}, 
Laporta developed a method to solve the system by 
substituting values for its parameters~\cite{Laporta:2001dd}. This method 
has been used to create generic programs that can handle integrals in a 
very flexible way~\cite{Anastasiou:2004vj,Smirnov:2008iw,Smirnov:2014hma,
Studerus:2009ye,vonManteuffel:2012np}. The drawback of these programs is 
that it is in essence a brute-force Gaussian reduction, that needs to 
reduce many subsystems that will drop out of the final answer. 
An extra complication is the fact that the system is 
riddled with `spurious poles' which are powers in $1/\epsilon=2/(4-D)$ that 
only cancel by the time all contributions to the coefficient of a master 
integral have been added. If it is not known in advance how many of these 
spurious poles will occur, one cannot safely perform a fixed expansions in 
$\epsilon$. In the three-loop \MINCER{} program the spurious poles could be 
avoided thanks to the resolved triangle formula by 
Tkachov~\cite{Tkachov:1984xk}, but for the all-$N$ calculation these 
spurious poles caused significant issues~\cite{Moch:2004pa,Vogt:2004mw,Vermaseren:2005qc}. 
In general, spurious pole avoidance is considered too 
complicated and is resorted to the very slow but exact arithmetic of 
rational polynomials.

Similar to our work, a method capable for a parametric reduction of massless four-loop propagator 
integrals has been developed by R. Lee in a series of 
papers~\cite{Lee:2012cn,Lee:2013mka}. 
It resulted in the LiteRed program, which is a Mathematica 
package that constructs reduction programs (also in Mathematica code). 
Although it is extremely elegant and as a method very powerful, the 
resulting four-loop propagator programs are too slow for most practical 
applications.

In this work we describe \FORCER{}, a \FORM{}~\cite{Tentyukov:2007mu,Kuipers:2012rf} 
program that is a hybrid between 
various approaches. We discuss the construction of a precomputed reduction graph
that describes how to reduce each topology and how to map topologies with missing 
propagators into
each other. Most topologies have
straightforward reduction rules due to known reducible substructures,
such as triangles or insertions. However, 21 special cases that 
could not be derived 
automatically in an efficient way have been constructed by means of hand-guided 
computer programs. We provide several heuristics and give an overview of how we solved
these complicated cases. During some manual derivations we ran into the previously
mentioned spurious pole problem.
For \FORCER{}, this means that it can work in either of two modes: with 
rational polynomials in $\epsilon$, or with a fixed expansion in which the 
depth has to be selected with great care.
%For \FORCER{}, this means that large coefficients arise and that expansions
%in $\epsilon$ have to be performed with great care.

\FORCER{} has already been used in some large scale computations, such as
new results for splitting functions at four loops \cite{Ruijl:2016pkm,Davies:2016jie},
four-loop propagator and vertex computations \cite{Ueda:2016sxw,Ueda:2016yjm,Ruijl:2017prop},
 and the five-loop beta
function \cite{Herzog:2017ohr}. We show some benchmarks for complicated cases.

The \FORCER{} source code can be found at 
\url{https://github.com/benruijl/forcer}.

The layout of the paper is as follows:
In section~\ref{sec:reductions} we discuss the reduction of simple substructures. Then in 
section~\ref{sec:solving} we discuss IBP identities and the way we manipulate them. 
Section~\ref{sec:fullreductions} presents all 
topologies that are considered to be irreducible by lacking simple substructures.
These involve the master 
integrals and a few more topologies that cannot be resolved in a single 
reduction step. In section~\ref{sec:superstructure} the framework of the program and its 
derivation are described. The usage of the program is discussed in section~\ref{sec:library} and in 
section~\ref{sec:qgrafforcer} we describe how to transform physical diagrams to input for \FORCER{}. 
In section~\ref{sec:performance} we show examples and we study the performance. Finally, 
section~\ref{sec:conclusions} presents the conclusions and applicability. Some technical details are 
discussed in the appendices. 

%--#] Introduction : 
%--#[ Reducing known substructures : 

\section{Reducing known substructures}
\label{sec:reductions}

\begin{figure}[ht]
\raisebox{0.8cm}{
\includegraphics[width=0.25\linewidth]{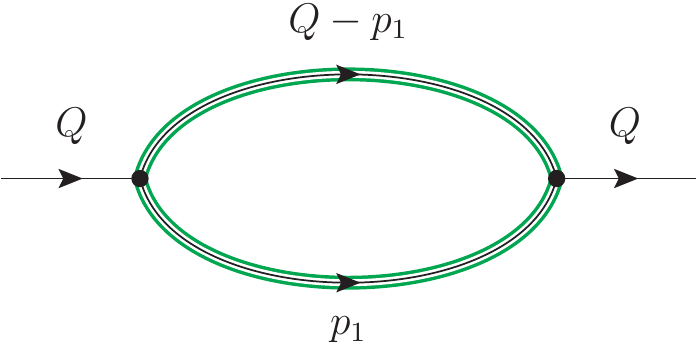}}
\includegraphics[width=0.3\linewidth]{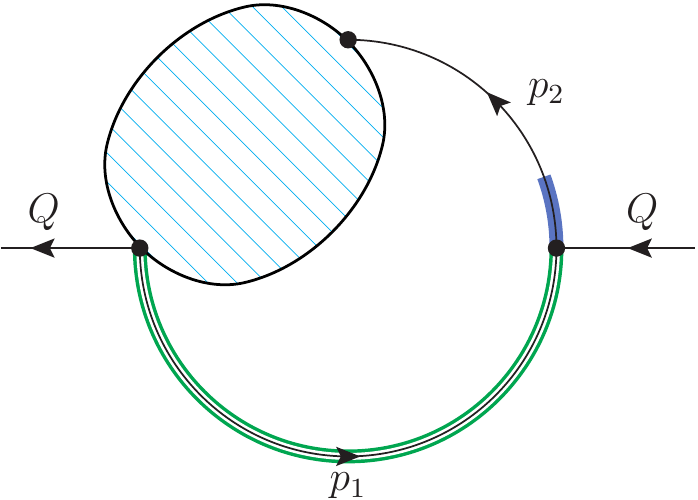}
\raisebox{-0.7cm}{
\includegraphics[width=0.39\linewidth]{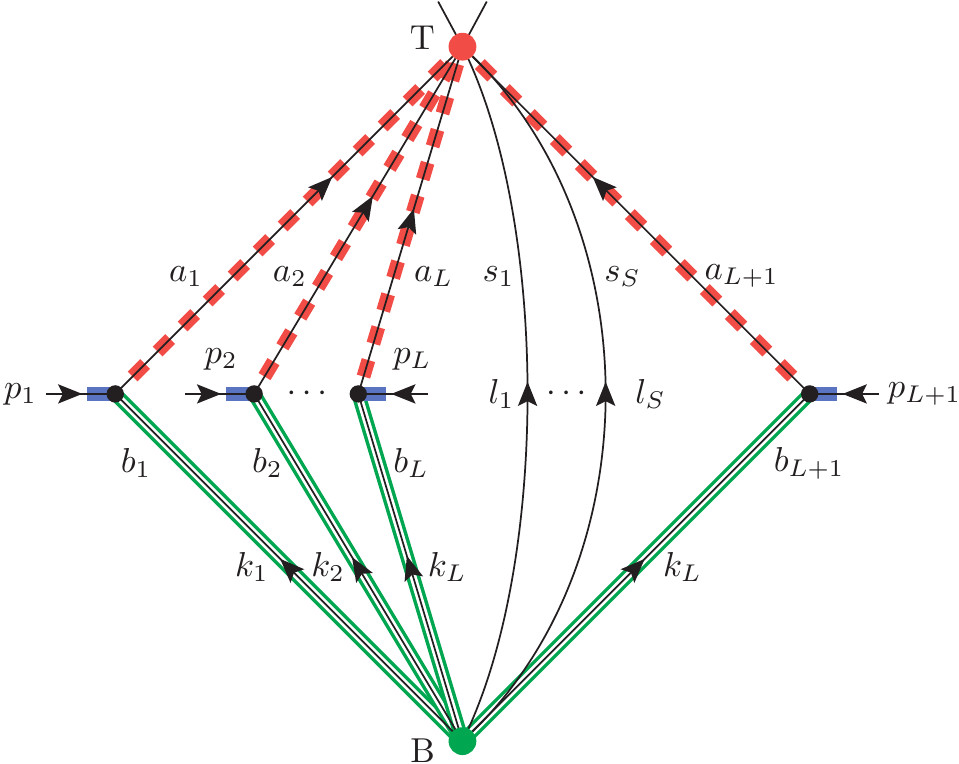}}
\caption{Left: two point function, centre: `carpet' rule, right: diamond rule. 
All are guaranteed to remove a green (triple) line or blue (external) line 
for massless propagator diagrams.}
\label{fig:substructures}
\end{figure}

For topologies with three specific substructures, there is no need to solve 
IBP equations, as efficient solutions are known. They are shown in 
figure~\ref{fig:substructures}. On the left we have a one-loop subgraph. 
If both propagators are massless, indicated by green (triple) lines, 
the loop can be integrated out and the result will be a single line with non-integer
power.
%Reduction programs like Reduze~\cite{Studerus:2009ye, vonManteuffel:2012np}
Reduction programs in non-\MINCER{} approaches do not consider such an 
operation and consider the integral to be a master integral if it has no 
reduction by means of IBP equations. For the rule to be simple, the dot 
products in the numerator must only depend on one of the two lines in the 
loop. In appendix~\ref{sec:oneloop} an efficient method is described to 
integrate the one-loop subtopology.

In the middle of figure~\ref{fig:substructures}, we have the `carpet' 
rule. It was first introduced in~\cite{Chetyrkin:1981qh} and later used in 
\MINCER{}~\cite{Larin:1991fz}. In diagrams where there is a propagator 
connecting one external line to the other, the loop momentum associated
to that line can be integrated out. We 
generalised the carpet rule to the four-loop case in appendix~\ref{sec:carpet}.

The two aforementioned rules were not derived from IBP identities. The last 
reduction rule is the diamond rule (the right one 
in figure~\ref{fig:substructures}). This recently found generic `diamond' shape is an 
extensible generalisation of the triangle rule~\cite{Ruijl:2015aca}. It can 
reduce a green (triple) line or blue (external $p_1$ to $p_{L+1}$) line. The dot 
products should be chosen in such a way that they do not depend on red 
(dashed) lines. As with the triangle rule, the diamond rule can be 
explicitly summed, so spurious poles can be avoided.

The combination of the above rules allows us to reduce 417 out of 438 
topologies in our classification at the four-loop level (see 
section~\ref{sec:superstructure}). 
Only 21 topologies do not have these substructures, out 
of which 17 contain master integrals. 
The other 4 can be reduced using IBP relations, 
but also require hand-guided solving of the IBP system. One is a four-loop diagram,
two are three-loop diagrams with one line having a non-integer power, and one 
is a two-loop diagram with two non-integer power lines.

The integrals that require a custom reduction are discussed in sections~\ref{sec:solving} 
and~\ref{sec:fullreductions}. Solving IBPs for diagrams with non-integer propagator
powers
is discussed in subsection~\ref{sec:insertions}.
The values of the master integrals are given in appendix~\ref{sec:masterintegrals}.

%--#] Reducing known substructures : 
%--#[ IBP identities :

\section{IBP identities}
\label{sec:solving}

In an $N$-loop propagator graph we have $N+1$ independent vectors: the 
external vector $Q$ and $N$ loop momenta $p_i$, where $i = 1,\ldots,N$. Together 
there are $(N+2)(N+1)/2$ independent variables. One of them, $Q^2$, can be 
used to set the scale. Hence there are $(N+2)(N+1)/2-1$ variables in the 
loops. Because there are at most $3N-1$ propagators, the 
remaining variables will be in the numerator and there is often quite some 
freedom as to which variables to choose. In topologies in which there are 
fewer propagators there will correspondingly be more variables in the 
numerator. The efficiency of the reduction depends critically on the 
selected numerators. In the \MINCER{} program the numerators were chosen to
be dot products, such as $2\ \!p_7 \cdot p_8$ for the ladder topology or $2\ \!Q \cdot p_2$ 
for the Benz topology. Alternatively, one could use extra
squared momenta such $p_9^2$ with $p_9 = p_7-p_8$.
%This is possible in
%Reduze~\cite{Studerus:2009ye,vonManteuffel:2012np} and 
%LiteRed~\cite{Lee:2012cn,Lee:2013mka}.
The advantage of the invariant method is that 
when rewriting the numerators to a new basis after a line removal, more invariants
of the old basis can be a part of the new basis.
The advantage of using dot products is that integration of one-loop subintegrals 
and the use of 
the rule of the triangle/diamond generates fewer terms compared to using invariants. 
Especially the simpler structure for integrating one-loop two-point subgraphs
is important, since we apply this rule as early as possible to reduce the number
of loops (and thus the number of parameters). Hence we choose to use 
dot products for the variables in the numerator in \FORCER{}.

In the reduction routines we represent the integrals by a function $Z$ with 
14 variables (for fewer than four loops there will naturally be fewer variables)
in which powers of variables in the denominators are given by 
positive numbers and powers in the numerator by negative numbers, as is commonly used.
For example:
\begin{equation}
  Z(1,1,1,1,1,1,1,1,1,1,-1,-2,-2,-2)
\end{equation}
is a four-loop integral with four dot products. One dot product has power one,
and the other three have two powers. Each of the 10 denominators has power 1.
Note that all information about the topology or the choice of dot products
is erased in this notation. Once the IBP relations are constructed, such 
information should be kept by different means.
%the basis is no longer needed.
We note that some indices may be associated with propagators that have 
non-integer powers if insertions are involved (see 
section~\ref{sec:insertions}).

We define the integral in which all denominators have power one (possibly with an 
extra multiple of $\epsilon$) and all numerators have power zero 
to have \emph{complexity} zero. For each extra power of a denominator or of a 
numerator the complexity is increased by one. When we construct the IBPs 
parametrically the variables are represented by parameters $n_1\ldots 
n_{14}$ in which at least three represent numerators. Now we 
define the integral with just $n_1\ldots n_{14}$ as arguments to have complexity zero 
and again raising the value of a denominator by one, or subtracting one 
from a numerator raises the complexity by one. To improve readability, we 
represent denominators by parameters $n$ and numerators by parameters $k$ in some examples.

We redefine $Z$ by adding \emph{minus} the complexity as the first argument.\footnote{
We use minus the complexity, so that \FORM{} prints the integrals with the highest 
complexity first. 
}
For example:
\begin{align}
  &Z(-3;2,1,1,1,1,1,1,1,1,1,0,-1,-1),\\
  &Z(-1;n_1,n_2,n_3+1,n_4,n_5,n_6,n_7,n_8,n_9,n_{10},n_{11}-1,k_{12}+1,k_{13}-1,k_{14}-1).
\end{align}

\begin{figure}[htb]
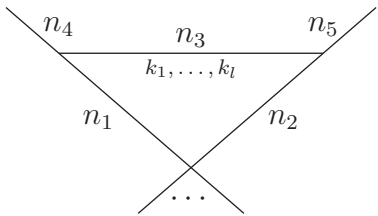

\centering
\begin{axopicture}{(100,100)(-50,-50)}
\Line(-50,0)(50,0)
\Line(20,-60.62)(-70,17.32)
\Line(-20,-60.62)(70,17.32)
\Text(-35,-25) {$n_1$}
\Text(35,-25) {$n_2$}
\Text(0,6) {$n_3$}
\Text(-50,10) {$n_4$}
\Text(50,10) {$n_5$}
\Text(0,-55) {$\cdots$}
\Text(0,-6) {{\scriptsize{$k_1,\ldots,k_l$}}}
\end{axopicture}
\caption{Triangle subtopology}
\label{fig:triangle}
\end{figure}

An example of a well-known solved parametric IBP system is the triangle rule for the 
triangle subtopology displayed in figure~\ref{fig:triangle}.
We will show the triangle rule below in our notation. 
We choose $n_3$ as the line from the triangle that can be reduced to $0$, 
and $k_1,\ldots,k_l$ as the dot products with the momentum of $n_3$:
\begin{equation}
\begin{split}
  Z(0;n_1,&n_2,n_3,n_4,n_5\ldots,k_1,\ldots,k_l,\ldots)
  =\\
  &\frac{1}{D + k_1 + \ldots k_l - n_1 - n_2 - 2 n_3}
  \Bigl[\\
    &+ n_1 Z(0;n_1+1,n_2,n_3-1,n_4,n_5,\ldots,k_1,\ldots,k_l,\ldots)\\
    &- n_1 Z(0;n_1+1,n_2,n_3,n_4-1,n_5,\ldots,k_1,\ldots,k_l,\ldots)\\
    &+ n_2 Z(0;n_1,n_2+1,n_3-1,n_4,n_5,\ldots,k_1,\ldots,k_l,\ldots)\\
    &- n_2 Z(0;n_1,n_2+1,n_3,n_4,n_5-1,\ldots,k_1,\ldots,k_l,\ldots)\\
  \Bigr]
  . 
  \label{eq:triangle-id}
\end{split}
\end{equation}
Repeated application of the rule will either bring $n_3,n_4$ or $n_5$ to 0.
The advantage of the triangle rule is that it generates few terms and only
depends on a local substructure. It is however not complexity reducing: for
each decreased propagator power, another is always raised.

In general, the goal is to construct a rule under which the basic complexity 0 integral
$Z(0;n_1,\ldots,n_{14})$ is expressed in terms of other complexity 0 integrals
or in terms of integrals with negative complexity. In the next section we present
several heuristics to find such rules.

\subsection{Solving parametric IBP identities}

For four-loop diagrams there are at first instance 20 unique IBP relations,
formed from the operation $\frac{\partial}{\partial p^\mu} q^\mu$, where $p$ is one of 
the four loop momenta and $q$ is one of the four loop momenta or the 
external momentum.
This set of equations can often be simplified by a Gaussian elimination of 
the more complex integrals. We call the simplified set of equations 
$S_0$. The most complex terms in $S_0$ have complexity 2 and
have one raised denominator and one raised irreducible numerator. 
This is a direct consequence of the IBP structure.
 
In the set $S_0$ one can distinguish several types of reduction identities. 
The nicest identities are the ones that lower the complexity, sometimes 
even by more than one unit. An example is
\begin{align}
    0 =~&Z(-2;\ldots,n+1,\ldots,k-1,\ldots) \cdot n \nonumber\\
       +&Z(0;\ldots,n,\ldots,k,\ldots) + \ldots \;,
\end{align}
where both a propagator and numerator are raised in the complexity 2 term.
By shifting $n \to n-1$ and $k \to k+1$, we find the reduction rule:
\begin{align}
    Z(0;\ldots,n,\ldots,k,\ldots) = \frac{-1}{n-1} \bigl[ Z(2;\ldots,n-1,\ldots,k+1,\ldots) + \ldots \bigr] \;.
\end{align}
Such identities are used for the simultaneous 
reduction of two variables. Since the equation will vanish once
$n=1$ or `overshoot' when $k=0$, it can only be used to 
speed up a reduction. Consequently, rules for the individual reduction of $n$ 
and $k$ are still required. 

In what follows we will omit the last step of shifting the equation such that
the highest complexity term becomes complexity 0. 
We also omit the coefficients of the $Z$ functions when 
they are deemed irrelevant and we do not consider integrals with lines missing to be $Z$-integrals.

We now study relations that raise only one coefficient:
\begin{align}
    0 =~&Z(-1;\ldots,n+1,\dots) \cdot n \nonumber\\
       +&Z(0;\ldots) + \ldots \;.
\end{align}
Repeated application of this relation will either take the variable $n$ down to one, or 
eventually create integrals in which one or more of the other lines are 
missing. For non-master topologies, at some point we find an equation
that looks like 
\begin{align}
    0 =~&Z(-1;\ldots,n+1,\ldots) \cdot (\epsilon + \ldots) \nonumber\\
       +&Z(0;\ldots) + \ldots \;.
\end{align}
This equation has an $\epsilon$ in the coefficient, which means it does not vanish
if $n=1$. As a result, it can be used to reduce $n$ to 0.

If after there is an equation in which the highest complexity is zero 
and the integral for which none of the parameters has been 
raised or lowered is present, there is a good chance that one can eliminate 
at least one line in that topology by repeated application of this 
identity, provided that there are no lines with a non-integer 
power. One example of such an equation is the rule of the triangle. The 
finding of more such equations while investigating the IBP systems of 
five-loop propagator diagrams
%the \FORCER{} methods during the construction of the \FORCER{} program
led to the discovery of the diamond rule~\cite{Ruijl:2015aca}.

\subsection{Reduction rules beyond \texorpdfstring{$S_0$}{S0}}
Even though the triangle and diamond rule can be derived from equations in
the set $S_0$, the set generally does not contain enough equations to reduce a topology
straight away. Therefore, we expand our system by taking
the set $S_0$ and constructing all equations in which 
either one denominator has been raised by one or one numerator lowered by 
one (which means that there is one more power of that variable because the 
numerators `count negative'). This set is called $S_1$, since the IBPs are 
constructed from a complexity one integral. In total, we now have $20+280$
equations. Similarly we could construct the set $S_2$ by raising the complexity 
of one of the variables in the set $S_1$ in all possible ways, generating 
an additional $2100$ equations. Usually $S_2$ is not needed. In some cases we 
may need a set like $S_{-1}$ in which the complexity of one of the 
variables has been lowered, or even $S_{1,-1}$ in which one has been raised 
and one has been lowered.

The essence of our method is to construct the combined sets $S_0$ and $S_1$ 
and use Gaussian elimination to remove all objects of complexities 3 and 2 
from the equations. The remaining equations only have objects of complexity 
one or lower. Out of these equations we construct an elimination scheme by 
defining an order of the variables, and we select for each variable an 
equation to eliminate it. For a denominator variable this is ideally an 
equation with a single term in which a variable $n$ has been raised and all 
other parameters are at their default values: $Z(-1,n_1,n_2,\ldots,n+1,\ldots)$
Once we have such an equation we can 
lower $n+m$, with $m$ being a positive integer, in all other equations to 
$n$. Since we know that after this either one of the other $n_i$ will be 0 
(meaning the reduction is done) or $n=1$, we can assume from this point on
that $n=1$. Thus, in all other equations we now set $n$ to 1, lowering the 
number of parameters by one. Similarly the numerator variables are worked up 
from $n-m$ to $n$ after which this variable is given the value zero. The 
order of elimination and the selection of the equations is critical: one of
our early carefully selected schemes resulted in a 
benchmark run of 53000 seconds, whereas a scheme with a different variable order
and a more sophisticated combination of the 
equations, performed the same test in 555 seconds.

Above we gave an example of a simple, useful equation. However, sometimes
these equations are not there. Below we discuss several other types of
equations one may encounter.
One example is if there are more integrals of complexity one:
\begin{align}
    0 =~&Z(-1;\ldots) \nonumber\\
       +&Z(-1;\ldots) \nonumber\\
       +&Z(-1;\ldots) \nonumber\\
       +&Z(-1;\ldots,n+1,\ldots) \cdot n \nonumber\\
       +&Z(0;\ldots) + \ldots \;,
\label{eq:multcompone}
\end{align}
where the last complexity one term can be used to eliminate the variable $n$ (provided 
that all other parameters have not been raised/lowered in this term), but 
this goes at the cost of increasing the number of terms with the same 
complexity. When the scheme is not carefully selected, the number of terms in 
the intermediate stages may become very large and the rational polynomials 
could become complicated.

A convenient subclass of the type shown in eq.~\eqref{eq:multcompone} is one that increases an 
index in only a single term in the equation, independent of the complexity:
\begin{align}
    0 =~&Z(-1;n_1+1,\ldots) \cdot n_1 \nonumber\\
       +&Z(-1;n_1,\ldots) \nonumber\\
       +&Z(0;n_1,\ldots) \nonumber\\
       +&Z(1;n_1,\ldots) \nonumber\\
       +&\ldots \;.
\end{align}
As a result, the equation can be used to lower the value of this variable 
at any level of complexity $c$:
\begin{align}
    0 =~&Z(-c;n_1+1,\ldots) \cdot n_1 \nonumber\\
       +&Z(-c;n_1,\ldots) \nonumber\\
       +&Z(-c+1;n_1,\ldots) \nonumber\\
       +&Z(-c+2;n_1,\ldots) \nonumber\\
       +&\ldots \;.
\label{eq:anycomp}
\end{align}
We emphasize that we apply these equations to any value of $n_1$, so also to terms
that look like $Z(-2,n_1+2,\ldots)$. In \FORM{} this can be done with a
pattern match: 
\begin{verbatim}
id Z(-c?,n1?,...,n14?) =  Z(-c,n1-1,...)/(n1-1)
                        + Z(-c+1,n1-1,...)/(n1-1)
                        + Z(-c+2,n1-1,...)/(n1-1)
                        + ...;
\end{verbatim}
These equations are convenient because after applying them and after setting
the variable to 1, there will not be a single term in the remaining 
equations in which there is a number greater than 1 in its position. We will
later see why this is desirable.

The next type of equations also has more than one term at complexity one, 
but there is no clean reduction of a given variable:
\begin{align}
    0 =~&Z(-1;n_1+1,n_2-1,n_3+1,\ldots)\cdot n_1 \nonumber\\
       +&Z(-1;n_1+1,\ldots)\cdot n_1 \nonumber\\
       +&Z(-1;n_1,\ldots) \nonumber\\
       +&Z(-1;n_1,\ldots) \nonumber\\
       +&Z(0;\ldots) +\ldots \;.
\label{eq:unclean}
\end{align}
In the numerical case, one just moves the second term to the left and either 
$n_1$ will be reduced to one or $n_2$ will eventually become zero. However, in the 
derivation of the scheme one needs to apply this equation inside 
other parametric equations and more care is called for. One should apply 
the equation as many times as needed until terms either have $n_1$ (or 
$n_1-1$, etc.) or the $n_2$ position has $n_2-1$. This means that 
for the integral
\begin{equation}
	Z(-1;n_1+1,n_2+2,\ldots)
\end{equation}
equation \eqref{eq:unclean} will have to be used up to three times. Once $n_1$ has been set 
equal to one, one may end up with terms such as
\begin{equation}
	Z(-1;2,n_2-1,\ldots) \;,
\label{eq:subl2}
\end{equation}
which are undesirable. The solution to this problem is to try to deal with 
$n_2$ immediately after $n_1$. Once we can put $n_2$ equal to one, $n_2-1$ 
becomes zero and hence it is an integral with a missing line. If one waits 
with the $n_2$ reduction and does another variable first, one risks 
that $n_2$ is raised because the equation for the other variable could 
have a term with $n_2+1$ and then one would end up with an integral of the 
type
\begin{equation}
	Z(-1;2,n_2,\ldots) \;.
\end{equation}
This introduces either unresolved integrals or loops in the reduction 
scheme. It is also possible that one has reductions with two such conditions as we 
saw above. This requires great care in the selection of the next equation. 
We have not run into impossible situations at this stage.

Another case is one where the coefficient limits its application. For example
\begin{align}
    0 =~&Z(-1;n_1+1,\ldots) \cdot (n_1-n_2) \nonumber\\
       +&Z(0;\ldots) + \ldots \;
\end{align}
cannot be applied when $n_1=n_2$. Such rules could be very compact and are
therefore used as a special case while the case in which $n_1$ is equal to 
$n_2$ is handled by a more general rule with less favourable properties but 
effectively one parameter less.

By far the most difficult equations are of the type
\begin{align}
    0 =~&Z(-1;n_1+1,n_2-1,n_3+1,\ldots) \cdot a(n_1,n_2,n_3) \nonumber\\
       +&Z(-1;n_1+1,n_2+1,n_3-1,\ldots) \cdot b(n_1,n_2,n_3) \nonumber\\
       +&Z(-1;n_1+1,\ldots) \nonumber\\
       +&Z(-1;n_1,\ldots) \nonumber\\
       +&Z(0;\ldots) +\ldots \;,
\label{eq:yoyo}
\end{align}
where $a$ and $b$ are coefficient functions.
We call this type a \emph{yoyo}. As a recursion it will never end, because 
the values of $n_2$ and $n_3$ will keep going up and down. There are 
various ways to resolve this. The first is to construct a new type of 
recursion. This is done by applying the equation twice:
\begin{eqnarray}
	Z(n_1+1,n_2,n_3,...)
		 & \rightarrow &
					+a(n_1,n_2,n_3) Z(n_1+1,n_2-1,n_3+1,...) \nonumber \\ &&
					+b(n_1,n_2,n_3) Z(n_1+1,n_2+1,n_3-1,...) + ... \nonumber \\
		 & \rightarrow &
					+a(n_1,n_2,n_3) a(n_1,n_2-1,n_3+1) Z(n_1+1,n_2-2,n_3+2,...)
						 \nonumber \\ &&
					+b(n_1,n_2,n_3) b(n_1,n_2+1,n_3-1) Z(n_1+1,n_2+2,n_3-2,...)
						\nonumber \\ &&
		+(a(n_1,n_2,n_3)b(n_1,n_2+1,n_3-1)+  \nonumber \\ &&
		\ \ \ \ \ \ \ +b(n_1,n_2,n_3)a(n_1,n_2-1,n_3+1))
				Z(n_1+1,n_2,n_3,...)
                    + ... \nonumber \\ &&
\end{eqnarray}
By moving the third term to the left one has a new recursion with a shift of two units.
This procedure can be repeated $i$ times until both 
$n_2-2^i$ and $n_3-2^i$ are less than one. The price to pay for this solution is high: 
fractions become enormously complicated and the number of terms could become
very large.

An improved solution is to find another equation with a similar yoyo and 
combine the equations in such a way that one of the yoyo terms is 
eliminated. After this, one has a regular condition. We call this `breaking 
the yoyo'. There is another way to break the yoyo that will be introduced 
below. We had to apply both methods of breaking the yoyo several times in 
the creation of the reduction schemes for the master topologies.

A final consideration is the structure of the coefficients of the 
integrals. In principle it is not very difficult to construct a reduction 
scheme from the available equations. The problem is that most schemes will 
end up with rational coefficients that take many megabytes to store because 
there are still quite a few variables in them. During the derivation this 
may cause problems with the limitations of the computer algebra 
system that is used (in our case \FORM). More importantly, the evaluation 
of such rational polynomials in the application of the reduction scheme to 
millions of integrals 
will render the reductions impossibly slow and hence useless for all 
practical purposes. Thus, if the coefficients are too large, an alternative reduction
has to be found.

%--#[ Identities for topologies with insertions :

\subsection{Identities for topologies with insertions}
\label{sec:insertions}

When a topology contains a line that does not have an integer power, the 
method of the previous section has to be slightly extended. Such cases occur either 
when the input diagram(s) can be written with a higher-order propagator in 
it, or when during the reduction a
two-point function can be integrated out. If the resulting topology needs 
a custom reduction, we not only have to lower powers of denominators and numerators, but we 
also have to bring the powers of the non-integer lines to a canonical 
value, which we take to be $1+m\epsilon$ for some positive integer $m$. 
As an example, we consider the two-loop \texttt{t1star05} topology (see 
also refs.~\cite{Gorishnii:1989gt,Larin:1991fz})

\begin{center}
\begin{axopicture}{(160,80)(0,0)}
\Line[arrow,arrowpos=0.5](40,40)(10,40)
\Arc[arrow,arrowpos=0.35](70,40)(30,90,180) % 1
\Arc[arrow,arrowpos=0.65](70,40)(30,180,270) % 4
\Line(70,70)(90,70)
\Line(70,10)(90,10)
\Line[arrow,arrowpos=0.65](80,70)(80,10) % 5
\Arc[arrow,arrowpos=0.35,flip](90,40)(30,270,360) % 3
\Arc[arrow,arrowpos=0.65](90,40)(30,0,90)  % 2
\Line[arrow,arrowpos=0.5](150,40)(120,40)
\Line(76.5,46.5)(83.5,53.5)
\Line(76.5,53.5)(83.5,46.5)
\Vertex(40,40){1.5}
\Vertex(80,70){1.5}
\Vertex(80,10){1.5}
\Vertex(120,40){1.5}
\Text(23,48)[b]{Q}
\Text(48,63)[rb]{$p_1$}
\Text(47,16)[rt]{$p_4$}
\Text(116,63)[lb]{$p_2$}
\Text(116,17)[lt]{$p_3$}
\Text(137,48)[b]{Q}
\Text(86,40)[l]{$p_5$}
\end{axopicture},
\end{center}
which has an $\epsilon$ in index 5, indicated by a single cross. We call such
a cross an \emph{insertion}.
We have the relation:
\begin{equation}
  Z_\text{t1star05}(n_1,n_2,n_3,n_4,n_5)
  =
  Z_\text{t1}(n_1,n_2,n_3,n_4,n_5+\epsilon) ,
\end{equation}
where the topology \texttt{t1} is the same two-loop topology but without any
implicit non-integer powers.
Since the $\epsilon$ can never be removed from the index during the reduction,
we suppress it in our notation for \texttt{t1star05}.
The IBPs for \texttt{t1star5} are generated from those of \texttt{t1}
by a substitution $n_5 \to n_5 + \epsilon$.
Typically, one tries to first reduce the integer indices $n_1,\dots,n_4$ to 1.
During these reductions, the contribution to the integral complexity from $n_5$ could be
taken as the absolute value of the difference to penalize any change of $n_5$,
or just be ignored to allow any change:
\newcommand{\Complexity}{\text{\textbf{Complexity}}}
\begin{equation}
  \Complexity(n_5+m_5) = |m_5| ,
  \qquad \text{or} \quad
  \Complexity(n_5+m_5) = 0 .
\end{equation}
After all $n_1,\dots,n_4$ are 1, we reduce the remaining
index $n_5$ to 1, which may be positive or negative at this point.
To derive a rule for the positive $n_5$ case, the complexity of $n_5$ can be
defined as usual for a propagator:
\begin{equation}
  \Complexity(n_5+m_5) = m_5,
  \qquad \text{for a rule with }
  n_5 > 1 .
\end{equation}
On the other hand, for the negative $n_5$ case, the complexity of $n_5$ can
be defined as usual for a numerator:
\begin{equation}
  \Complexity(n_5+m_5) = - m_5 ,
  \qquad \text{for a rule with }
  n_5 < 1 .
\end{equation}
In this way, all integrals belonging to \texttt{t1star05} can be reduced to
the master integral $Z_\text{t1star05}(1,1,1,1,1)$ and integrals with simpler
topologies.

%--#] Identities for topologies with insertions : 

\subsection{Solving strategy}
The heuristics for `solving' a topology can now be outlined.
\begin{enumerate}
\item Select a numerator basis. The quality of the IBPs will depend on
 this choice.
\item Construct the IBP identities.
\item (Important) Use a type of Gaussian elimination to simplify the IBP 
identities, minimizing the number of terms with the highest complexity. 
We call this set $S_0$. Most of the time this simplification can be 
done in an automated way. Only for the most difficult cases we have applied 
manual interference to obtain better results.
\item Construct the set $S_1$ by generating all possible options of raising an index
in $S_0$.
This gives terms of complexity 2 and 3. Use 
Gaussian elimination to eliminate all those terms. The remaining set of 
equations has terms of at most complexity 1.
\item (Important) Use the equations of the set $S_0$ (applying it to
any complexity and configuration as in eq.~\eqref{eq:anycomp}) to 
eliminate as many complexity one terms as possible. This can simplify the 
following task and results in simpler formulas in the final 
reduction program. It also breaks up some difficult yoyos.
\item Determine an order of elimination of the variables. Often the first 
variables are rather obvious from the presence of simple reduction equations. 
Some variables may not be so obvious 
and one may have to experiment. The resulting programs may differ by 
orders of magnitude in their efficiency. Here is where either human
intelligence, or a cleverly written AI program can help.
\item In many cases, one cannot find a decent equation for the last 
variable. This can be because either the results have become extremely 
lengthy, or one has discarded some long equations that seemed of no further 
relevance. In that case, the almost complete 
reduction scheme is applied to the set $S_0$. This will give a number of 
varieties of the final reduction(s). One can select the shortest one.
\item (Checking) Now apply the custom reduction scheme to the set $S_0$ with 
numbers for the variables and make sure that master integrals are indeed
irreducible, and that the program does not get caught in loops. 
There may be equations remaining which only consist of integrals with missing lines. 
We did not take relations between those into 
account.
\item Combine all reductions and useful double reduction equations 
(equations that need at least two variables that are above their minimal 
complexity) based on $S_0$ or substitutions made during the Gaussian 
elimination. Together this forms the reduction procedure for the given 
topology.
\end{enumerate}

In some cases the resulting schemes were still deemed too slow and more 
exhaustive methods were asked for. In such cases the sets $S_2$ and 
$S_{-1}$ were also constructed and many different ways of combining the 
equations were tried automatically. Such programs could take much time 
because of the very complicated rational polynomials in the parameters of 
the integrals, but they eventually did result in a number of shorter reductions. 

A number of \FORM{} procedures has been constructed to execute the above 
steps. The most laborious step is to determine a proper order for the 
elimination sequence, and which equations to use for each. Furthermore, we had a 
case (the bebe topology of section~\ref{sec:fullreductions}) in which there were no good 
reductions for two of the variables, unless we used two of the equations in 
the set $S_0$ to eliminate them with a complexity raising operation. It 
also reduced the number of remaining equations to 18 and hence left fewer 
options during the remaining parts of the derivation.

There are two major reasons why some reduction rules perform faster than 
others. The first reason is that even though a rule may have only one 
$Z(-1;...)$ term, it could be that the sub-leading terms increase the value 
of a variable that was set to 1 in one of the early steps of the scheme
(see eq.~\eqref{eq:subl2}). 
This forces the program back to an earlier reduction rule of the scheme, even though 
now at a lower complexity. The second reason is the coefficient growth: if 
a rule has a particularly complicated overall coefficient, it multiplies 
every term in the RHS and all subsequent terms will have rather lengthy 
rational polynomials in $\epsilon$. Expanding in $\epsilon$ (see 
appendix~\ref{sec:Expansions}) can alleviate some of these problems, 
provided one expands deep enough to avoid issues with spurious poles.

Determining the order of elimination seems suited for AI techniques, such as 
Monte Carlo Tree Search (see e.g. \cite{Browne2012}). One could use the number of top complexity 
terms, the number of lower complexity terms, the number of spectators and 
the size of the most complicated rational polynomial as parameters for an 
evaluation function for a given scheme and then use this in a MCTS to find 
an optimal scheme. This is currently under investigation. It should be 
noted that such type of use of AI for precisely this purpose was already 
hinted at in ref.~\cite{Vermaseren:2004mc}.

%--#] IBP identities : 
%--#[ The topologies that need custom reductions :

\section{The topologies that need custom reductions}
\label{sec:fullreductions}

Because we integrate over one-loop two-point functions, our classification
of the master integrals differs from 
refs.~\cite{Baikov:2010hf,Lee:2011jt}. In general, any diagram that factorizes
we do not consider a master topology. The master diagrams that contain
one-loop two-point functions that cannot be factorized, will have slightly
different values, since we integrate out the bubble.
The full list with the values of the master integrals in our convention
are given in appendix~\ref{sec:masterintegrals}.

There are eight four-loop master integrals, excluding the diagrams in which
a 2-point function can be integrated out. 
For these master integrals we have to design a custom scheme in which 
the parameters are reduced, one by one, to the value they have in the master 
integral. In addition (and perhaps surprisingly) there are four non-master
topologies that need such a 
custom reduction. Only when all but a few parameters are set to 1, do we find
a relation to reduce an edge to 0. In this category there is one at the four-loop 
level, two at the three-loop level (with one non-integer edge) and one at the two-loop level
(with two non-integer edges). In total we 
need 21 custom reduction schemes. All other topologies can be dealt with 
using generic formulas that can either eliminate a line or integrate out a 
loop (see section~\ref{sec:reductions}). We list all topologies that need a custom reduction 
scheme:

\begin{longtable}[c]{m{0.54\textwidth} m{.4\textwidth}}
\toprule\\
\centering
\scalebox{0.9}{
\begin{axopicture}{(190,80)(0,0)}
\Line[arrow,arrowpos=0.5](40,40)(10,40)
\Arc[arrow,arrowpos=0.45](70,40)(30,90,180)
\Line(75,70)(70,70)
\Arc[arrow,arrowpos=0.55,flip](70,40)(30,180,270)  % 6
\Line(70,10)(75,10)
\Line[arrow,arrowpos=0.5,flip](75,40)(75,70) % 7
\Line[arrow,arrowpos=0.5](75,10)(75,40)
\Line[arrow,arrowpos=0.5](115,70)(75,70)
\Line[arrow,arrowpos=0.5](75,10)(115,10)
\Line[arrow,arrowpos=0.5,flip](115,40)(75,40) % 11
\Line[arrow,arrowpos=0.5](115,70)(115,40)   % 9
\Line[arrow,arrowpos=0.5,flip](115,40)(115,10)  % 10
\Line(120,70)(115,70)
\Line(115,10)(120,10)
\Arc[arrow,arrowpos=0.55](120,40)(30,270,360)
\Arc[arrow,arrowpos=0.45](120,40)(30,0,90)
\Line[arrow,arrowpos=0.5](180,40)(150,40)
\Vertex(40,40){1.5}
\Vertex(75,70){1.5}
\Vertex(75,40){1.5}
\Vertex(75,10){1.5}
\Vertex(115,70){1.5}
\Vertex(115,40){1.5}
\Vertex(115,10){1.5}
\Vertex(150,40){1.5}
\Text(23,48)[b]{Q}
\Text(167,48)[b]{Q}
\Text(45,63)[rb]{$p_1$}
\Text(45,17)[rt]{$p_6$}
\Text(146,63)[lb]{$p_3$}
\Text(146,17)[lt]{$p_4$}
\Text(95,65)[t]{$p_2$}
\Text(95,15)[b]{$p_5$}
\Text(95,45)[b]{$p_{11}$}
\Text(71,55)[r]{$p_7$}
\Text(71,25)[r]{$p_8$}
\Text(120,55)[l]{$p_9$}
\Text(120,25)[l]{$p_{10}$}
\end{axopicture}
}
&
Topology name: \textbf{haha}, master.

Momenta: $p_1, p_2, p_4, p_5$.

Numerators: $2\ \!Q\cdot p_2, 2\ \!Q\cdot p_5, 2\ \!p_1\cdot p_4$\\
%% no1
\centering
\scalebox{0.9}{
\begin{axopicture}{(220,80)(0,0)}
\Line[arrow,arrowpos=0.5,flip](150,10)(70,70) % 11
\SetWidth{4} \SetColor{White}
\Line(110,70)(70,10)
\Line(150,70)(110,10)
\SetWidth{0.5} \SetColor{Black}
\Line[arrow,arrowpos=0.5](110,70)(70,10)
\Line[arrow,arrowpos=0.5](150,70)(110,10)
\Line[arrow,arrowpos=0.5](40,40)(10,40)
\Arc[arrow,arrowpos=0.5](70,40)(30,90,180)
\Arc[arrow,arrowpos=0.5,flip](70,40)(30,180,270) % 8
\Line[arrow,arrowpos=0.5](110,70)(70,70) % 2
\Line[arrow,arrowpos=0.5,flip](70,10)(110,10) % 7
\Line[arrow,arrowpos=0.5](150,70)(110,70) %3
\Line[arrow,arrowpos=0.5,flip](110,10)(150,10) % 6
\Arc[arrow,arrowpos=0.5](150,40)(30,0,90)
\Arc[arrow,arrowpos=0.5,flip](150,40)(30,270,360) % 5
\Line[arrow,arrowpos=0.5](210,40)(180,40)
\Vertex(40,40){1.5}
\Vertex(70,70){1.5}
\Vertex(70,10){1.5}
\Vertex(110,70){1.5}
\Vertex(110,10){1.5}
\Vertex(150,70){1.5}
\Vertex(150,10){1.5}
\Vertex(180,40){1.5}
\Text(23,48)[b]{Q}
\Text(197,48)[b]{Q}
\Text(47,63)[rb]{$p_1$}
\Text(47,15)[rt]{$p_8$}
\Text(177,63)[lb]{$p_4$}
\Text(174,15)[lt]{$p_5$}
\Text(95,66)[t]{$p_2$}
\Text(128,66)[t]{$p_3$}
\Text(93,15)[b]{$p_7$}
\Text(125,15)[b]{$p_6$}
\Text(86,40)[r]{$p_{9}$}
\Text(138,40)[l]{$p_{10}$}
\Text(108,36)[t]{$p_{11}$}
\end{axopicture}
}
&Topology name: \textbf{no1}, master.

Momenta: $p_1, p_2, p_3, p_4$.

Numerators: $2\ \!p_2\cdot p_4, 2\ \!Q\cdot p_2, 2\ \!Q\cdot p_3$.
\\
%% no2
\centering
\scalebox{0.9}{
\begin{axopicture}{(220,80)(0,0)}
\Line[arrow,arrowpos=0.70,flip](150,10)(70,70) % 9
\Line[arrow,arrowpos=0.30](150,70)(70,10)
\SetWidth{8} \SetColor{White}
\Line(110,70)(110,10)
\SetWidth{0.5} \SetColor{Black}
\Line[arrow,arrowpos=0.5](110,10)(110,70)
\Line[arrow,arrowpos=0.5](40,40)(10,40)
\Arc[arrow,arrowpos=0.5](70,40)(30,90,180)
\Arc[arrow,arrowpos=0.5,flip](70,40)(30,180,270) % 8
\Line[arrow,arrowpos=0.5](110,70)(70,70)
\Line[arrow,arrowpos=0.5,flip](70,10)(110,10) % 7
\Line[arrow,arrowpos=0.5](150,70)(110,70)
\Line[arrow,arrowpos=0.5](110,10)(150,10)
\Arc[arrow,arrowpos=0.5](150,40)(30,0,90)
\Arc[arrow,arrowpos=0.5,flip](150,40)(30,270,360) % 5
\Line[arrow,arrowpos=0.5](210,40)(180,40)
\Vertex(40,40){1.5}
\Vertex(70,70){1.5}
\Vertex(70,10){1.5}
\Vertex(110,70){1.5}
\Vertex(110,10){1.5}
\Vertex(150,70){1.5}
\Vertex(150,10){1.5}
\Vertex(180,40){1.5}
\Text(23,48)[b]{Q}
\Text(197,48)[b]{Q}
\Text(47,63)[rb]{$p_1$}
\Text(47,15)[rt]{$p_8$}
\Text(177,63)[lb]{$p_4$}
\Text(174,15)[lt]{$p_5$}
\Text(95,66)[t]{$p_2$}
\Text(126,66)[t]{$p_3$}
\Text(95,15)[b]{$p_7$}
\Text(126,15)[b]{$p_6$}
\Text(90,50)[r]{$p_9$}
\Text(100,38)[r]{$p_{10}$}
\Text(132,48)[l]{$p_{11}$}
\end{axopicture}
}
&
Topology name: \textbf{no2}, master.

Momenta: $p_1, p_2, p_3, p_4$.

Numerators: $2\ \!Q\cdot p_2, 2\ \!p_1\cdot p_4, 2\ \!Q\cdot p_3$.\\
%%% no6
\centering
\scalebox{0.9}{
\begin{axopicture}{(220,80)(0,0)}
\Line[arrow,arrowpos=0.5,flip](100,40)(150,70)
\SetWidth{5} \SetColor{White}
\Line(120,10)(110,70)
\SetWidth{0.5} \SetColor{Black}
\Line[arrow,arrowpos=0.45,flip](120,10)(110,70)
\Line[arrow,arrowpos=0.5](40,40)(10,40)
\Arc[arrow,arrowpos=0.5](70,40)(30,90,180)  % 8
\Arc[arrow,arrowpos=0.8,flip](70,40)(30,180,270)
\Line[arrow,arrowpos=0.5](110,70)(70,70) % 7
\Line[arrow,arrowpos=0.5](150,70)(110,70) % 10
\Arc[arrow,arrowpos=0.5](150,40)(30,0,90)
\Arc[arrow,arrowpos=0.2,flip](150,40)(30,270,360)
\Line[arrow,arrowpos=0.5](210,40)(180,40)
\Line(70,10)(120,10)
\Line(120,10)(150,10)
\Line[arrow,arrowpos=0.5,flip](120,10)(100,40)
\Line[arrow,arrowpos=0.5,flip](100,40)(70,70)  % 7
\Vertex(40,40){1.5}
\Vertex(70,70){1.5}
\Vertex(110,70){1.5}
\Vertex(150,70){1.5}
\Vertex(180,40){1.5}
\Vertex(120,10){1.5}
\Vertex(100,40){1.5}
%\Vertex(70,10){1.5}
%\Vertex(150,10){1.5}
\Text(23,48)[b]{Q}
\Text(197,48)[b]{Q}
\Text(47,15)[rt]{$p_1$}
\Text(177,63)[lb]{$p_4$}
\Text(174,15)[lt]{$p_5$}
\Text(47,63)[rb]{$p_8$}
\Text(95,66)[t]{$p_7$}
\Text(126,67)[t]{$p_{10}$}
\Text(120,37)[l]{$p_6$}
\Text(127,52)[tl]{$p_3$}
\Text(105,25)[r]{$p_2$}
\Text(83,52)[rt]{$p_{9}$}
\end{axopicture}
}
&Topology name: \textbf{no6}, master.

Momenta: $p_1, p_2, p_3, p_4$.

Numerators: $2\ \!p_1\cdot p_2, 2\ \!p_2\cdot p_4,$ 
$2\ \!Q\cdot p_2,2\ \!Q\cdot p_3$.\\
%% lala
\centering
\scalebox{0.9}{
\begin{axopicture}{(210,80)(0,0)}
\SetScale{0.96}
\Line[arrow,arrowpos=0.5](40,40)(10,40)
\Arc[arrow,arrowpos=0.5](70,40)(30,90,180)
\Arc[arrow,arrowpos=0.6,flip](70,40)(30,180,270) % 6
\Line(70,10)(90,10)
\Line[arrow,arrowpos=0.5](120,70)(70,70)
\Line[arrow,arrowpos=0.5](90,10)(140,10)
\Line(140,70)(120,70)
\Arc[arrow,arrowpos=0.6](140,40)(30,0,90)
\Arc[arrow,arrowpos=0.5](140,40)(30,270,360)
\Line[arrow,arrowpos=0.5](90,10)(70,70)
\Line[arrow,arrowpos=0.5](90,10)(120,70)
\Line[arrow,arrowpos=0.5,flip](120,70)(140,10) % 7
\Line[arrow,arrowpos=0.5](200,40)(170,40)
\Vertex(40,40){1.5}
\Vertex(70,70){1.5}
\Vertex(90,10){1.5}
\Vertex(120,70){1.5}
\Vertex(140,10){1.5}
\Vertex(170,40){1.5}
\Text(23,48)[b]{Q}
\Text(187,48)[b]{Q}
\Text(45,63)[rb]{$p_1$}
\Text(95,65)[t]{$p_2$}
%
% Old notation, used with the invariants.
%%\Text(45,17)[rt]{$p_8$}
%%\Text(166,63)[lb]{$p_4$}
%%\Text(166,17)[lt]{$p_5$}
%%%\Text(140,65)[t]{$p_3$}
%%%\Text(100,15)[b]{$p_7$}
%%\Text(115,15)[b]{$p_6$}
%%\Text(75,40)[r]{$p_9$}
%%\Text(102,43)[r]{$p_{10}$}
%%\Text(135,42)[l]{$p_{11}$}
%
% New notation, used with the dot products.
\Text(166,63)[lb]{$p_3$}
\Text(166,17)[lt]{$p_4$}
\Text(115,15)[b]{$p_5$}
\Text(45,17)[rt]{$p_6$}
\Text(135,42)[l]{$p_7$}
\Text(102,43)[r]{$p_8$}
\Text(75,40)[r]{$p_9$}
\end{axopicture}
}
&Topology name: \textbf{lala}, master.

Momenta: $p_1, p_2,p_4,p_5$.

Numerators: $2\ \!Q\cdot p_5, 2\ \!Q\cdot p_2, 2\ \!p_1\cdot p_4,$

$2\ \!p_1\cdot p_5, 2\ \!p_2\cdot p_4$.\\
\centering
%%% nono
\scalebox{0.9}{
\begin{axopicture}{(190,80)(0,0)}
\Line[arrow,arrowpos=0.3](120,10)(75,70)
\SetWidth{4} \SetColor{White}
\Line(120,70)(75,10)
\SetWidth{0.5} \SetColor{Black}
\Line[arrow,arrowpos=0.3](120,70)(75,10)
\Line[arrow,arrowpos=0.5](40,40)(10,40)
\Arc[arrow,arrowpos=0.45](70,40)(30,90,180)
\Line(75,70)(70,70)
\Arc[arrow,arrowpos=0.55,flip](70,40)(30,180,270) % 6
\Line(70,10)(75,10)
\Line[arrow,arrowpos=0.5,flip](75,10)(75,70) % 7
\Line[arrow,arrowpos=0.5](120,70)(75,70)
\Line[arrow,arrowpos=0.5,flip](75,10)(120,10) % 5
\Arc[arrow,arrowpos=0.55,flip](120,40)(30,270,360) % 4
\Arc[arrow,arrowpos=0.45](120,40)(30,0,90)  % 3
\Line[arrow,arrowpos=0.5](180,40)(150,40)
\Vertex(40,40){1.5}
\Vertex(75,70){1.5}
\Vertex(75,10){1.5}
\Vertex(120,70){1.5}
\Vertex(120,10){1.5}
\Vertex(150,40){1.5}
\Text(23,48)[b]{Q}
\Text(71,40)[r]{$p_7$}
\Text(45,63)[rb]{$p_1$}
\Text(45,17)[rt]{$p_6$}
\Text(146,63)[lb]{$p_3$}
\Text(146,17)[lt]{$p_4$}
\Text(167,48)[b]{Q}
\Text(98,65)[t]{$p_2$}
\Text(98,15)[b]{$p_5$}
\Text(113,53)[lt]{$p_9$}
\Text(113,27)[lb]{$p_8$}
\end{axopicture}
}
&Topology name: \textbf{nono}, master.

Momenta: $p_1, p_2, p_3, p_{10}=p_2+p_8$.

Numerators: $2\ \!p_2\cdot p_8, 2\ \!p_6\cdot p_7, 2\ \!Q\cdot p_2,$

$2\ \!p_1\cdot p_2, 2\ \!p_7\cdot p_9$.\\
\centering
%% cross
\scalebox{0.9}{
\begin{axopicture}{(180,80)(0,0)}
\Line[arrow,arrowpos=0.5](40,40)(10,40)
\Arc[arrow,arrowpos=0.3](70,40)(30,90,180)
\Arc[arrow,arrowpos=0.7,flip](70,40)(30,180,270)
\Line(90,70)(70,70)
\Line(110,70)(90,70)
\Line(90,10)(70,10)
\Line(110,10)(90,10)
\Line[arrow,arrowpos=0.5](90,40)(40,40)
\Line[arrow,arrowpos=0.5](140,40)(90,40)
\Line[arrow,arrowpos=0.5](90,40)(90,70)
\Line[arrow,arrowpos=0.5](90,40)(90,10)
\Arc[arrow,arrowpos=0.3,flip](110,40)(30,270,360)
\Arc[arrow,arrowpos=0.7](110,40)(30,0,90)
\Line[arrow,arrowpos=0.5](170,40)(140,40)
\Vertex(40,40){1.5}
\Vertex(90,70){1.5}
\Vertex(90,40){1.5}
\Vertex(90,10){1.5}
\Vertex(140,40){1.5}
\Text(23,48)[b]{Q}
\Text(45,63)[rb]{$p_1$}
\Text(45,17)[rt]{$p_3$}
\Text(136,63)[lb]{$p_2$}
\Text(135,17)[lt]{$p_4$}
\Text(157,48)[b]{Q}
\Text(65,45)[b]{$p_5$}
\Text(115,45)[b]{$p_6$}
\Text(95,55)[l]{$p_7$}
\Text(95,25)[l]{$p_8$}
\end{axopicture}
}
&Topology name: \textbf{cross}, master.

Momenta: $p_1, p_2, p_3, p_4$.

Numerators: $2\ \!Q\cdot p_1, 2\ \!Q\cdot p_2, 2\ \!Q\cdot p_3,$

$2\ \!Q\cdot p_4, 2\ \!p_1\cdot p_4, 2\ \!p_2\cdot p_3$.\\
\centering
%%% bebe
\scalebox{0.9}{
\begin{axopicture}{(175,80)(0,0)}
\Line[arrow,arrowpos=0.5](40,40)(10,40)
\Arc[arrow,arrowpos=0.45](70,40)(30,90,180)
\Line(75,70)(70,70)
\Arc[arrow,arrowpos=0.55,flip](70,40)(30,180,270)
\Line(70,10)(75,10)
\Line[arrow,arrowpos=0.5](75,10)(75,70)
\Line(105,70)(75,70)
\Line(105,10)(75,10)
\Arc[arrow,arrowpos=0.8](105,40)(30,0,90)
\Arc[arrow,arrowpos=0.2,flip](105,40)(30,270,360)
\Line[arrow,arrowpos=0.5](105,40)(75,70)
\Line[arrow,arrowpos=0.5](105,40)(75,10)
\Line[arrow,arrowpos=0.5](135,40)(105,40)
\Line[arrow,arrowpos=0.5](165,40)(135,40)
\Vertex(40,40){1.5}
\Vertex(75,70){1.5}
\Vertex(75,10){1.5}
\Vertex(105,40){1.5}
\Vertex(135,40){1.5}
\Text(23,48)[b]{Q}
\Text(152,48)[b]{Q}
\Text(47,63)[rb]{$p_1$}
\Text(47,17)[rt]{$p_3$}
\Text(130,63)[lb]{$p_2$}
\Text(130,17)[lt]{$p_4$}
\Text(120,45)[b]{$p_5$}
\Text(95,56)[lb]{$p_6$}
\Text(95,24)[lt]{$p_7$}
\Text(70,40)[r]{$p_8$}
\end{axopicture}
}
&Topology name: \textbf{bebe}, master.

Momenta: $p_1, p_2, p_4, p_6$.

Numerators: $2\ \!Q\cdot p_2, 2\ \!Q\cdot p_4, 2\ \!Q\cdot p_6,$

$2\ \!p_1\cdot p_2, 2\ \!p_2\cdot p_6, 2\ \!p_1\cdot p_4$.\\
\centering
%% bubu
\scalebox{0.9}{
\begin{axopicture}{(180,80)(0,0)}
\Line[arrow,arrowpos=0.7](110,10)(75,70)
\SetWidth{4} \SetColor{White}
\Line(110,40)(75,40)
\SetWidth{0.5} \SetColor{Black}

\Arc[arrow,arrowpos=0.45](70,40)(30,90,180)
\Line(75,70)(70,70)
\Arc[arrow,arrowpos=0.55,flip](70,40)(30,180,270) % 7
\Line(70,10)(75,10)
\Line[arrow,arrowpos=0.5](75,10)(75,40)
\Line[arrow,arrowpos=0.5](75,40)(75,70)
\Line[arrow,arrowpos=0.55](75,10)(105,10)
\Line(105,10)(110,10)
\Line(110,70)(75,70)
\Arc[arrow,arrowpos=0.45](110,40)(30,270,360)
\Arc[arrow,arrowpos=0.85](110,40)(30,0,90)
\Line[arrow,arrowpos=0.6](75,40)(140,40)
\Line[arrow,arrowpos=0.5](40,40)(10,40)
\Line[arrow,arrowpos=0.5](170,40)(140,40)
\Vertex(40,40){1.5}
\Vertex(75,70){1.5}
\Vertex(75,40){1.5}
\Vertex(75,10){1.5}
\Vertex(110,10){1.5}
\Vertex(140,40){1.5}
\Text(23,48)[b]{Q}
\Text(157,48)[b]{Q}
\Text(45,63)[rb]{$p_4$}
\Text(45,17)[rt]{$p_7$}
\Text(93,15)[b]{$p_8$}
\Text(71,55)[r]{$p_6$}
\Text(71,25)[r]{$p_9$}
\Text(130,66)[lb]{$p_1$}
\Text(136,17)[lt]{$p_2$}
\Text(93,51)[lb]{$p_5$}
\Text(113,35)[t]{$p_3$}
\end{axopicture}
}
&
Topology name: \textbf{bubu}, not a master.

Momenta: $p_2, p_3, p_8, p_9$.

Numerators: $2\ \!Q\cdot p_2, 2\ \!Q\cdot p_8, 2\ \!p_2\cdot p_3,$

$2\ \!p_2\cdot p_9, 2\ \!p_3\cdot p_8$.\\
%% nostar5
\centering
\scalebox{0.9}{
\begin{axopicture}{(190,80)(0,0)}
\Line[arrow,arrowpos=0.3](75,70)(120,10) % 7
\SetWidth{4} \SetColor{White}
\Line(120,70)(75,10)
\SetWidth{0.5} \SetColor{Black}
\Line[arrow,arrowpos=0.3](120,70)(75,10) %8
\Line[arrow,arrowpos=0.5](40,40)(10,40)
\Arc[arrow,arrowpos=0.45](70,40)(30,90,180)
\Line(75,70)(70,70)
\Arc[arrow,arrowpos=0.55](70,40)(30,180,270) % 6
\Line(70,10)(75,10)
\Line[arrow,arrowpos=0.5](120,70)(75,70)
\Line[arrow,arrowpos=0.35](75,10)(120,10) % 5
% 106.5
\Line(103,6.5)(110,13.5)
\Line(103,13.5)(110,6.5)
\Arc[arrow,arrowpos=0.55,flip](120,40)(30,270,360) % 4
\Arc[arrow,arrowpos=0.45](120,40)(30,0,90)  % 3
\Line[arrow,arrowpos=0.5](180,40)(150,40)
\Vertex(40,40){1.5}
\Vertex(75,70){1.5}
\Vertex(75,10){1.5}
\Vertex(120,70){1.5}
\Vertex(120,10){1.5}
\Vertex(150,40){1.5}
\Text(23,48)[b]{Q}
\Text(45,63)[rb]{$p_1$}
\Text(45,17)[rt]{$p_6$}
\Text(146,63)[lb]{$p_3$}
\Text(146,17)[lt]{$p_4$}
\Text(167,48)[b]{Q}
\Text(98,65)[t]{$p_2$}
\Text(98,15)[b]{$p_5$}
\Text(113,53)[lt]{$p_8$}
\Text(84,53)[rt]{$p_7$}
\end{axopicture}
}
&Topology name: \textbf{nostar5}, master.

Momenta: $p_1, p_2, p_3$.

Numerators: $2\ \!Q\cdot p_2,$\\
\centering
%%nostar6
\scalebox{0.9}{
\begin{axopicture}{(190,80)(0,0)}
\Line[arrow,arrowpos=0.3](75,70)(120,10) % 7
\SetWidth{4} \SetColor{White}
\Line(120,70)(75,10)
\SetWidth{0.5} \SetColor{Black}
\Line[arrow,arrowpos=0.3](120,70)(75,10) %8
\Line[arrow,arrowpos=0.5](40,40)(10,40)
\Arc[arrow,arrowpos=0.45](70,40)(30,90,180)
\Line(75,70)(70,70)
\Arc[arrow,arrowpos=0.45](70,40)(30,180,270) % 6
%\Line(61.2905,6.9558)(68.2905,13.9558) % (64.7905,10.4558)
%\Line(61.2905,13.9558)(68.2905,6.9558) %
\Line(56.2394,8.3092)(63.2394,15.3092) % (59.7394,11.8092)
\Line(63.2394,8.3092)(56.2394,15.3092) %
\Line(70,10)(75,10)
\Line[arrow,arrowpos=0.5](120,70)(75,70)
\Line[arrow,arrowpos=0.5](75,10)(120,10) % 5
\Arc[arrow,arrowpos=0.55,flip](120,40)(30,270,360) % 4
\Arc[arrow,arrowpos=0.45](120,40)(30,0,90)  % 3
\Line[arrow,arrowpos=0.5](180,40)(150,40)
\Vertex(40,40){1.5}
\Vertex(75,70){1.5}
\Vertex(75,10){1.5}
\Vertex(120,70){1.5}
\Vertex(120,10){1.5}
\Vertex(150,40){1.5}
\Text(23,48)[b]{Q}
\Text(45,63)[rb]{$p_1$}
\Text(45,17)[rt]{$p_6$}
\Text(146,63)[lb]{$p_3$}
\Text(146,17)[lt]{$p_4$}
\Text(167,48)[b]{Q}
\Text(98,65)[t]{$p_2$}
\Text(98,15)[b]{$p_5$}
\Text(113,53)[lt]{$p_8$}
\Text(84,53)[rt]{$p_7$}
\end{axopicture}
}
&Topology name: \textbf{nostar6}, master.

Momenta: $p_1, p_2, p_3$.

Numerators: $2\ \!Q\cdot p_2,$\\
\centering
%% lastar5
\scalebox{0.9}{
\begin{axopicture}{(190,80)(0,0)}
\Line[arrow,arrowpos=0.5](75,70)(75,10) % 7
\Line[arrow,arrowpos=0.5](120,70)(120,10) %8
\Line[arrow,arrowpos=0.5](40,40)(10,40)
\Arc[arrow,arrowpos=0.45](70,40)(30,90,180)
\Line(75,70)(70,70)
\Arc[arrow,arrowpos=0.55](70,40)(30,180,270) % 6
\Line(70,10)(75,10)
\Line[arrow,arrowpos=0.5](120,70)(75,70)
\Line[arrow,arrowpos=0.35](75,10)(120,10) % 5
% 106.5
\Line(103,6.5)(110,13.5)
\Line(103,13.5)(110,6.5)
\Arc[arrow,arrowpos=0.55,flip](120,40)(30,270,360) % 4
\Arc[arrow,arrowpos=0.45](120,40)(30,0,90)  % 3
\Line[arrow,arrowpos=0.5](180,40)(150,40)
\Vertex(40,40){1.5}
\Vertex(75,70){1.5}
\Vertex(75,10){1.5}
\Vertex(120,70){1.5}
\Vertex(120,10){1.5}
\Vertex(150,40){1.5}
\Text(23,48)[b]{Q}
\Text(45,63)[rb]{$p_1$}
\Text(45,17)[rt]{$p_6$}
\Text(146,63)[lb]{$p_3$}
\Text(146,17)[lt]{$p_4$}
\Text(167,48)[b]{Q}
\Text(98,65)[t]{$p_2$}
\Text(98,15)[b]{$p_5$}
\Text(126,40)[l]{$p_8$}
\Text(71,40)[r]{$p_7$}
\end{axopicture}
}
&Topology name: \textbf{lastar5}, not a master.

Momenta: $p_1, p_2, p_3$.

Numerators: $2\ \!p_1\cdot p_3,$\\
%% fastar1
\centering
\scalebox{0.9}{
\begin{axopicture}{(190,80)(0,0)}
\Line[arrow,arrowpos=0.5](75,70)(97.5,10) % 6
\Line[arrow,arrowpos=0.5](120,70)(97.5,10) % 7
\Line[arrow,arrowpos=0.5](40,40)(10,40)
\Arc[arrow,arrowpos=0.60](70,40)(30,90,180) % 1
\Line(56.2394,64.6908)(63.2394,71.6908) % (59.7394.68.1908)
\Line(56.2394,71.6908)(63.2394,64.6908)
\Line(75,70)(70,70)
\Arc[arrow,arrowpos=0.70](70,40)(30,180,270) % 5
\Line[arrow,arrowpos=0.5](120,70)(75,70)
\Line(70,10)(120,10) % 5
\Arc[arrow,arrowpos=0.35,flip](120,40)(30,270,360) % 4
\Arc[arrow,arrowpos=0.45](120,40)(30,0,90)  % 3
\Line[arrow,arrowpos=0.5](180,40)(150,40)
\Vertex(40,40){1.5}
\Vertex(75,70){1.5}
\Vertex(120,70){1.5}
\Vertex(97.5,10){1.5}
\Vertex(150,40){1.5}
\Text(23,48)[b]{Q}
\Text(45,63)[rb]{$p_1$}
\Text(45,17)[rt]{$p_5$}
\Text(146,63)[lb]{$p_3$}
\Text(146,17)[lt]{$p_4$}
\Text(167,48)[b]{Q}
\Text(98,64)[t]{$p_2$}
\Text(115,40)[l]{$p_7$}
\Text(82,40)[r]{$p_6$}
\end{axopicture}
}
&Topology name: \textbf{fastar1}, not a master.

Momenta: $p_1, p_2, p_3$.

Numerators: $2\ \!p_1\cdot p_3, 2\ \!Q\cdot p_2,$\\
%% fastar2
\centering
\scalebox{0.9}{
\begin{axopicture}{(190,80)(0,0)}
\Line[arrow,arrowpos=0.5](75,70)(97.5,10) % 6
\Line[arrow,arrowpos=0.5](120,70)(97.5,10) % 7
\Line[arrow,arrowpos=0.5](40,40)(10,40)
\Arc[arrow,arrowpos=0.45](70,40)(30,90,180)
\Line(75,70)(70,70)
\Arc[arrow,arrowpos=0.70](70,40)(30,180,270) % 5
\Line[arrow,arrowpos=0.65](120,70)(75,70)
% 106.5
\Line(103,66.5)(110,73.5)
\Line(103,73.5)(110,66.5)
\Line(70,10)(120,10) % 5
\Arc[arrow,arrowpos=0.35,flip](120,40)(30,270,360) % 4
\Arc[arrow,arrowpos=0.45](120,40)(30,0,90)  % 3
\Line[arrow,arrowpos=0.5](180,40)(150,40)
\Vertex(40,40){1.5}
\Vertex(75,70){1.5}
\Vertex(120,70){1.5}
\Vertex(97.5,10){1.5}
\Vertex(150,40){1.5}
\Text(23,48)[b]{Q}
\Text(45,63)[rb]{$p_1$}
\Text(45,17)[rt]{$p_5$}
\Text(146,63)[lb]{$p_3$}
\Text(146,17)[lt]{$p_4$}
\Text(167,48)[b]{Q}
\Text(98,64)[t]{$p_2$}
\Text(115,40)[l]{$p_7$}
\Text(82,40)[r]{$p_6$}
\end{axopicture}
}
&Topology name: \textbf{fastar2}, master.

Momenta: $p_1, p_2, p_3$.

Numerators: $2\ \!p_1\cdot p_3, 2\ \!Q\cdot p_2,$\\
%% bustar5
\centering
\scalebox{0.9}{
\begin{axopicture}{(160,80)(0,0)}
\Line[arrow,arrowpos=0.5](40,40)(10,40)
\Arc[arrow,arrowpos=0.35](70,40)(30,90,180) % 1
\Arc[arrow,arrowpos=0.65](70,40)(30,180,270) % 3
\Line[arrow,arrowpos=0.5](80,40)(40,40) % 2
\Line(70,70)(90,70)
\Line(70,10)(90,10)
\Line[arrow,arrowpos=0.5](80,40)(80,70) % 4
\Line[arrow,arrowpos=0.7](80,40)(80,10) % 5
\Line(76.5,27.5)(83.5,34.5) % (80,31)
\Line(76.5,34.5)(83.5,27.5)
\Arc[arrow,arrowpos=0.45,flip](90,40)(30,270,360) % 7
\Arc[arrow,arrowpos=0.55](90,40)(30,0,90)  % 6
\Line[arrow,arrowpos=0.5](150,40)(120,40)
\Vertex(40,40){1.5}
\Vertex(80,70){1.5}
\Vertex(80,10){1.5}
\Vertex(80,40){1.5}
\Vertex(120,40){1.5}
\Text(23,48)[b]{Q}
\Text(48,63)[rb]{$p_1$}
\Text(47,16)[rt]{$p_3$}
\Text(116,63)[lb]{$p_6$}
\Text(116,17)[lt]{$p_7$}
\Text(137,48)[b]{Q}
\Text(64,44)[b]{$p_2$}
\Text(86,55)[l]{$p_4$}
\Text(86,25)[l]{$p_5$}
\end{axopicture}
}
&Topology name: \textbf{bustar5}, not a master.

Momenta: $p_4, p_5, p_6$.

Numerators: $2\ \!Q\cdot p_4, 2\ \!Q\cdot p_5,$\\
%% t1star55
\centering
\scalebox{0.9}{
\begin{axopicture}{(160,80)(0,0)}
\Line[arrow,arrowpos=0.5](40,40)(10,40)
\Arc[arrow,arrowpos=0.35](70,40)(30,90,180) % 1
\Arc[arrow,arrowpos=0.65](70,40)(30,180,270) % 4
\Line(70,70)(90,70)
\Line(70,10)(90,10)
\Line[arrow,arrowpos=0.5](80,70)(80,10) % 5
\Arc[arrow,arrowpos=0.35,flip](90,40)(30,270,360) % 3
\Arc[arrow,arrowpos=0.65](90,40)(30,0,90)  % 2
\Line[arrow,arrowpos=0.5](150,40)(120,40)
\Line(76.5,51.5)(83.5,58.5)
\Line(76.5,58.5)(83.5,51.5)
\Line(76.5,21.5)(83.5,28.5)
\Line(76.5,28.5)(83.5,21.5)
\Vertex(40,40){1.5}
\Vertex(80,70){1.5}
\Vertex(80,10){1.5}
\Vertex(120,40){1.5}
\Text(23,48)[b]{Q}
\Text(48,63)[rb]{$p_1$}
\Text(47,16)[rt]{$p_4$}
\Text(116,63)[lb]{$p_2$}
\Text(116,17)[lt]{$p_3$}
\Text(137,48)[b]{Q}
\Text(86,40)[l]{$p_5$}
\end{axopicture}
}
&Topology name: \textbf{t1star55}, master.

Momenta: $p_1, p_2$.\\
% t1star24
\centering
\scalebox{0.9}{
\begin{axopicture}{(160,80)(0,0)}
\Line[arrow,arrowpos=0.5](40,40)(10,40)
\Arc[arrow,arrowpos=0.35](70,40)(30,90,180) % 1
\Arc[arrow,arrowpos=0.40](70,40)(30,180,270) % 4
\Line(70,70)(90,70)
\Line(70,10)(90,10)
\Line[arrow,arrowpos=0.5](80,70)(80,10) % 5
\Arc[arrow,arrowpos=0.35,flip](90,40)(30,270,360) % 3
\Arc[arrow,arrowpos=0.40](90,40)(30,0,90)  % 2
\Line[arrow,arrowpos=0.5](150,40)(120,40)
\Line(95.7705,65.0317)(102.7705,72.0317) % (79.2705,68.5317)
\Line(95.7705,72.0317)(102.7705,65.0317)
\Line(57.2295,7.9683)(64.2295,14.9683) % (60.7295,11.4683)
\Line(57.2295,14.9683)(64.2295,7.9683)
\Vertex(40,40){1.5}
\Vertex(80,70){1.5}
\Vertex(80,10){1.5}
\Vertex(120,40){1.5}
\Text(23,48)[b]{Q}
\Text(48,63)[rb]{$p_1$}
\Text(47,16)[rt]{$p_4$}
\Text(116,63)[lb]{$p_2$}
\Text(116,17)[lt]{$p_3$}
\Text(137,48)[b]{Q}
\Text(86,40)[l]{$p_5$}
\end{axopicture}
}
&Topology name: \textbf{t1star24}, master.

Momenta: $p_1, p_2$.\\
%% t1star34
\centering
\scalebox{0.9}{
\begin{axopicture}{(160,80)(0,0)}
\Line[arrow,arrowpos=0.5](40,40)(10,40)
\Arc[arrow,arrowpos=0.35](70,40)(30,90,180) % 1
\Arc[arrow,arrowpos=0.40](70,40)(30,180,270) % 4
\Line(70,70)(90,70)
\Line(70,10)(90,10)
\Line[arrow,arrowpos=0.5](80,70)(80,10) % 5
\Arc[arrow,arrowpos=0.60,flip](90,40)(30,270,360) % 3
\Arc[arrow,arrowpos=0.65](90,40)(30,0,90)  % 2
\Line[arrow,arrowpos=0.5](150,40)(120,40)
\Line(95.7705,7.9683)(102.7705,14.9683) % (79.2705,11.4683)
\Line(95.7705,14.9683)(102.7705,7.9683)
\Line(57.2295,7.9683)(64.2295,14.9683) % (60.7295,11.4683)
\Line(57.2295,14.9683)(64.2295,7.9683)
\Vertex(40,40){1.5}
\Vertex(80,70){1.5}
\Vertex(80,10){1.5}
\Vertex(120,40){1.5}
\Text(23,48)[b]{Q}
\Text(48,63)[rb]{$p_1$}
\Text(47,16)[rt]{$p_4$}
\Text(116,63)[lb]{$p_2$}
\Text(116,17)[lt]{$p_3$}
\Text(137,48)[b]{Q}
\Text(86,40)[l]{$p_5$}
\end{axopicture}
}
&Topology name: \textbf{t1star34}, not a master.

Momenta: $p_1, p_2$.\\
%% t1star45
\centering
\scalebox{0.9}{
\begin{axopicture}{(160,80)(0,0)}
\Line[arrow,arrowpos=0.5](40,40)(10,40)
\Arc[arrow,arrowpos=0.35](70,40)(30,90,180) % 1
\Arc[arrow,arrowpos=0.40](70,40)(30,180,270) % 4
\Line(70,70)(90,70)
\Line(70,10)(90,10)
\Line[arrow,arrowpos=0.65](80,70)(80,10) % 5
\Arc[arrow,arrowpos=0.35,flip](90,40)(30,270,360) % 3
\Arc[arrow,arrowpos=0.65](90,40)(30,0,90)  % 2
\Line[arrow,arrowpos=0.5](150,40)(120,40)
\Line(76.5,46.5)(83.5,53.5)
\Line(76.5,53.5)(83.5,46.5)
\Line(57.2295,7.9683)(64.2295,14.9683) % (60.7295,11.4683)
\Line(57.2295,14.9683)(64.2295,7.9683)
\Vertex(40,40){1.5}
\Vertex(80,70){1.5}
\Vertex(80,10){1.5}
\Vertex(120,40){1.5}
\Text(23,48)[b]{Q}
\Text(48,63)[rb]{$p_1$}
\Text(47,16)[rt]{$p_4$}
\Text(116,63)[lb]{$p_2$}
\Text(116,17)[lt]{$p_3$}
\Text(137,48)[b]{Q}
\Text(86,40)[l]{$p_5$}
\end{axopicture}
}
&Topology name: \textbf{t1star45}, master.

Momenta: $p_1, p_2$.\\
%% no
\centering
\scalebox{0.9}{
\begin{axopicture}{(190,80)(0,0)}
\Line[arrow,arrowpos=0.3](75,70)(120,10) % 7
\SetWidth{4} \SetColor{White}
\Line(120,70)(75,10)
\SetWidth{0.5} \SetColor{Black}
\Line[arrow,arrowpos=0.3](120,70)(75,10) %8
\Line[arrow,arrowpos=0.5](40,40)(10,40)
\Arc[arrow,arrowpos=0.45](70,40)(30,90,180)
\Line(75,70)(70,70)
\Arc[arrow,arrowpos=0.55](70,40)(30,180,270) % 6
\Line(70,10)(75,10)
\Line[arrow,arrowpos=0.5](120,70)(75,70)
\Line[arrow,arrowpos=0.5](75,10)(120,10) % 5
\Arc[arrow,arrowpos=0.55,flip](120,40)(30,270,360) % 4
\Arc[arrow,arrowpos=0.45](120,40)(30,0,90)  % 3
\Line[arrow,arrowpos=0.5](180,40)(150,40)
\Vertex(40,40){1.5}
\Vertex(75,70){1.5}
\Vertex(75,10){1.5}
\Vertex(120,70){1.5}
\Vertex(120,10){1.5}
\Vertex(150,40){1.5}
\Text(23,48)[b]{Q}
\Text(45,63)[rb]{$p_1$}
\Text(45,17)[rt]{$p_6$}
\Text(146,63)[lb]{$p_3$}
\Text(146,17)[lt]{$p_4$}
\Text(167,48)[b]{Q}
\Text(98,65)[t]{$p_2$}
\Text(98,15)[b]{$p_5$}
\Text(113,53)[lt]{$p_8$}
\Text(84,53)[rt]{$p_7$}
\end{axopicture}
}
&Topology name: \textbf{no}, master.

Momenta: $p_1, p_2, p_3$.

Numerators: $2\ \!Q\cdot p_2,$

Remarks: Already in \MINCER{}.\\
% t1star05
\centering
\scalebox{0.9}{
\begin{axopicture}{(160,80)(0,0)}
\Line[arrow,arrowpos=0.5](40,40)(10,40)
\Arc[arrow,arrowpos=0.35](70,40)(30,90,180) % 1
\Arc[arrow,arrowpos=0.65](70,40)(30,180,270) % 4
\Line(70,70)(90,70)
\Line(70,10)(90,10)
\Line[arrow,arrowpos=0.65](80,70)(80,10) % 5
\Arc[arrow,arrowpos=0.35,flip](90,40)(30,270,360) % 3
\Arc[arrow,arrowpos=0.65](90,40)(30,0,90)  % 2
\Line[arrow,arrowpos=0.5](150,40)(120,40)
\Line(76.5,46.5)(83.5,53.5)
\Line(76.5,53.5)(83.5,46.5)
\Vertex(40,40){1.5}
\Vertex(80,70){1.5}
\Vertex(80,10){1.5}
\Vertex(120,40){1.5}
\Text(23,48)[b]{Q}
\Text(48,63)[rb]{$p_1$}
\Text(47,16)[rt]{$p_4$}
\Text(116,63)[lb]{$p_2$}
\Text(116,17)[lt]{$p_3$}
\Text(137,48)[b]{Q}
\Text(86,40)[l]{$p_5$}
\end{axopicture}
}
&Topology name: \textbf{t1star05}, master.

Momenta: $p_1, p_2$.

Remarks: Already present in \MINCER{}.\\
\bottomrule
\caption{Table of all the topologies that require a custom reduction.}
\label{tab:customred}
\end{longtable}

In order to choose the best reduction schemes for the topologies in table~%
\ref{tab:customred}, we measure the performance of a complete calculation
of the integrals with all indices raised by 1 (a complexity 14 integral at four loops).
By performing a complete calculation, we confirm that the number of terms with
a simpler topology created by the reduction rules does not cause bottlenecks.
Additionally, we confirm that for the case where all indices are raised by 
2 (a complexity 28 integral at four loops), the
reduction is still performing well.

We note that the ordering of variables in the reductions scheme is not the only relevant 
parameter. The choice of numerators can influence the presence of 
non-leading terms, which after the Gaussian elimination become leading 
terms. Such terms can spoil the efficiency of certain reduction rules. 
In particular the three complicated topologies 
\texttt{nono}, \texttt{bebe}, and \texttt{no2} are sensitive to the choice of dot products.

Most schemes could be derived using the heuristics introduced in section~\ref{sec:solving},
by selecting the reduction variable that corresponds to the shortest reduction rule.
However, there are a few derivations that need more care. 
For \texttt{nono}, one needs to avoid a circular path in a special way. 
The formulas for the last two variables, $n_4$ and $n_8$, can only be 
obtained by reusing the original set $S_0$. At this point one uses either 
combinations of nearly all equations to obtain very lengthy formulas 
($>1000$ lines) or one uses a relatively short formula with a term that 
sends the reduction back to a previous rule, because it contains a 
term with $n_{11}=-1$. This would normally introduce a loop, but by sending 
only this term through the unfinished scheme and combining the result 
with the remaining part of the formula, we obtain a compact reduction 
formula for $n_4$ (39 terms).

The \texttt{bebe} reduction is more complicated as it does not yield a regular 
reduction for $n_1$ and $n_3$. However, in the set $S_0$ there are 
equations that can be used for their reduction, provided we are willing to 
raise the complexity. This does not agree with the automatic nature of our 
derivation tools, and hence some work needs to be guided by hand. 
Furthermore, we can no longer use a number of equations from the $S_0$ set for 
generating reduction rules for other variables. As a consequence, we are left with far fewer 
equations after the Gaussian eliminations, although their number is still 
sufficient for the next 11 variables. Eventually the $n_2$ variable has to 
be obtained again from the $S_0$ set.

For the construction of a reduction scheme the \texttt{bubu} topology is by far the 
most complicated, even though it is not a master topology. There 
are five different numerators and the elimination of the last numerator needs 
to be split into several cases, each with a rather lengthy formula 
involving complicated rational polynomials. 
In order to prevent a blow-up of terms, the order of elimination of the 
variables is critical, as well as using the equations obtained 
during the Gaussian elimination that give a direct reduction of the 
complexity. It took more
than two months to find a first suitable reduction scheme.

We use the $S_0$ set and equations that come from the Gaussian elimination 
before we start with the 14 reduction identities of the complete schemes. 
This speeds up the reduction enormously, because these equations are 
usually much more compact and will often reduce the complexity immediately. 
It turns out that the final result is very sensitive to how we use these 
equations, because sometimes there are options when there is more than one 
term with the highest complexity, and also the order in which they are 
applied is relevant. Additionally one has to be careful with this ordering 
to avoid loops in the reduction. Unfortunately, it is not always possible to 
indicate which ordering is optimal, because some orderings may yield a 
faster scheme at the cost of more spectator terms and/or higher powers of 
$\epsilon$ in the rational polynomials.

Considering the amount of work involved in deriving the schemes, it is not 
excluded that better schemes will be found. It seems to be a good candidate 
for the application of automated AI techniques.

%--#] The topologies that need custom reductions : 
%--#[ Superstructure :

\section{The \FORCER{} framework}
\label{sec:superstructure}

In essence, the \FORCER{} program provides a method to reduce each topology to simpler ones.
There is quite some freedom: sometimes multiple reduction rules can be applied, sometimes it is
best to use a different set of independent momenta, etc.
In order to obtain the best performance, all decisions in the \FORCER{} program are
precomputed by a script: for each topology the
action is hard-coded and the momentum rewrites are known. The advantage of this method is that
costly optimizations, such as choosing an optimal basis for each topology, have no
runtime cost.

\subsection{Reduction graph generation}

Before going into details, we first give an overview of the program. The program structure
can be viewed as a directed acyclic graph (DAG), the \emph{reduction graph}, where the nodes are topologies
and each edge indicates a transition from one topology to another when a propagator is removed.
As a result, each node may have more than one parent. The root nodes of the 
reduction graph are the top-level topologies, which 
are topologies that only have three-point vertices.
All tadpole topologies will be zero, so they are not included in the graph.
To reduce the number of topologies, propagators with the same momentum are
always merged.

Each node represents a topology, which consists of a graph with a certain
fixed labelling of all the propagators, including momentum directions, and a
fixed set of irreducible numerators. 
Each topology also has an action that determines how it can be 
reduced. They are, in order of preference: integration of a two-point function, carpet 
rule, triangle/diamond rule, or a custom reduction. Each topology
contains transitions to other topologies for all removable edges (edges
with integer power). Even though
the specific rule may not be able to nullify any propagator in the graph, 
the dot product
rewrites may, so therefore we generate all possible transitions.
If there are lines missing, in most cases the topology action is not executed
and the topology is automatically rewritten to another.
The exception is for 
integrating insertions: insertions are guaranteed to reduce the number of 
loops, which simplifies the dot product basis. Thus, first rewriting the dot 
products to a new topology would be wasteful.

\begin{figure}[htb]
\centering
\includegraphics[scale=0.4]{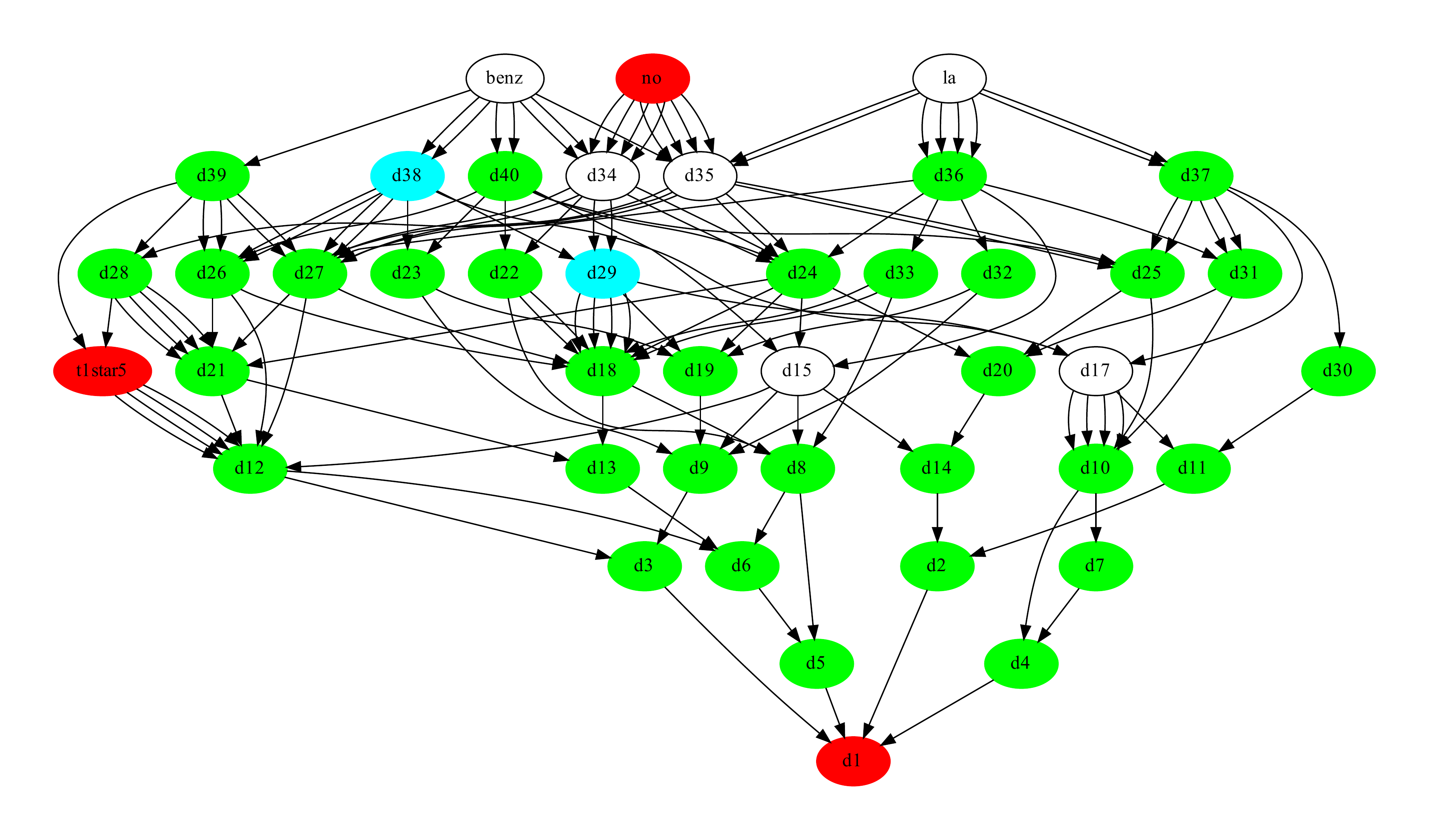}
\caption{The three-loop reduction graph. Each node represents a topology, and 
each arrow a transition if a certain line is removed. The colour defines the 
topology action: white means the triangle or diamond rule, cyan the carpet 
rule, green the insertion rule, and red a custom reduction.}
\label{fig:reductionflow}
\end{figure}

In figure~\ref{fig:reductionflow} the reduction graph is displayed for 
three-loop massless propagator graphs. The names of the topologies are 
automatically generated. Every arrow denotes a transition that occurs when a 
propagator is removed. Multiply arrows could point to the same node if the 
resulting diagram is isomorphic. An example of this is \texttt{t1star5} 
(same as \texttt{t1star05} in section~\ref{sec:fullreductions}), where 
removing any of the four outside lines results in the same topology. The central 
line cannot be removed, since it has a non-integer power. The four-loop 
reduction graph, with over 400 nodes, is far too large to display. 

The reduction graph is generated from the top-level topologies down.
For every topology, a new one is generated where a particular line is missing,
For this new topology, we determine its action. Next, we generate a dot product basis that
is compatible with the action, e.g. for the insertion rule all dot products should
only involve at most one of the two momenta. We also determine its automorphisms (graph symmetries), so that
we can map every topology instance to the same unique form (we will go into more detail
about this in the next section). 

The dot product basis is chosen according to the following three rules:
1) it is compatible with the action,
2) it minimizes the number of terms created when rewriting from the parent topology. As a criterion
we choose the sum of the square of the number of terms that are created in rewriting each dot product.
3) The dot products are chosen in line with the symmetries of the topology.

We summarize the generation of the reduction graph in algorithm~\ref{alg:graphgen}:
\begin{center}
\begin{minipage}{.7\linewidth}
\begin{algorithm}[H]
\SetAlgoLined
\SetKwInOut{Input}{Input}
\SetKwInOut{Output}{Output}
\Input{top-level topologies $T$}
\Output{reduction tree $T_{\text{all}}$}

$T_{\text{all}} \leftarrow T$ \;
\lForEach{$t \in T$}{determine action and automorphisms}

\While{$T \neq \emptyset$}{
pop $t \in T$\;

\ForEach{propagator $p \in t$}{
  $h \leftarrow$ new topology without $p$ \;
  \eIf{$h' \in T_{\text{all}}$ isomorphic to $h$}{
   construct mapping from $h \to h'$\;
}{
determine action for $h$\;
generate dot product basis for $h$\;
generate automorphisms for $h$\;
generate mapping of dot products from $t$ to $h$\;
$T \leftarrow T + \{h\}$\;
$T_{\text{all}} \leftarrow T_{\text{all}} + \{h\}$\;
}
}
}
\caption{Reduction graph generation.}
\label{alg:graphgen}
\end{algorithm}
\end{minipage}
\end{center}

The reduction graph is generated with a Python script, using \texttt{igraph}~\cite{igraph} for a basic
graph representation of the topologies and for the isomorphism algorithm. 
Since by default only simple graphs (without self-edges and duplicate edges)
are supported by the isomorphism algorithm, we merge all double edges and use a custom function to determine if the topologies
are truly isomorphic (one could view duplicate edges with possible insertions as a special edge colouring). 
This function enforces that the number of duplicate edges is the same, and that the distribution of insertions
over duplicate edges is the same. Additionally, we generate all possible permutations over similar duplicate edges, 
to generate the edge isomorphisms. 
The reduction graph contains 438 topologies and requires $40\,000$ lines of \FORM{} code.

\subsection{Reduction graph execution}

So far, we have discussed the generation of the reduction graph. Now we consider how the graph
is processed in runtime.

As input, we have integrals that are labelled by the name of their topology in a symbol.
In contrast to \MINCER{}, the input expressions can contain multiple topologies.
In \FORCER{}, every topology is put in a separate expression and is hidden.
The topologies are processed one by one, in the order of the number of edges.
When a topology is treated, the expression is unhidden, 
the integrals are symmetrized using automorphisms, the 
topology action is executed, and finally, the resulting integrals are 
rewritten to their new topology. The topologies in the output are either 
master integrals, which require no further reductions, or topologies with 
fewer lines. These topologies will be merged into the designated expression 
for that topology. All the masters integrals are stored in their own 
expression.

After rewriting dot products, multiple edges could have vanished. Some of the integrals that
remain could have become massless tadpoles, which are zero in dimensional regularisation. A table is 
used that maps the topology and a list of missing edges to zero if the resulting topology is a tadpole.

The execution of the reduction graph is summarized in algorithm~\ref{alg:graphex}:
\begin{center}
\begin{minipage}{.7\linewidth}
\begin{algorithm}[H]
\SetAlgoLined
\SetKwInOut{Input}{Input}
\SetKwInOut{Output}{Output}
\Input{input integrals $I$}
\Input{reduction graph $T_{\text{all}}$}

convert $I$ to \FORCER{} topologies \;
\lForEach{$t \in I$}{ put in its own expression $E_t$ and deactivate}

\For{$l = 11$ to $1$}{
  \ForEach{$t \in T_{\text{all}}$ with $l$ edges}{
  activate expression with topologies $t$ \;
  symmetrize terms (apply automorphisms) \;
  perform reduction operation (triangle, carpet, etc.) \;
  rewrite result with missing lines to \FORCER{} topologies $h_i \in T_{\text{all}}$ \;
  move the terms with topology $h_i$ to $E_{h_i}$\;
}
}
\caption{Reduction graph execution.}
\label{alg:graphex}
\end{algorithm}
\end{minipage}
\end{center}

\subsection{Example}
Below we give an example of the treatment of a topology. 
The topology is depicted in figure~\ref{fig:Md366}, and is internally called \texttt{d366}.

\begin{figure}[htb]
\begin{minipage}{0.4\textwidth}
\centering
\begin{axopicture}{(160,60)(-50,-30)}
\Arc[arrow,flip](10,0)(20,0,180)
\Arc[arrow](10,0)(20,180,360)
\Arc[arrow,flip](60,0)(30,0,90)
\Arc[arrow,flip](60,0)(30,90,180)
\Arc[arrow](60,0)(30,180,230)
\Arc[arrow](60,0)(30,230,320)
\Arc[arrow,flip](60,0)(30,320,360)
\CArc[arrow,arrowpos=0.75](90,-60)(60,90,143)
\CCirc(70,-2){8}{White}{White}
\CArc[arrow,arrowpos=0.25](120,30)(60,180,233)
\Line[arrow](-30,0)(-10,0)
\Line[arrow](90,0)(110,0)
\Vertex(-10,0){1.5}
\Vertex(30,0){1.5}
\Vertex(90,0){1.5}
\Vertex(60,30){1.5}
\Vertex(84,-18){1.5}
\Vertex(42,-24){1.5}
\Text(10,12) {$p_1$}
\Text(10,-12) {$p_2$}
\Text(35,28) {$p_3$}
\Text(94,25) {$p_4$}
\Text(70,15) {$p_5$}
\Text(50,-2) {$p_6$}
\Text(30,-20) {$p_7$}
\Text(60,-37) {$p_8$}
\Text(97,-10) {$p_9$}
\end{axopicture}
\end{minipage}
\qquad
\begin{minipage}{0.45\textwidth}
\centering
\begin{align*}
n_{10} &= Q \cdot p_4 \qquad n_{13} = p_1 \cdot p_4\\
n_{11} &= Q \cdot p_6 \qquad n_{14} = p_1 \cdot p_6\\
n_{12} &= p_1 \cdot p_3\\
\end{align*}
\end{minipage}
\caption{\FORCER{} topology \texttt{d366}.}
\label{fig:Md366}
\end{figure}

In the input, the integral is represented by a compact notation in terms of symbols only:
\begin{verbatim}
	Md366/i1/i2/i3/i4^2/i5/i6/i7/i8/i9*i10*i11*i13;
\end{verbatim}
where \texttt{Md366} is the marker of the topology and the powers of $\texttt{i}_n$ represent the propagator and
numerator powers. In this example we have three additional powers: 
$1/p_4^2$, $Q \cdot p_4$, $p_1 \cdot p_6$, and $p_1 \cdot p_4$.
Since all rules are precomputed, the information of the topology such as the vertex structure, momentum flow, non-integer
powers of lines and which dot products are in the basis, is never stored in the terms that are processed.
Instead, the topology marker \texttt{Md366} will be used to call the correct routines.

When treating topology \texttt{d366}, we first apply symmetries to make sure that similar configurations of \texttt{d366}
are merged. We use the automorphisms of the graph, of which there are four:
$(p_1 \leftrightarrow p_2) \times (p_4 \leftrightarrow p_6, p_3 \leftrightarrow p_7, p_7 \leftrightarrow p_8)$. 
However, since there may be dot products in these momenta, the symmetry may be broken unless the set of dot products
maps into itself. For the symmetry $(p_1 \leftrightarrow p_2)$, the dot products $p_1 \cdot p_3$,
$p_1 \cdot p_4$, $p_1 \cdot p_6$ should be absent. The other symmetry can only be applied when $p_1 \cdot p_3$ is absent.

To find the smallest isomorphism, we hash the powers of the \texttt{i}, and take the smallest.
In code we have:
{\footnotesize
\begin{verbatim}
if (match(Md366*<1/i1^n1?$n1>*...*<1/i14^n14?$n14>));
  if (($n12==0)&&($n13==0)&&($n14==0));
    #call hash(0,$n14,$n13,$n12,$n11,$n10,$n9,$n8,$n7,$n6,$n5,$n4,$n3,$n1,$n2)
    #call hash(1,$n14,$n13,$n12,$n10,$n11,$n9,$n5,$n3,$n4,$n8,$n6,$n7,$n1,$n2)
  endif;
  if (($n12==0));
    #call hash(2,$n14,$n13,$n12,$n11,$n10,$n9,$n8,$n7,$n6,$n5,$n4,$n3,$n2,$n1)
    #call hash(3,$n13,$n14,$n12,$n10,$n11,$n9,$n5,$n3,$n4,$n8,$n6,$n7,$n2,$n1)
  endif;
* stores best hash in $bestiso
  #call smallesthash(0,1,2,3)
  if ($bestiso == 0); Multiply replace_(i1,i2,i2,i1);
  elseif ($bestiso == 1); Multiply sign_($n10+$n11+$n13+$n14)
      *replace_(i1,i2,i2,i1,i3,i7,i4,i6,i5,i8,i6,i4,i7,i3,i8,i5,i10,i11,i11,i10);
  elseif ($bestiso == 3); Multiply sign_($n10+$n11+$n13+$n14)
      *replace_(i3,i7,i4,i6,i5,i8,i6,i4,i7,i3,i8,i5,i10,i11,i11,i10,i13,i14,i14,i13);
  endif;
endif;
\end{verbatim}
}

The action that will be performed in \texttt{d366} is the integration of
the left bubble, $p_1$ and $p_2$. As can be seen in figure~\ref{fig:Md366}, all relevant dot products are written only in terms of
$p_1$ and none in terms of $p_2$, in alignment with the insertion rule. The dot products that involve $p_1$ can
all be re-expressed in terms of inverse propagators after integrating the insertion. 
The two dot products that remain, $Q \cdot p_4$,
and $Q \cdot p_6$ (represented by \texttt{i10} and \texttt{i11} respectively) have to be rewritten to the new topology.

The new topology is called \texttt{d118}:

\begin{minipage}{0.4\textwidth}
\centering
\begin{axopicture}{(100,80)(-20,-40)}
\Arc[arrow,flip](60,0)(30,0,90)
\Arc[arrow,flip](60,0)(30,90,180)
\Arc[arrow](60,0)(30,180,230)
\Arc[arrow,flip](60,0)(30,230,320)
\Arc[arrow,flip](60,0)(30,320,360)
\CArc[arrow,arrowpos=0.75](90,-60)(60,90,143)
\CCirc(70,-2){8}{White}{White}
\CArc[arrow,arrowpos=0.3](120,30)(60,180,233)
\Line[arrow](10,0)(30,0)
\Line[arrow](90,0)(110,0)
\Vertex(30,0){1.5}
\Vertex(90,0){1.5}
\Vertex(60,30){1.5}
\Vertex(84,-18){1.5}
\Vertex(42,-24){1.5}
\Text(35,28) {$p_2$}
\Text(94,25) {$p_1$}
\Text(70,15) {$p_7$}
\Text(50,-2) {$p_6$}
\Text(30,-20) {$p_3$}
\Text(60,-37) {$p_4$}
\Text(97,-10) {$p_5$}
\end{axopicture}
\end{minipage}
\begin{minipage}{0.4\textwidth}
\centering
\begin{align*}
n_{8} = Q \cdot p_4\\
n_{9} = Q \cdot p_7\\
\end{align*}
\end{minipage}
\vspace{0.5cm}

where we have suppressed the $\epsilon$ power of the external line.

Below is the mapping from \texttt{d366} to \texttt{d118}, which includes rewriting the old dot products:
\begin{verbatim}
Multiply replace_(i3,j2,i4,j1,i5,j7,i6,j6,i7,j3,i8,j4,i9,j5);
id i10 =  Q^2/2+j2/2-j3/2-j9;
id i11 = -Q^2/2+j2/2-j3/2+j8;
Multiply replace_(Md366,Md118,<j1,i1>,...,<j7,i7>,j8,-i8,j9,i9);
\end{verbatim}

%--#] Superstructure : 
%--#[ The library :

\section{Usage}
\label{sec:library}

The \FORCER{} program can be downloaded from
\url{https://github.com/benruijl/forcer}. 
Currently, the latest development version of \FORM{} is required, which can be
obtained from \url{https://github.com/vermaseren/form}.
The generation scripts require Python~2.7, Python~3 or higher
as well igraph~\cite{igraph}, numpy~\cite{DBLP:journals/corr/abs-1102-1523} and
sympy~\cite{sympy}.

An example of \FORCER{} input is the following program:
{\small
\begin{verbatim}
#-
#include forcer.h

L F =
 +1/<p1.p1>/.../<p6.p6>*Q.p3*Q.p4*vx(Q,p1,p5,p6)*vx(-p1,p2,p3)
                       *vx(-p5,-p6,p4)*vx(-Q,-p2,-p3,-p4)
 +1/<p1.p1>/.../<p5.p5>*vx(-Q,p2,p3)*vx(p1,-p2,p5)*vx(-p1,p4,Q)
                       *vx(-p3,-p4,-p5)*ex(p1,p4)
;

#call Forcer(msbarexpand=4)
B ep;
P +s;
.end
\end{verbatim}
}

After \texttt{forcer.h} is included, the input integral can be defined. 
This is done by specifying the vertex structure using \texttt{vx}. The external
momentum should be called \texttt{Q}. The propagators and momenta can simply be multiplied in,
as shown in the example above. Insertions on lines can be specified using the \texttt{ex} function.
In the second integral above \texttt{ex(p1,p4)} means that there is a single $\epsilon$ on the propagator
associated with momentum $p_1$, and one on $p_4$. The topologies will automatically be matched to
\FORCER{}'s internal topologies. The dot products will also
automatically be rewritten (see subsection~\ref{sec:momsubs}).

By calling \texttt{Forcer}, the integrals{} are computed. The optional argument \texttt{msbarexpand}
can give the (unrenormalized) answer expanded in \MSbar{}. Otherwise, the result will be given exactly in terms of
the master integrals and rational coefficients (see appendix~\ref{sec:masterintegrals}).
Other options include \texttt{polyratfunexpand=div} and
\texttt{polyratfunexpand=maxpow},
which enable the expansions of rational coefficients in $\epsilon$
at intermediate steps
using the \FORM{} statement \texttt{PolyRatFun}
(see appendix.~\ref{sec:Expansions}).

Alternatively, one could execute each of the tree transition steps to \FORCER{} individually: first, the topologies need to be matched. Next, the momentum substitution to the \FORCER{} topology basis for a given topology \texttt{`TOPO'} has
to be performed. Finally, the reduction is executed. The steps are sketched below:
\begin{verbatim}
#call loadTopologies()
#call matchTopologies(1)

#call momsubs(`TOPO',1,1)

#call DoForcer()
\end{verbatim}

%--#] The library : 
%--#[ From Qgraf to Forcer :

\section{From physical diagrams to \FORCER{}}
\label{sec:qgrafforcer}

The interface provided in the previous section expects scalar integrals
as input.
In order to compute Feynman diagrams, process-specific preprocessing has to
be performed.
Below we discuss some general optimizations that could be applied there. 
Since the actual implementation is highly dependent
on conventions, we will only sketch certain parts.

The program Qgraf~\cite{QGRAF} provides a convenient way to generate the Feynman graphs 
that are needed for the actual calculations, because it can generate \FORM{} 
compatible output. However, the challenge remains of converting the 
diagrams as presented by Qgraf to something that the \FORCER{} program can 
deal with. This involves mapping the topology and momenta of the 
diagrams to \FORCER{}'s internal notation. 
For this purpose, the Python program that generates the reduction graph also 
generates a file called \texttt{notation.h} which contains a specification of all 
topologies in such a way that a conversion program can use it for
\begin{enumerate}
\item topology recognition,
\item labelling the momenta and their directions for each line,
\item using symmetry transformations.
\end{enumerate}
Each topology is represented by a term in \FORM{} notation. Two typical 
terms are
\begin{verbatim}
  +vx(-Q,p4,p5)
   *vx(p3,-p4,p11)
   *vx(p6,p7,p10)
   *vx(p2,-p3,-p10)
   *vx(p1,-p2,p9)
   *vx(-p5,-p6,-p9)
   *vx(-p7,p8,-p11)
   *vx(-p1,-p8,Q)
   *SYM()
   *SYM(Q,-Q,p1,-p5,p2,p6,p3,-p7,p4,-p8,p5,-p1,p6,p2,p7,-p3,p8,-p4
                          ,p9,-p9,p10,-p10,p11,-p11)
   *SYM(Q,-Q,p1,-p4,p2,-p3,p3,-p2,p4,-p1,p5,-p8,p6,p7,p7,p6,p8,-p5
                          ,p9,p11,p11,p9)
   *SYM(p1,p8,p2,p7,p3,-p6,p4,p5,p5,p4,p6,-p3,p7,p2,p8,p1,p9,-p11
                          ,p10,-p10,p11,-p9)
   *TOPO(Mno2)
  
  +vx(-Q,p3,p4)
   *vx(p2,-p3,p7)
   *vx(p1,-p2,p6)
   *vx(-p1,p5,Q)
   *vx(-p4,-p5,-p6,-p7)
   *ex(p2)
   *SYM()
   *SYM(Q,-Q,p1,-p3,p2,-p2,p3,-p1,p4,p5,p5,p4,p6,p7,p7,p6)
   *TOPO(Mfastar2)
\end{verbatim}
The first term indicates the \texttt{no2} topology. The function \texttt{vx} indicates the 
vertices and the momenta belonging to that vertex. Negative momenta are 
incoming. The function \texttt{TOPO} has a symbol as an argument that indicates 
the topology. In the \FORCER{} program terms that are in the notation of a 
given topology are labelled with one power of the corresponding symbol. The 
function SYM describes a symmetry operation of the topology. The \FORM{} 
statement
\begin{verbatim}
    id,once,SYM(?a) = replace_(?a);
\end{verbatim}
will execute such an operation. In practice one could use it in the 
following way:
\begin{verbatim}
    id  vx(?a) = f1(vx(?a));
    repeat id f1(x1?)*f1(x2?) = f1(x1*x2);
    repeat id SYM(?a)*f1(x?) = f1(x)*f2(x*replace_(?a));
    id  f1(x1?)*f2(x2?) = x2;
    id  f2(x?) = 1;
\end{verbatim}
This process makes for each occurrence of the function \texttt{SYM} a copy of the 
contents of the function \texttt{f1} in which the corresponding symmetry operation 
has been applied. Because the normal ordering algorithm of \FORM{} puts the 
smallest of the functions \texttt{f2} first, we end up with the smallest 
representation of the term. If this is applied at a later stage in the 
program more statements may be needed, because there may be more objects 
than \texttt{vx}.

The notation file includes more topologies than actually exist in
the \FORCER{} reduction graph, since physical diagrams can have duplicate momenta.
If this is the case, the term in the notation file also contains a function
\texttt{ID}, for example \texttt{ID(p4,-p5)}, indicating that $p_4$ and $-p_5$ are actually
the same momentum.
After the topology is matched and the labelling is done, the \texttt{ID} function
can be applied: 
\begin{verbatim}
id ID(p1?,p2?) = replace_(p1,p2);
\end{verbatim}

The first step in determining the topology of a diagram is to read the 
\texttt{notation.h} file, number its topologies, and store each 
of them in a dollar variable with a name that is labelled by this number. 
We also store the names of the topologies in such an array of dollar 
variables. The topology of a diagram can now be determined by trying 
whether one of the topologies can be substituted in the term. If this 
pattern matching involves wildcards, and the match of the wildcards is 
stored inside dollar variables we can use this to relabel the diagram 
itself and bring it to the notation of the topology. The main problem is 
creating the match structure, since we need wildcards for all the momenta 
followed by the name of a dollar variable. This issue is 
resolved with the dictionary feature of \FORM{}. The essential part of the 
code is:
\begin{verbatim}
    #OpenDictionary wildmom
        #do i = 1,`$MAXPROPS'
            #add p`i': "p`i'?$p`i'"
        #enddo
    #CloseDictionary

    #do i = 1,`$MAXPROPS'
        $p`i' = p`i';
    #enddo

    #UseDictionary wildmom($)
    #do i = 1,`$numtopo'
        if ( match(`$topo`i'') );
            $toponum = `i';
            goto caught;
        endif;
    #enddo
    #CloseDictionary
    label caught;

    Multiply replace_(Q,Q,<$p1,p1>,...,<$p`$MAXPROPS',p`$MAXPROPS'>)*
                  topo($toponum);
\end{verbatim}
When we try to match, the printing of the \verb:`$topo`i'': variable 
will result in objects like \verb:vx(p1?$p1,p2?$p2,p3?$p3)*...: rather than 
the \verb:vx(p1,p2,p3)*...: that it actually contains. This way the 
\$-variables get the value of the momenta in the diagram for which we want to 
determine the topology and the notation. The final \texttt{replace} substitutes these 
momenta by the value they have in the topology file.

It is possible to speed up the process considerably by hashing the 
topologies by the number of vertices and by first stripping the signs of 
the momenta. These signs can be recovered in a later step.

\subsection{Self-energy filtering}
Another optimization is to filter self-energy insertions from the Qgraf output.
Here we present an algorithm that can detect one particle {\it 
reducible} propagator insertions.
\begin{enumerate}
\item Select a representative for a one-loop propagator. A representative 
is a single diagram that occurs in this propagator. For the ghost and the 
quark propagators this is trivial, since there is only a single diagram. For 
the gluon we select the diagram with the ghost loop (not forgetting the 
minus sign).
\item In the propagators we indicate the number of loops with an extra 
parameter. Adjacent loop representatives are combined and their number of 
loops is the sum of those parameters. This means that the representative of 
a three-loop gluon propagator is a chain of three one-loop diagrams, each 
with a ghost loop.
\item Next we make a copy of all remaining vertices into a function \texttt{acc}. In 
this function we remove all vertices that have an external line.
\item In the function \texttt{acc} we start selecting one vertex in all possible 
ways.
\item If this special vertex has more than two lines, it `consumes' in all 
possible ways one of its neighbouring vertices, removing the connecting 
momentum. If the same momentum connects twice to the new vertex, it is 
removed as well.
\item We keep doing this until either the super-vertex in one of the terms 
has two lines remaining in which case we can eliminate the whole diagram as 
it is part of a propagator, or we cannot remove any more lines. If all 
possibilities end in the last way we keep the diagram.
\end{enumerate}
Let us show this diagrammatically for a non-trivial diagram:

%--#[ Example :
\begin{center}
\begin{axopicture}(100,60)(0,0)
\Line(5,30)(15,30)
\Line(85,30)(95,30)
\Line(40,55)(60,55)
\Line(40,5)(60,5)
\Line(50,55)(50,45)
\Line(50,5)(50,15)
\Line(35,30)(65,30)
\Arc(40,30)(25,90,270)
\Arc(60,30)(25,270,90)
\Arc(50,30)(15,0,360)
\end{axopicture}
\begin{axopicture}(30,60)(0,0)
\Line[arrow,arrowpos=1](5,30)(25,30)
\Text(15,35)[b]{Step 3}
\end{axopicture}
\begin{axopicture}(50,60)(0,0)
\Line(5,55)(45,55)
\Line(5,5)(45,5)
\Line(25,55)(25,45)
\Line(25,5)(25,15)
\Line(10,30)(40,30)
\Arc(25,30)(15,0,360)
\end{axopicture}
\end{center}

\begin{center}
\begin{axopicture}(40,60)(0,0)
\Line[arrow,arrowpos=1](0,30)(17,30)
\Text(8.5,36)[b]{Step 4}
\Text(25,30)[l]{$2\times$}
\end{axopicture}
\begin{axopicture}(50,60)(0,0)
\Line(5,55)(45,55)
\Line(5,5)(45,5)
\Line(25,55)(25,45)
\Line(25,5)(25,15)
\Line(10,30)(40,30)
\Arc(25,30)(15,0,360)
\Vertex(25,55){2}
\end{axopicture}
\begin{axopicture}(20,60)(0,0)
\Text(1,30)[l]{$+2\times$}
\end{axopicture}
\begin{axopicture}(50,60)(0,0)
\Line(5,55)(45,55)
\Line(5,5)(45,5)
\Line(25,55)(25,45)
\Line(25,5)(25,15)
\Line(10,30)(40,30)
\Arc(25,30)(15,0,360)
\Vertex(25,45){2}
\end{axopicture}
\begin{axopicture}(20,60)(0,0)
\Text(1,30)[l]{$+2\times$}
\end{axopicture}
\begin{axopicture}(50,60)(0,0)
\Line(5,55)(45,55)
\Line(5,5)(45,5)
\Line(25,55)(25,45)
\Line(25,5)(25,15)
\Line(10,30)(40,30)
\Arc(25,30)(15,0,360)
\Vertex(10,30){2}
\end{axopicture}
\end{center}

\begin{center}
\begin{axopicture}(40,60)(0,0)
\Line[arrow,arrowpos=1](0,30)(17,30)
\Text(8.5,36)[b]{Step 5,1}
\Text(25,30)[l]{$4\times$}
\end{axopicture}
\begin{axopicture}(50,60)(0,0)
\Line(5,55)(45,55)
\Line(5,5)(45,5)
\Line(25,5)(25,25)
\Line(10,40)(40,40)
\Arc(25,40)(15,0,360)
\Vertex(25,55){2}
\end{axopicture}
\begin{axopicture}(20,60)(0,0)
\Text(1,30)[l]{$+8\times$}
\end{axopicture}
\begin{axopicture}(50,60)(0,0)
\Line(5,55)(45,55)
\Line(5,5)(45,5)
\Line(25,55)(25,45)
\Line(25,5)(25,15)
\Arc(25,30)(15,0,360)
\Arc(40,45)(15,180,270)
\Vertex(25,45){2}
\end{axopicture}
\begin{axopicture}(20,60)(0,0)
\Text(1,30)[l]{$+2\times$}
\end{axopicture}
\begin{axopicture}(50,60)(0,0)
\Line(5,55)(45,55)
\Line(5,5)(45,5)
\Line(25,55)(25,45)
\Line(25,5)(25,15)
\Arc(25,37.5)(7.5,0,360)
\Arc(25,22.5)(7.5,0,360)
\Vertex(25,30){2}
\end{axopicture}
\end{center}

\begin{center}
\begin{axopicture}(40,60)(0,0)
\Line[arrow,arrowpos=1](0,30)(17,30)
\Text(8.5,36)[b]{Step 5,2}
\Text(25,30)[l]{$8\times$}
\end{axopicture}
\begin{axopicture}(50,60)(0,0)
\Line(5,55)(45,55)
\Line(5,5)(45,5)
\Line(25,5)(25,25)
\Arc(25,40)(15,0,360)
\Arc(40,55)(15,180,270)
\Vertex(25,55){2}
\end{axopicture}
\begin{axopicture}(26,60)(0,0)
\Text(1,30)[l]{$+24\times$}
\end{axopicture}
\begin{axopicture}(50,60)(0,0)
\Line(5,55)(45,55)
\Line(5,5)(45,5)
\Line(25,55)(25,40)
\Line(25,5)(25,20)
\Arc(25,30)(10,0,360)
\Vertex(25,40){2}
\end{axopicture}
\begin{axopicture}(20,60)(0,0)
\Text(1,30)[l]{$+8\times$}
\end{axopicture}
\begin{axopicture}(50,60)(0,0)
\Line(5,55)(45,55)
\Line(5,5)(45,5)
\Line(20,55)(20,5)
\Line(20,30)(40,30)
\Arc(30,30)(10,0,360)
\Vertex(20,30){2}
\end{axopicture}
\end{center}

\begin{center}
\begin{axopicture}(45,60)(0,0)
\Line[arrow,arrowpos=1](0,30)(17,30)
\Text(8.5,36)[b]{Step 5,3}
\Text(25,30)[l]{$56\times$}
\end{axopicture}
\begin{axopicture}(50,60)(0,0)
\Line(5,55)(45,55)
\Line(5,5)(45,5)
\Line(25,5)(25,35)
\Arc(25,45)(10,0,360)
\Vertex(25,55){2}
\end{axopicture}
\begin{axopicture}(26,60)(0,0)
\Text(1,30)[l]{$+32\times$}
\end{axopicture}
\begin{axopicture}(50,60)(0,0)
\Line(5,30)(45,30)
\Line(5,5)(45,5)
\Line(25,5)(25,50)
\Arc(25,40)(10,0,360)
\Vertex(25,30){2}
\end{axopicture}
\begin{axopicture}(26,60)(0,0)
\Text(1,30)[l]{$+32\times$}
\end{axopicture}
\begin{axopicture}(50,60)(0,0)
\Line(5,30)(45,30)
\Line(5,5)(25,30)
\Line(45,5)(25,30)
\Line(25,30)(25,50)
\Arc(25,40)(10,0,360)
\Vertex(25,30){2}
\end{axopicture}
\begin{axopicture}(26,60)(0,0)
\Text(1,30)[l]{$+72\times$}
\end{axopicture}
\begin{axopicture}(50,60)(0,0)
\Line(5,45)(45,45)
\Line(5,15)(45,15)
\Line(25,45)(25,15)
\Vertex(25,30){2}
\end{axopicture}
\end{center}

\begin{center}
\begin{axopicture}(50,40)(0,0)
\Line[arrow,arrowpos=1](0,20)(20,20)
\Text(8.5,26)[b]{Step 5,4}
\end{axopicture}
\begin{axopicture}(50,40)(0,0)
\Line[arrow,arrowpos=1](0,20)(17,20)
\Text(8.5,26)[b]{Step 5,5}
\Text(25,20)[l]{$304\times$}
\end{axopicture}
\begin{axopicture}(50,40)(0,0)
\Line(5,35)(25,20)
\Line(25,20)(45,35)
\Line(5,5)(25,20)
\Line(25,20)(45,5)
\Vertex(25,20){2}
\end{axopicture}
\begin{axopicture}(26,40)(0,0)
\Text(1,20)[l]{$+72\times$}
\end{axopicture}
\begin{axopicture}(50,40)(0,0)
\Line(5,35)(45,35)
\Line(5,5)(45,5)
\Line(25,35)(25,5)
\Vertex(25,20){2}
\end{axopicture}
\end{center}

%--#] Example : 

In the example, the diagram can be eliminated at the moment the super-vertex 
with just two lines appears. This is at step 5,3. We did not stop at that 
point because we wanted to show how the other possibilities develop for 
diagrams that would survive.

The above algorithm can be programmed rather easily in \FORM{} with the new 
id,all option of the id statement. For instance step 4 is just the 
statement
\begin{verbatim}
    id,all,v(?a) = w(?a);
\end{verbatim}
in which \texttt{v} represents the vertices and \texttt{w} is the super-vertex. This is 
followed by a symmetrization to reduce the number of different diagrams.
A complete procedure that works for all types of diagrams, independent of 
the number of external lines or loops contains 30 \FORM{} statements. The 
elimination of insertions simplifies the calculation considerably, because 
multi-loop gluon propagator insertions have many diagrams. This is 
particularly important when calculating moments of splitting and 
coefficient functions in DIS.

\subsection{Colour split-off}
We split each diagram in its colour part and its `Lorentz' part
before applying the Feynman rules. The 4-gluon vertex is split up
into three terms with their own overall colour factor.
Technically it is not required to do the split-off at this stage,
but the remaining program will be considerably faster when the colour
is a global factor.

To compute the colour factor we use a modified version of the \texttt{color} package 
of ref.~\cite{vanRitbergen:1998pn}, which is available on the \FORM{} 
pages (http://www.nikhef.nl/$\sim$form). 
It has been observed that even when one may have $100\,000$ diagrams or more, 
there are usually at most a few hundred different colour factors to be worked out. 
Hence the way to process these factors is by pulling all colour objects into 
a function \texttt{color} and then, after using colour projectors on the external 
lines, only working out the colour bracket:
\begin{verbatim}
   Normalize color;
   B   color;
   .sort: Prepare color;
   Keep brackets;
   Argument color;
      #call color
      #call simpli
   EndArgument;    
\end{verbatim}

By replacing every \texttt{.sort} by 
\begin{verbatim}
   #procedure SORT(text)
   EndArgument;
   B   color;
   .sort:`text';
   Keep Brackets;
   Argument color;
   #endprocedure
\end{verbatim}
we guarantee that each different colour object is worked out only once.

\subsection{Diagram database}
Diagrams with the same topology and colour factor are grouped together in
superdiagrams.
The superdiagrams provide a convenient way to distribute the work over
multiple computers. This grouping can speed up the calculation by a 
modest factor (typically ${\cal O}(3)$).

We use the minos database program provided (with its source code) in the 
\FORM{} pages to store the superdiagrams. After each superdiagram is computed,
it is multiplied with its colour factors. Finally, the values of all superdiagrams
are added. Only at this stage do we substitute 
the formulas for the insertion propagators and the master integrals. Up 
until the substitution of the master integrals the results are exact to all 
orders in $\epsilon$ if one uses rational polynomials in $\epsilon$ for the 
coefficients of the terms.

\subsection{Momentum substitutions}
\label{sec:momsubs}

After the Feynman rules have been applied, the integrals are in a form
in which they can be converted to \FORCER{}'s basis for the topologies. 
The reduction to this basis needs to be done with great care as it is very easy to 
generate an extremely large number of terms. This process is split up in two
components: rewriting the momenta to a momentum basis and rewriting the dot products
to \FORCER{}'s basis. 

The momentum basis should contain all the momenta of the irreducible dot products belonging to this
\FORCER{} topology. The other basis elements are obtained by an exhaustive search that tries to 
minimize the number of terms that will be created when rewriting to the basis. The optimization criterion is the 
sum of the square of the number of terms that get created for all the momentum and dot product rewrites.

In order to prevent a blow-up in the number of terms, we create a layered rewrite of momenta. This layering
is constructed automatically and makes the momentum rewrites order dependent:

\begin{center}
\begin{tabular}{c c c}
\hspace{5mm}
\begin{minipage}{0.45\textwidth}
\begin{verbatim}
p9.p?!{p9,}=+p2.p+p7.p+p11.p-Q.p;
p5.p?!{p5,}=-p11.p-p3.p+Q.p;
p6.p?!{p6,}=-p2.p+p3.p-p7.p;
p1.p?!{p1,p4}=+Q.p-p8.p;
p10.p?!{p10,}=+p2.p-p3.p;
p4.p?!{p4,p1}=+p11.p+p3.p;
p7.p?!{p7,}=+p8.p-p11.p;
\end{verbatim}
\end{minipage}
&
$\displaystyle \rightarrow$
&
\begin{minipage}{0.40\textwidth}
\begin{verbatim}
p9.p?!{p9,}=-p6.p-p5.p;
p5.p?!{p5,}=-p4.p+Q.p;
p6.p?!{p6,}=-p10.p-p7.p;
p1.p?!{p1,p4}=+Q.p-p8.p;
p10.p?!{p10,}=+p2.p-p3.p;
p4.p?!{p4,p1}=+p11.p+p3.p;
p7.p?!{p7,}=+p8.p-p11.p;
\end{verbatim}
\end{minipage}
\end{tabular}
\end{center}

Because some terms will merge during the momentum rewrites, the layered approach is much faster.
Note that dot products will not be rewritten if they are elements of the dot product basis.

Finally, the dot products are rewritten, straight to the internal \FORCER{} notation:
\begin{center}
\begin{minipage}{13cm}
\begin{verbatim}
id Q.p1  = Q^2/2+i1/2-i8/2;
id p1.p2 = i1/2+i2/2-i9/2;
id p2.p3 = -i10/2+i2/2+i3/2;
id Q.p4  = Q^2/2+i4/2-i5/2;
id p3.p4 = -i11/2+i3/2+i4/2;
id p1.p3 = -Q^2/2+i11/2+i13+i14-i4/2+i5/2-i7/2+i8/2;
id p2.p4 = -Q^2/2-i1/2+i12+i13+i5/2-i6/2+i8/2+i9/2;
\end{verbatim}
\end{minipage}
\end{center}

We note that in the actual code there will be \texttt{.sort} statements between the
id statements and that there are extra optimizations in place to prevent excessive term
generation.

%--#] From Qgraf to Forcer : 
%--#[ Examples and performance :

\section{Examples and performance}
\label{sec:performance}

The \FORCER{} program has recently been used in many large calculations. As a first demonstration
of its capabilities, the four-loop QCD beta function has been recomputed~\cite{Ueda:2016sxw,Ueda:2016yjm},
and it agrees with refs.~\cite{vanRitbergen:1997va,Czakon:2004bu}.

Since \FORCER{} can compute the finite pieces to any power of $\epsilon$, the QCD propagators
and three-point vertices with one nullified momentum could straightforwardly be computed
\cite{Ruijl:2017prop}. The most expensive computation was the triple-gluon vertex,
which took one week on a single machine with 24 cores.

The \FORCER{} program has also been extensively used in the computation of Mellin moments
of structure functions~%
\cite{Ruijl:2016pkm,Davies:2016jie,Moch2017}. For the non-singlet case, up to six 
Mellin moments have been computed.
For the simpler leading and sub-leading $n_f$ diagrams, up to the 40th Mellin moment was computed, which
allowed for an analytical reconstruction. Even though most of these diagrams are essentially three loops,
the complexity of the integrals, as defined in section~\ref{sec:reductions}, is more than 80.
Using the OPE method, the moments of the non-singlet splitting functions have
been computed up to $N=16$; in the large-$N_c$ limit $N=19$ has been reached.
The latter result has proven sufficient for
reconstructing and validating their all-$N$ form~%
\cite{Moch2017,RuijlZurich}. From this quantity
the four-loop cusp anomalous dimension in the large-$N_c$ limit has been computed, and
it agrees with the calculation of the same quantity via the quark form factor~\cite{Lee:2016ixa}.
The high complexities integrals involved in the aforementioned computations exceed what
current Laporta-style methods can compute by a large margin. Even with \FORCER{}, some computations can take weeks.
In order to improve performance, the momentum basis was chosen in such a way that it aligns
with the $P$-flow through the diagram. Additionally, expansions in $\epsilon$ were used (see appendix~\ref{sec:Expansions}). 

Most recently, the \FORCER{} program was essential in computing the five-loop beta function for
Yang-Mills theory with fermions~\cite{Herzog:2017ohr},
which confirmed the SU(3) result of~\cite{Baikov:2016tgj}.
To compute the poles of five-loop diagrams, infrared rearrangement was used to rewrite any integral to a carpet
integral~\cite{Herzog:2017bjx}. This process generates a large number of counterterms with dot products. 
The complete computation took three days on a cluster.

Below we demonstrate some benchmarks of the \FORCER{} program. We start with some specific configurations,
displayed in table~\ref{tab:bench1}. We have chosen top-level topologies for the benchmark, since these are the most time-consuming ones. In their reduction,
many other master topologies (and thus custom reductions) are encountered. The topology la4 is the four-loop ladder
topology.
\begin{table}[ht]
\centering
\begin{tabular}{c|c|c}
ID & Configuration & Time (s)\\
\hline
no1   & $Z(-14;2,2,2,2,2,2,2,2,2,2,2,-1,-1,-1)$ & 10476\\
no2   & $Z(-14;2,2,2,2,2,2,2,2,2,2,2,-1,-1,-1)$ & 147 \\
haha  & $Z(-14;2,2,2,2,2,2,2,2,2,2,2,-1,-1,-1)$ & 338 \\
la4   & $Z(-14;2,2,2,2,2,2,2,2,2,2,2,-1,-1,-1)$ & 68 \\
no2   & $Z(-17;2,2,2,2,2,2,2,2,2,2,2,-2,-2,-2)$ & 370 \\
la4   & $Z(-20;2,2,3,2,2,2,2,3,2,2,2,-2,-2,-3)$ & 2848 \\
haha  & $Z(-20;2,1,2,2,1,2,1,2,2,2,2,-4,-4,-4)$ & 12943 \\
la4   & $Z(-20;2,1,2,2,1,2,1,2,2,2,2,-4,-4,-4)$ & 117906 \\
\end{tabular}
\caption{Benchmark for several specific configurations, using 4 cores.}
\label{tab:bench1}
  % tform -w4, maxtermsize=100K, on lippe with faststore,
  % FORM 4.1 (Apr 16 2017, v4.1-20131025-333-g41642ab) 64-bits,
  % Forcer 3359e5c.
\end{table}

\begin{figure}[tb]
\centering
\includegraphics[scale=0.8]{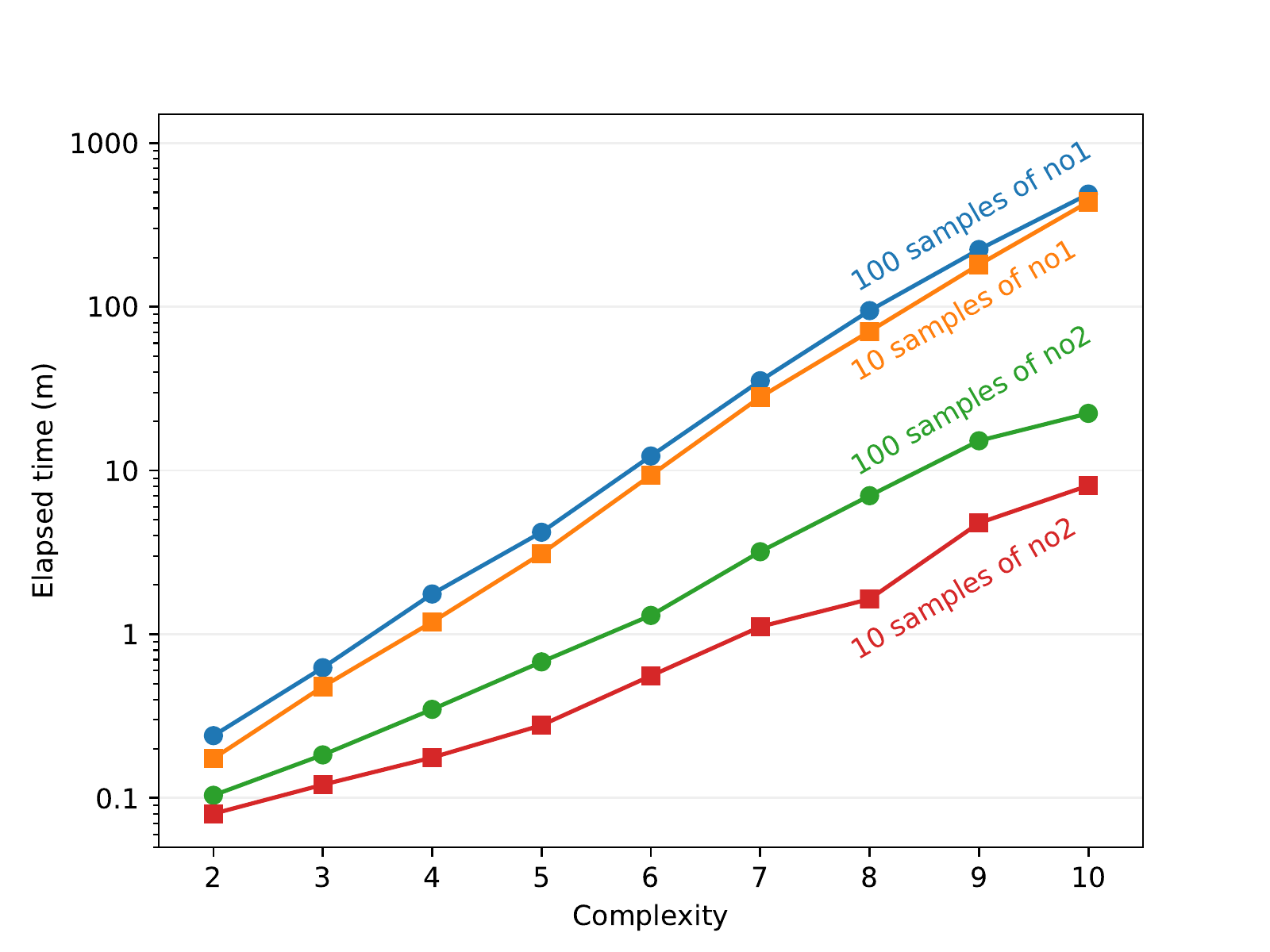} % TODO: do lines or just data point and a fit?
\caption{A benchmark for the complete reduction of \texttt{no1} and \texttt{no2} configurations, using 4 cores. The line with the
dots indicates the joint computation time of 100 sampled configurations, the line with the squares
the computation time of 10 samples. Even though 10 times more integrals are computed, the computation time
is only 20\% longer. The scaling in complexity is exponential: each increase in complexity increases the computation time by
$2.5$.}
\label{fig:bench}
\end{figure}

\begin{figure}[tb]
  \centering
  \includegraphics[width=0.8\textwidth]{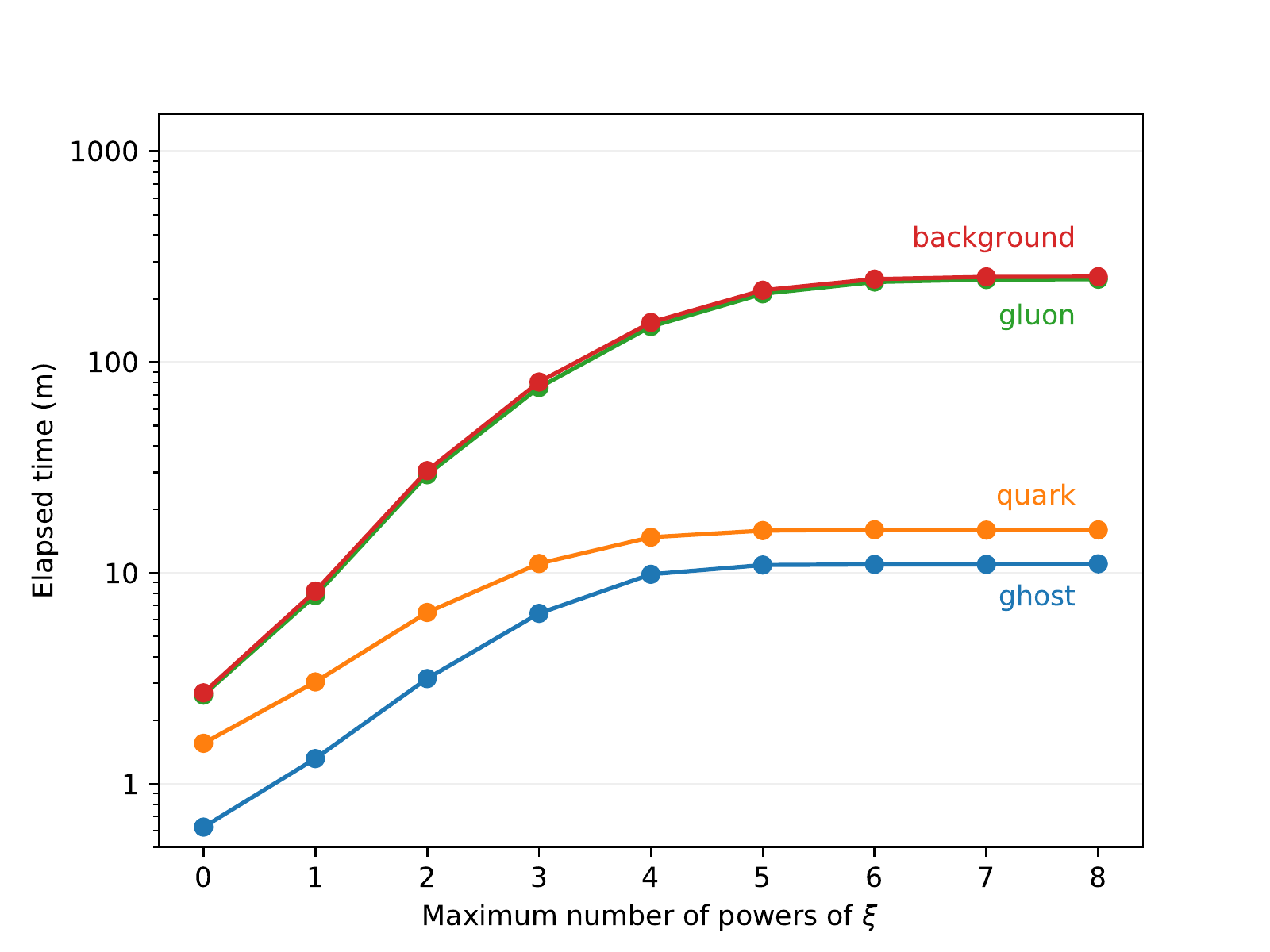}
  \caption{%
    A benchmark for computing four-loop QCD self-energies (ghost, quark, gluon
    and background-gluon) on a 32-core machine.
    The two curves for the gluon and background-gluon almost coincide.
  }
  % 8x4, on lippe, with faststore,
  % FORM 4.1 (Apr  4 2017, v4.1-20131025-320-g5b55fdb) 64-bits.
  \label{fig:bench-props}
\end{figure}

Next, we compute samples of configurations with a specific complexity 
of the top-level non-planar master integral \texttt{no1} and \texttt{no2}.
In figure~\ref{fig:bench}, we show the total wall-clock time of computing 10 and 100 samples for a given complexity
at the same time, using 4 cores. We observe that even though the difference in number of samples is a factor 10, 
the computation time increases only by about 20\%. 
This demonstrates that the \FORCER{} program makes use of symmetries and grouping,
which cause shared configurations
deeper in the reduction process to merge. 
Additionally, the graph shows that the computation time scales exponentially in the complexity
with a base of about $2.5$.

Finally, figure~\ref{fig:bench-props} shows the timings for computing four-loop QCD
self-energies for a certain maximum power of the (unrenormalized) gauge parameter $\xi$.
Here $\xi=0$ corresponds to the Feynman gauge.
In our setup, all techniques discussed in section~\ref{sec:qgrafforcer} are
applied.

The background field propagator in figure~\ref{fig:bench-props}
can be used to obtain the beta function without computing an
additional propagator and a vertex~\cite{Abbott80,AbbottGS83}.
Interestingly, the curve for the background-gluon is quite similar to
that for the gluon, even though one may expect that the background-gluon to be more
time consuming than the gluon propagator because of extra vertices.
The high performance can be understood by the fact that we are using superdiagrams;
as we have seen in figure~\ref{fig:bench}, the increase of the number of terms
does not matter much, provided complexities of integrals are similar, and
there are many chances for merges and cancellations of coefficients of
the integrals at intermediate stages in the reduction.

Using the background field method, we are able to compute the 
four-loop beta function for Yang-Mills theory with fermions
in less than three minutes in the Feynman gauge, and in about four hours for
all powers of the gauge parameter on a single machine with 32 cores.

%--#] Examples and performance : 
%--#[ Conclusions :
\section{Conclusions and outlook}
\label{sec:conclusions}

We have shown how the \FORCER{} program has been constructed, what algorithms 
it uses and demonstrated its performance. In addition, we have 
summarized how \FORCER{} may be used for computing physical diagrams. The 
predominantly automatic construction of the program is not limited 
to four loops. We have run the fully automatic pieces for a five-loop 
program. Even though there are over 200 missing topology actions that require a custom
reduction (and the masters integrals have not yet been determined), 
about 30\% of the diagrams of the gluon propagator can be computed with this
five-loop \FORCER{} equivalent.
This means that if in the future the master topologies can be worked out 
automatically as well, a five-loop program could be constructed shortly after. 
The idea is quite challenging: the number of parameters that have
to be reduced, grows from 14 at four loops to 20 at five loops.

The \FORCER{} program has already been used for some large calculations
at the four- and five-loop levels. For Yang-Mills theory with fermions, 
the propagators and
three-point functions with one vanishing momentum have been determined
exactly in terms of master integrals and with a rational coefficient in $\epsilon$%
~\cite{Ruijl:2017prop}.
Recently, the \FORCER{} program was utilized in the computation of the 
five-loop beta function with generic colour group \cite{Herzog:2017bjx}. 
This computation took three
days on a cluster.
Additionally, \FORCER{} has been used for the determination of a number of Mellin 
moments of splitting and coefficient functions in DIS~\cite{Ruijl:2016pkm,Davies:2016jie,Moch2017}. 
When we use the same methods as in ref.~\cite{Larin:1996wd} for the three-loop 
moments, the calculations can become very demanding when the moment is six or higher. 
With the use of operator vertices, the 
calculations are considerably less needy of resources and already some 
rather high moments ($N\geq16$) for the non-singlet splitting functions have been 
obtained~\cite{Moch2017,RuijlZurich}. We hope to be able to do this for the 
singlet splitting functions as well. Eventually it should contribute to a 
more precise determination of Higgs boson production at the LHC.

%--#] Conclusions : 
%--#[ Acknowledgements :

\section*{Acknowledgements}

This work has been supported by the ERC Advanced Grant no.~320651, HEPGAME.

We are much indebted to A. Vogt for helping with the testing of the 
programs and making many useful suggestions about the efficiency.

%--#] Acknowledgements : 
%--#[ Appendix :
\appendix
%	 #[ One-loop sub integrals :

\section{One-loop sub integrals}
\label{sec:oneloop}

In programs such as Reduze and LiteRed reductions to master integrals are
performed only through IBP identities. 
The \FORCER{} approach is similar 
to the \MINCER{} approach in which one reduces the integrals to a point where 
one-loop sub integrals can be performed. For these, one has the 
formula~\cite{Tkachov:1981wb}
\begin{eqnarray}
\label{eq:oneloop}
\lefteqn{
   \int \frac{d^D P}{(2\pi)^D} \frac{{\cal P}_n(P)}
   {P^{2\alpha}(P-Q)^{2\beta}}  = 
} \nonumber \\
 &  \displaystyle{
    \frac{1}{(4\pi)^2}(Q^2)^{D/2-\alpha-\beta}\sum_{\sigma \ge 0}^{[n/2]}
    G(\alpha,\beta,n,\sigma)Q^{2\sigma} \left\{ \frac{1}{\sigma!}
    \left( \frac {\Box} {4} \right) ^\sigma {\cal P}_n(P)\right\}_{P=Q},
    }
\end{eqnarray}
in which
\begin{equation}
    {\cal P}_n(P) = P_{\mu_1} P_{\mu_2} \cdots P_{\mu_n}.
\end{equation}
$D$ is the dimension of space-time and is also given by $D = 4 - 
2\epsilon$, $\Box = \partial^2/\partial P_\mu \partial P_\mu$ and $G$ 
can be expressed in terms of $\Gamma$-functions:
\begin{equation}
\label{eq:Gamma}
    G(\alpha,\beta,n,\sigma) = 
    (4\pi)^\epsilon\,
    \frac{\Gamma(\alpha+\beta-\sigma-D/2)
    \Gamma(D/2-\alpha+n-\sigma)\Gamma(D/2-\beta+\sigma)}
    {\Gamma(\alpha)\Gamma(\beta)\Gamma(D-\alpha-\beta+n)}.
\end{equation}

In the presence of powers of the loop momentum $P$ in the numerator, it is 
far less work to write out this formula than to continue with recursions. 
This holds in particular when one works with the \FORM{} system, because its 
instruction set allows the evaluation of powers of the d'Alembertians with 
perfect efficiency. This means that each term gets generated with the proper 
combinatoric factor and hence never gets generated more than once. This is 
thanks to the combinatoric functions \texttt{distrib\_} and \texttt{dd\_} as in (running 
on the laptop of one of the authors):
\begin{verbatim}
    Tensor Ptensor,del;
    Vector P,Q,p1,p2,p3,p4;
    Symbols dAlembertian,j;
    Local F = dAlembertian^15*P.p1^15*P.p2^15*P.p3^15*P.p4^15;
    ToTensor,P,Ptensor;
    id	dAlembertian^j?*Ptensor(?a) = distrib_(1,2*j,del,Ptensor,?a);
    ToVector,Ptensor,Q;
    id	del(?a) = dd_(?a);
    Print +f +s;
    .end

Time =       3.09 sec    Generated terms =    1133616
               F         Terms in output =    1133616
                         Bytes used      =  140937744

   F =
       + 3092470075094400000*Q.p1*Q.p2*Q.p3^13*Q.p4^15*p1.p1*
      p1.p2^10*p1.p3^2*p2.p2^2
            .....
       + 1451044943048200500000*Q.p1^7*Q.p2^10*Q.p3^13*p1.p1^3*
      p1.p4^2*p2.p2*p2.p3^2*p2.p4*p4.p4^6
         etc.
\end{verbatim}
This code is fully explained in the \FORM{} courses in the \FORM{} web site. It 
is essential when one is interested in higher Mellin 
moments. The d'Alembertians are used both in the one-loop integrals and the 
harmonic projections when one calculates moments of structure functions.

The three drawbacks of this method are: 1) rewriting the 
dot products or invariants in the numerator to such a form that they are 
usable for the above formula is slow, 2) rewriting the resulting dot products to
the basis of the lower loop integrals is slow, and 3) one has to generate
(once) reduction algorithms for lower loop integrals with one or more 
denominators that may have a non-integer power. However, this is a small 
price to pay for the amount of computer resources that are saved.

The function $G$ is normalized to a function $G$ in which the powers of the 
denominators are one plus potentially a multiple of $\epsilon$ and in which 
there are no numerators. The difference is a number of Pochhammer symbols 
in $\epsilon$ which can either be expressed as rational polynomials or can 
be expanded in terms of $\epsilon$, depending on what is needed. When 
finite expansions are used it is easy to generate tables of these 
Pochhammer symbols. The remaining function $G$ is basically part of the 
master integral and kept for the end of the program when the master 
integrals are substituted.

%  #] One-loop sub integrals : 
%	 #[ Generalised carpet rule :

\section{Generalised carpet rule}
\label{sec:carpet}

For integrals where a subgraph is embedded in an outer one-loop 
graph, scaling and Lorenz invariance argument~\cite{Chetyrkin:1981qh} 
allow us to integrate out the outer one first:
\def\Lsub{{L_\text{sub}}}
\def\Nsub{{N_\text{sub}}}
\begin{equation}
\begin{split}
  \MoveEqLeft[3]
  \int \frac{d^D p}{(2\pi)^D}
  \frac{1}{(p^2)^\alpha\bigl[(p-q)^2\bigr]^\beta}
  \biggl[ \prod_{i=1}^\Lsub \int \frac{d^D l_i}{(2\pi)^D} \biggr]
  \biggl[ \prod_{i=1}^\Nsub \frac{1}{(p_i^2)^{a_i}} \biggr]
  \mathcal{P}_n(\{p_i\},q)
  \\
  = {} &
  \frac{1}{(4\pi)^2}
  (q^2)^{D/2-\alpha-\beta}
  \sum_{\sigma=0}^{\lfloor n/2 \rfloor}
  \biggl( \frac{D}{2} + n - \sigma \biggr)_{-\sigma}
  \\
  & \times
  G\biggl(\alpha+\sum_{i=1}^\Nsub a_i-\frac{D}{2} \Lsub -\sigma,
          \beta,n-2\sigma,0\biggr)
  \\
  & \times
  \sum_{j=0}^{\lfloor n/2 -\sigma \rfloor}
  (-1)^j
  \biggl( \frac{D}{2} + n - 2\sigma - 1 \biggr)_{-j}
  (q^2)^{\sigma+j}
  \\
  & \times
  \biggl[ \prod_{i=1}^\Lsub \int \frac{d^D l_i}{(2\pi)^D} \biggr]
  \biggl[ \prod_{i=1}^\Nsub \frac{1}{(p_i^2)^{a_i}} \biggr]
  \biggl[
    \frac{1}{\sigma!j!}
    \left(\frac{\Box_q}{4}\right)^{\sigma+j}
    \mathcal{P}_n(\{p_i\},q)
  \biggr]_{p=q}
  .
\end{split}
\end{equation}
Here $L_\text{sub}$ is the number of loops in the embedded subgraph.
The integrand of the subgraph consists of two parts:
a product of $1/(p_i^2)^{a_i}$ and $\mathcal{P}_n(\{p_i\},q)$.
Each $p_i^2$ indicates not only a squared propagator in the subgraph but also
any quadratic Lorentz scalar that becomes $p^2$ after the integrations in the
subgraph, e.g., $p_i \cdot p_j$ and $p_i \cdot p$.
If $\mathcal{P}_n(\{p_i\},q)=1$ (and $n=0$), the formula just describes that
the knowledge of the dimension of the subgraph is sufficient to write down the
result of the outer loop integral.
On the other hand, $\mathcal{P}_n(\{p_i\},q)$ is a polynomial with degree $n$
both in $p_i$ and $q$, which are taken as dot products of $p_i \cdot q$
in \FORCER{}.
%After the subloop integration,
%$H_q^{(n-2\sigma)} (\Box_q)^\sigma \mathcal{P}_n$ is harmonic
%both in $p$ and $q$.
In the right-hand side of the formula, $(x)_n = \Gamma(x+n) / \Gamma(x)$ is
the Pochhammer symbol and
the function $G$ is given by eq.~\eqref{eq:Gamma}.
The d'Alembertian $\Box_q = \partial^2 / \partial q_\mu \partial q^\mu$ can be
efficiently implemented by \texttt{distrib\_} and \texttt{dd\_} functions in
\FORM{} as explained in appendix~\ref{sec:oneloop}.

%	 #] Generalised carpet rule : 
%	 #[ The master integrals :

\section{The master integrals}
\label{sec:masterintegrals}

The four-loop master integrals are copied from refs.~\cite{Baikov:2010hf,%
Lee:2011jt}. The first reference gives the integrals to a sufficient power 
in $\epsilon$ for four-loop calculations. More powers can be found in 
the second reference. For the master integrals that involve insertions, we 
have adopted a different notation. We set the line with the 
insertion to have the power $1+m\epsilon$, whereas in the literature
the $m$-loop bubbles are kept with propagators of power 1. 
This means that a one-loop insertion refers to a 
propagator with power $\epsilon$, compared to $1+\epsilon$ in our convention. 
Hence, for completeness we provide all 
master integrals as we have used them. Here we truncate the 
expansions in $\epsilon$ at the point where the precision would be enough 
for a five-loop version of the program, although the program contains the full 
precision of ref.~\cite{Lee:2011jt}. All integrals are normalized
to powers of the fundamental one-loop integral 
multiplied by $\epsilon$ to keep the conversion factor finite. This is 
called the G-scheme.
\begin{center}
\begin{minipage}{2cm} haha \end{minipage}
	\begin{minipage}{13cm} $     +1/\epsilon (-10 \zeta_5)
      -25 \zeta_6+50 \zeta_5-10 \zeta_3^2
      +\epsilon (\frac{19}{2} \zeta_7+125 \zeta_6+90 \zeta_5-30 \zeta_3 \zeta_4+50 \zeta_3^2)
      +\epsilon^2 (\frac{324}{5} \zeta_{5,3}-\frac{621}{10} \zeta_8+\frac{2229}{2} \zeta_7+225 \zeta_6-1750 \zeta_5+1240 \zeta_3 \zeta_5+
         150 \zeta_3 \zeta_4+234 \zeta_3^2)
      +\epsilon^3 (-\frac{2916}{5} \zeta_{5,3}+\frac{21637}{3} \zeta_9+\frac{50039}{10} \zeta_8-\frac{24279}{2} \zeta_7-4375 \zeta_6+
         10250 \zeta_5+1374 \zeta_4 \zeta_5+3150 \zeta_3 \zeta_6-1200 \zeta_3 \zeta_5+702 \zeta_3 \zeta_4-3910 \zeta_3^2+
         \frac{3440}{3} \zeta_3^3)$
	\end{minipage}
\vspace{2mm} \\
\begin{minipage}{2cm} no1 \end{minipage}
	\begin{minipage}{13cm} $      +1/\epsilon (-5 \zeta_5)
      +\frac{161}{2} \zeta_7-\frac{25}{2} \zeta_6+45 \zeta_5-41 \zeta_3^2
      +\epsilon (\frac{3132}{5} \zeta_{5,3}-\frac{24641}{20} \zeta_8-\frac{1065}{2} \zeta_7+\frac{225}{2} \zeta_6-195 \zeta_5+1730 \zeta_3 \zeta_5
         -123 \zeta_3 \zeta_4+225 \zeta_3^2)
      +\epsilon^2 (\frac{5724}{5} \zeta_{5,3}+\frac{111709}{36} \zeta_9-\frac{187769}{40} \zeta_8-\frac{1713}{2} \zeta_7-\frac{975}{2} \zeta_6+
         625 \zeta_5-2103 \zeta_4 \zeta_5+4325 \zeta_3 \zeta_6-294 \zeta_3 \zeta_5+675 \zeta_3 \zeta_4+273 \zeta_3^2+1526/
         3 \zeta_3^3) $
	\end{minipage}
\vspace{2mm} \\
\begin{minipage}{2cm} no2 \end{minipage}
\begin{minipage}{13cm} $     +1/\epsilon (-10 \zeta_5)
      -70 \zeta_7-25 \zeta_6+130 \zeta_5-10 \zeta_3^2
      +\epsilon (432 \zeta_{5,3}-1289 \zeta_8+\frac{831}{2} \zeta_7+325 \zeta_6-870 \zeta_5+400 \zeta_3 \zeta_5-30 \zeta_3 \zeta_4+
         970 \zeta_3^2)
      +\epsilon^2 (-\frac{16524}{5} \zeta_{5,3}-\frac{58460}{9} \zeta_9+\frac{95021}{10} \zeta_8+\frac{22461}{2} \zeta_7-2175 \zeta_6+
         4250 \zeta_5-2640 \zeta_4 \zeta_5+1000 \zeta_3 \zeta_6-2340 \zeta_3 \zeta_5+2910 \zeta_3 \zeta_4-10734 \zeta_3^2
         +\frac{4528}{3} \zeta_3^3) $
	\end{minipage}
\vspace{2mm} \\
\begin{minipage}{2cm} no6 \end{minipage}
\begin{minipage}{13cm} $      +1/\epsilon (-5 \zeta_5)
      -\frac{25}{2} \zeta_6+45 \zeta_5-17 \zeta_3^2
      +\epsilon (-\frac{85}{2} \zeta_7+\frac{225}{2} \zeta_6-195 \zeta_5-51 \zeta_3 \zeta_4+153 \zeta_3^2)
      +\epsilon^2 (-\frac{8532}{5} \zeta_{5,3}+\frac{158967}{40} \zeta_8+\frac{765}{2} \zeta_7-\frac{975}{2} \zeta_6+625 \zeta_5-3118 \zeta_3
          \zeta_5+459 \zeta_3 \zeta_4-663 \zeta_3^2)
      +\epsilon^3 (\frac{76788}{5} \zeta_{5,3}+\frac{16232}{3} \zeta_9-\frac{1430703}{40} \zeta_8-\frac{3315}{2} \zeta_7+\frac{3125}{2} \zeta_6-
         1875 \zeta_5+8121 \zeta_4 \zeta_5-7710 \zeta_3 \zeta_6+28062 \zeta_3 \zeta_5-1989 \zeta_3 \zeta_4+2125 \zeta_3^2
         -\frac{4310}{3} \zeta_3^3) $
	\end{minipage}
\vspace{2mm}\\
\begin{minipage}{2cm} lala \end{minipage}
\begin{minipage}{13cm} $     +\frac{441}{8} \zeta_7
      +\epsilon (-\frac{81}{5} \zeta_{5,3}+\frac{18567}{80} \zeta_8-\frac{1323}{4} \zeta_7-135 \zeta_3 \zeta_5)
      +\epsilon^2 (\frac{486}{5} \zeta_{5,3}+\frac{4583}{2} \zeta_9-\frac{55701}{40} \zeta_8+\frac{1323}{2} \zeta_7-81 \zeta_4 \zeta_5-\frac{675}{2} 
         \zeta_3 \zeta_6+810 \zeta_3 \zeta_5-267 \zeta_3^3) $
	\end{minipage}
\vspace{2mm} \\
\begin{minipage}{2cm} nono \end{minipage}
\begin{minipage}{13cm} $     +36 \zeta_3^2
      +\epsilon (-378 \zeta_7+108 \zeta_3 \zeta_4)
      +\epsilon^2 (\frac{3024}{5} \zeta_{5,3}-\frac{26901}{10} \zeta_8+2844 \zeta_3 \zeta_5+432 \zeta_3^2)
      +\epsilon^3 (-\frac{42458}{3} \zeta_9-4536 \zeta_7-270 \zeta_4 \zeta_5+6930 \zeta_3 \zeta_6+1296 \zeta_3 \zeta_4+2304 
         \zeta_3^2-732 \zeta_3^3) $
	\end{minipage}
\vspace{2mm} \\
\begin{minipage}{2cm} cross \end{minipage}
\begin{minipage}{13cm} $     +1/\epsilon (5 \zeta_5)
      +\frac{25}{2} \zeta_6-5 \zeta_5-7 \zeta_3^2
      +\epsilon (\frac{127}{2} \zeta_7-\frac{25}{2} \zeta_6+35 \zeta_5-21 \zeta_3 \zeta_4+7 \zeta_3^2)
      +\epsilon^2 (\frac{972}{5} \zeta_{5,3}-\frac{12387}{40} \zeta_8-\frac{127}{2} \zeta_7+\frac{175}{2} \zeta_6+135 \zeta_5+22 \zeta_3 \zeta_5+
         21 \zeta_3 \zeta_4-49 \zeta_3^2)
      +\epsilon^3 (-\frac{972}{5} \zeta_{5,3}+\frac{346}{3} \zeta_9+\frac{12387}{40} \zeta_8+\frac{889}{2} \zeta_7+\frac{675}{2} \zeta_6+675 \zeta_5-
         1425 \zeta_4 \zeta_5+90 \zeta_3 \zeta_6-22 \zeta_3 \zeta_5-147 
         \zeta_3 \zeta_4-189 \zeta_3^2+\frac{1742}{3} \zeta_3^3) $
	\end{minipage}
\vspace{2mm} \\
\begin{minipage}{2cm} bebe \end{minipage}
\begin{minipage}{13cm} $     +1/\epsilon^2 (\frac{1}{2} \zeta_3)
      +1/\epsilon (\frac{3}{4} \zeta_4+\frac{3}{2} \zeta_3)
      -\frac{23}{2} \zeta_5+\frac{9}{4} \zeta_4+\frac{19}{2} \zeta_3
      +\epsilon (-30 \zeta_6-\frac{69}{2} \zeta_5+\frac{57}{4} \zeta_4+\frac{103}{2} \zeta_3+\frac{29}{2} \zeta_3^2)
      +\epsilon^2 (-\frac{1105}{4} \zeta_7-90 \zeta_6-\frac{437}{2} \zeta_5+\frac{309}{4} \zeta_4+\frac{547}{2} \zeta_3+\frac{87}{2} \zeta_3 \zeta_4+87/
         2 \zeta_3^2)
      +\epsilon^3 (\frac{486}{5} \zeta_{5,3}-\frac{84627}{80} \zeta_8-\frac{3315}{4} \zeta_7-570 \zeta_6-\frac{2369}{2} \zeta_5+\frac{1641}{4} 
         \zeta_4+\frac{2863}{2} \zeta_3+1153 \zeta_3 \zeta_5+\frac{261}{2} \zeta_3 \zeta_4+\frac{551}{2} \zeta_3^2)
      +\epsilon^4 (\frac{1458}{5} \zeta_{5,3}-5144 \zeta_9-\frac{253881}{80} \zeta_8-\frac{20995}{4} \zeta_7-3090 \zeta_6-12581/
         2 \zeta_5+\frac{8589}{4} \zeta_4+\frac{2001}{2} \zeta_4 \zeta_5+\frac{14827}{2} \zeta_3+2810 \zeta_3 \zeta_6+3459 \zeta_3 \zeta_5+
         \frac{1653}{2} \zeta_3 \zeta_4+\frac{2987}{2} \zeta_3^2-967 
         \zeta_3^3) $
	\end{minipage}
\vspace{2mm} \\
\begin{minipage}{2cm} nostar6 \end{minipage}
\begin{minipage}{13cm} $     +20 \zeta_5
      +\epsilon (50 \zeta_6-80 \zeta_5+80 \zeta_3^2)
      +\epsilon^2 (625 \zeta_7-200 \zeta_6+80 \zeta_5+240 \zeta_3 \zeta_4-320 \zeta_3^2)
      +\epsilon^3 (-3240 \zeta_{5,3}+\frac{144241}{12} \zeta_8-2500 \zeta_7+200 \zeta_6-\frac{3373}{2} \zeta_4^2-4480 \zeta_3
          \zeta_5-960 \zeta_3 \zeta_4+320 \zeta_3^2)
      +\epsilon^4 (12960 \zeta_{5,3}+\frac{109895}{6} \zeta_9-\frac{1146935}{24} \zeta_8+2500 \zeta_7+17580 \zeta_4 \zeta_5+
         \frac{25985}{4} \zeta_4^2-11600 \zeta_3 \zeta_6+17920 \zeta_3 
         \zeta_5+960 \zeta_3 \zeta_4-2920 \zeta_3^3) $
	\end{minipage}
\vspace{2mm} \\
\begin{minipage}{2cm} nostar5 \end{minipage}
\begin{minipage}{13cm} $     +20 \zeta_5
      +\epsilon (50 \zeta_6-80 \zeta_5+86 \zeta_3^2)
      +\epsilon^2 (\frac{5777}{8} \zeta_7-200 \zeta_6+80 \zeta_5+258 \zeta_3 \zeta_4-344 \zeta_3^2)
      +\epsilon^3 (-\frac{22977}{5} \zeta_{5,3}+\frac{6622129}{480} \zeta_8-\frac{5777}{2} \zeta_7+200 \zeta_6-\frac{1541}{16} \zeta_4^2
         -6646 \zeta_3 \zeta_5-1032 \zeta_3 \zeta_4+344 \zeta_3^2)
      +\epsilon^4 (\frac{91908}{5} \zeta_{5,3}+\frac{198659}{8} \zeta_9-\frac{6571099}{120} \zeta_8+\frac{5777}{2} \zeta_7+\frac{48993}{2} 
         \zeta_4 \zeta_5+\frac{83}{4} \zeta_4^2-17045 \zeta_3 \zeta_6+26584 
         \zeta_3 \zeta_5+1032 \zeta_3 \zeta_4-4314 \zeta_3^3) $
	\end{minipage}
\vspace{2mm} \\
\begin{minipage}{2cm} no \end{minipage}
\begin{minipage}{13cm} $     +20 \zeta_5
      +\epsilon (50 \zeta_6-80 \zeta_5+68 \zeta_3^2)
      +\epsilon^2 (450 \zeta_7-200 \zeta_6+80 \zeta_5+204 \zeta_3 \zeta_4-272 \zeta_3^2)
      +\epsilon^3 (-\frac{9072}{5} \zeta_{5,3}+\frac{59633}{10} \zeta_8-1800 \zeta_7+200 \zeta_6-2448 \zeta_3 \zeta_5-816 \zeta_3 
         \zeta_4+272 \zeta_3^2)
      +\epsilon^4 (\frac{36288}{5} \zeta_{5,3}+\frac{88036}{9} \zeta_9-\frac{119266}{5} \zeta_8+1800 \zeta_7+9936 \zeta_4 \zeta_5-
         6460 \zeta_3 \zeta_6+9792 \zeta_3 \zeta_5+816 \zeta_3 
         \zeta_4-\frac{4640}{3} \zeta_3^3) $
	\end{minipage}
\vspace{2mm} \\
\begin{minipage}{2cm} fastar2 \end{minipage}
\begin{minipage}{13cm} $     +20 \zeta_5
      +\epsilon (50 \zeta_6-80 \zeta_5-4 \zeta_3^2)
      +\epsilon^2 (639 \zeta_7-200 \zeta_6+80 \zeta_5-12 \zeta_3 \zeta_4+16 \zeta_3^2)
      +\epsilon^3 (\frac{648}{5} \zeta_{5,3}+\frac{795539}{480} \zeta_8-2556 \zeta_7+200 \zeta_6+\frac{1103}{16} \zeta_4^2-456 
         \zeta_3 \zeta_5+48 \zeta_3 \zeta_4-16 \zeta_3^2)
      +\epsilon^4 (-\frac{2592}{5} \zeta_{5,3}+\frac{142462}{9} \zeta_9-\frac{795539}{120} \zeta_8+2556 \zeta_7-1656 \zeta_4 \zeta_5-
         \frac{1103}{4} \zeta_4^2-1120 \zeta_3 \zeta_6+1824 \zeta_3 
         \zeta_5-48 \zeta_3 \zeta_4+\frac{328}{3} \zeta_3^3) $
	\end{minipage}
\vspace{2mm} \\
\begin{minipage}{2cm} t1star55 \end{minipage}
\begin{minipage}{13cm} $     +6 \zeta_3
      +\epsilon (9 \zeta_4-12 \zeta_3)
      +\epsilon^2 (192 \zeta_5-18 \zeta_4)
      +\epsilon^3 (465 \zeta_6-384 \zeta_5-168 \zeta_3^2)
      +\epsilon^4 (4509 \zeta_7-930 \zeta_6-504 \zeta_3 \zeta_4+336 \zeta_3^2)
      +\epsilon^5 (648 \zeta_{5,3}+\frac{295483}{24} \zeta_8-9018 \zeta_7-\frac{529}{4} \zeta_4^2-6492 \zeta_3 \zeta_5+1008 
         \zeta_3 \zeta_4)
      +\epsilon^6 (-1296 \zeta_{5,3}+98490 \zeta_9-\frac{295483}{12} \zeta_8-14598 \zeta_4 \zeta_5+\frac{529}{2} \zeta_4^2-
         15390 \zeta_3 \zeta_6+12984 \zeta_3 \zeta_5+2676 \zeta_3^3) $
	\end{minipage}
\vspace{2mm} \\
\begin{minipage}{2cm} t1star15 \end{minipage}
\begin{minipage}{13cm} $     +6 \zeta_3
      +\epsilon (9 \zeta_4-12 \zeta_3)
      +\epsilon^2 (157 \zeta_5-18 \zeta_4)
      +\epsilon^3 (\frac{755}{2} \zeta_6-314 \zeta_5-179 \zeta_3^2)
      +\epsilon^4 (\frac{26657}{8} \zeta_7-755 \zeta_6-537 \zeta_3 \zeta_4+358 \zeta_3^2)
      +\epsilon^5 (243 \zeta_{5,3}+\frac{38195}{4} \zeta_8-\frac{26657}{4} \zeta_7-\frac{1743}{8} \zeta_4^2-6736 \zeta_3 \zeta_5+1074
          \zeta_3 \zeta_4)
      +\epsilon^6 (-486 \zeta_{5,3}+\frac{1657525}{24} \zeta_9-\frac{616223}{32} \zeta_8-\frac{23853}{2} \zeta_4 \zeta_5+\frac{9159}{16}
          \zeta_4^2-15945 \zeta_3 \zeta_6+13472 \zeta_3 
          \zeta_5+\frac{8776}{3} \zeta_3^3) $
	\end{minipage}
\vspace{2mm} \\
\begin{minipage}{2cm} t1star13 \end{minipage}
\begin{minipage}{13cm} $     +6 \zeta_3
      +\epsilon (9 \zeta_4-12 \zeta_3)
      +\epsilon^2 (127 \zeta_5-18 \zeta_4)
      +\epsilon^3 (\frac{605}{2} \zeta_6-254 \zeta_5-173 \zeta_3^2)
      +\epsilon^4 (\frac{18989}{8} \zeta_7-605 \zeta_6-519 \zeta_3 \zeta_4+346 \zeta_3^2)
      +\epsilon^5 (\frac{243}{5} \zeta_{5,3}+\frac{549939}{80} \zeta_8-\frac{18989}{4} \zeta_7-\frac{639}{4} \zeta_4^2-6082 \zeta_3 \zeta_5+
         1038 \zeta_3 \zeta_4)
      +\epsilon^6 (-\frac{486}{5} \zeta_{5,3}+\frac{1084927}{24} \zeta_9-\frac{1100823}{80} \zeta_8-\frac{18975}{2} \zeta_4 \zeta_5+2637/
         8 \zeta_4^2-14340 \zeta_3 \zeta_6+12164 \zeta_3 
         \zeta_5+\frac{8554}{3} \zeta_3^3) $
	\end{minipage}
\vspace{2mm} \\
\begin{minipage}{2cm} t1star5 \end{minipage}
\begin{minipage}{13cm} $     +6 \zeta_3
      +\epsilon (9 \zeta_4-12 \zeta_3)
      +\epsilon^2 (102 \zeta_5-18 \zeta_4)
      +\epsilon^3 (240 \zeta_6-204 \zeta_5-78 \zeta_3^2)
      +\epsilon^4 (1413 \zeta_7-480 \zeta_6-234 \zeta_3 \zeta_4+156 \zeta_3^2)
      +\epsilon^5 (\frac{648}{5} \zeta_{5,3}+\frac{148157}{40} \zeta_8-2826 \zeta_7-\frac{363}{4} \zeta_4^2-1812 \zeta_3 \zeta_5+468 
         \zeta_3 \zeta_4)
      +\epsilon^6 (-\frac{1296}{5} \zeta_{5,3}+18918 \zeta_9-\frac{148157}{20} \zeta_8-3690 \zeta_4 \zeta_5+\frac{363}{2} \zeta_4^2-
         4140 \zeta_3 \zeta_6+3624 \zeta_3 \zeta_5+588 \zeta_3^3) $
	\end{minipage}
\end{center}

To convert to \MSbar{}, one has to multiply the results 
with the appropriate number of powers of the following conversion factor:
%\begin{align}
%C_\MSbar & = 1  \nonumber \\
%  &  +\epsilon^3 (-\frac{7}{3} \zeta_3) \nonumber \\
%  &  +\epsilon^4 (-\frac{13}{4} \zeta_4) \nonumber \\
%  &  +\epsilon^5 (-\frac{31}{5} \zeta_5) \nonumber \\
%  &  +\epsilon^6 (\frac{49}{18} \zeta_3^2-\frac{61}{6} \zeta_6) \nonumber \\
%  &  +\epsilon^7 (\frac{91}{12} \zeta_4 \zeta_3-\frac{127}{7} \zeta_7) \nonumber \\
%  &  +\epsilon^8 (\frac{169}{32} \zeta_4^2+\frac{217}{15} \zeta_5 \zeta_3-\frac{253}{8} \zeta_8) \nonumber \\
%  &  +\epsilon^9 (-\frac{343}{162} \zeta_3^3+\frac{403}{20} \zeta_5 \zeta_4
%			+\frac{427}{18} \zeta_6 \zeta_3-\frac{511}{9} \zeta_9) .
%\end{align}
\begin{align}
  C_\MSbar & = \frac{1}{1-2\epsilon} \biggl[
  1  \nonumber \\
  &  +\epsilon^3 \Bigl(-\frac{7}{3} \zeta_3\Bigr) \nonumber \\
  &  +\epsilon^4 \Bigl(-\frac{13}{4} \zeta_4\Bigr) \nonumber \\
  &  +\epsilon^5 \Bigl(-\frac{31}{5} \zeta_5\Bigr) \nonumber \\
  &  +\epsilon^6 \Bigl(\frac{49}{18} \zeta_3^2-\frac{61}{6} \zeta_6\Bigr) \nonumber \\
  &  +\epsilon^7 \Bigl(\frac{91}{12} \zeta_4 \zeta_3-\frac{127}{7} \zeta_7\Bigr) \nonumber \\
  &  +\epsilon^8 \Bigl(\frac{169}{32} \zeta_4^2+\frac{217}{15} \zeta_5 \zeta_3-\frac{253}{8} \zeta_8\Bigr) \nonumber \\
  &  +\epsilon^9 \Bigl(-\frac{343}{162} \zeta_3^3+\frac{403}{20} \zeta_5 \zeta_4
                        +\frac{427}{18} \zeta_6 \zeta_3-\frac{511}{9} \zeta_9\Bigr)
  + \mathcal{O}(\epsilon^{10} )\biggr].
\end{align}
Here we have already dropped $\zeta_2$, which never appears in physical results
of massless propagators.

%	 #] The master integrals : 
%	 #[ Expansions :

\section{Expansions}
\label{sec:Expansions}

In principle the coefficients of the integrals can be kept as rational 
polynomials in $D$ or $\epsilon$. However, the nature of the reductions is
such that these polynomials can contain very high powers in their numerators 
and denominators. Adding such rational polynomials is easily the most 
costly operation during the reductions. During the development of the 
\FORCER{} program, we have encountered polynomials with powers of $\epsilon$ 
that went over 700, and that was not even for a complete reduction. In practice 
one needs such `precision' only in very rare cases, such as when one needs 
to change dimensions during or after the reduction. In our program this is 
not necessary, and hence a better strategy is to expand these polynomials 
to a finite power of $\epsilon$. The main problem is that we do not know in 
advance how many powers are needed. The reductions will at times generate 
extra powers of $1/\epsilon$ (spurious poles) that will only cancel near the 
end of the reduction when all terms that contribute to a given master 
integral are added. 
An exact solution for the spurious problem 
is a denominator notation~\cite{Vermaseren2014}, but to make this 
workable \FORM{} still needs supporting facilities.

We have opted for a method in which the reduction formulas still use 
rational polynomials, but after each step they are expanded to sufficient 
depth. It is possible to make a special trial run to determine how many 
powers are needed. In this trial run only the minimum power of $\epsilon$ 
is kept with the coefficient one, to avoid that such terms can cancel. 
Avoiding all calculations, such a run can be relatively fast, provided that 
the main computational effort is in the \FORCER{} part of the program (it 
usually is). After the run, one can see how deep the expansions have to be. We 
usually take the worst value that we encounter for all diagrams and add one 
`guard power'. With this value the program 
generates the proper tables for the various Pochhammer symbols and other 
objects that may need expansions. Then during the actual reductions the 
rational polynomials will be expanded to the proper depth.

A simpler and safer method is to run the whole calculation twice with 
different settings for the expansion depth and observe at which power of 
$\epsilon$ the coefficients change. This is similar to running numerical 
programs with different floating point precisions to study the numerical 
instabilities.

\FORM{} has options to use expansions in its coefficients.
The command \path{PolyRatFun,rat(divergence,variable)} keeps only 
the lowest power of $\epsilon$. Generally, the program
is quite fast in this mode. To expand, the statement
\path{PolyRatFun,rat(expand,variable,maxpow)} can be used. These commands 
are implemented in the latest development version of \FORM{}.

%	 #] Expansions : 
%--#[ Bibliography :
\bibliographystyle{JHEP}
\bibliography{references}

\providecommand{\url}[1]{#1}\providecommand{\href}[2]{#2}\begingroup\raggedright\begin{thebibliography}{10}

\bibitem{Anastasiou:2015ema}
C.~Anastasiou, C.~Duhr, F.~Dulat, F.~Herzog and B.~Mistlberger, \emph{{Higgs
  Boson Gluon-Fusion Production in QCD at Three Loops}},
  \href{https://dx.doi.org/10.1103/PhysRevLett.114.212001}{\emph{Phys. Rev.
  Lett.} {\bfseries 114} (2015) 212001}
  [\href{https://arxiv.org/abs/1503.06056}{{\ttfamily arXiv:1503.06056}}].
%%CITATION = ARXIV:1503.06056;%%

\bibitem{Anastasiou:2016cez}
C.~Anastasiou, C.~Duhr, F.~Dulat, E.~Furlan, T.~Gehrmann, F.~Herzog,
  A.~Lazopoulos and B.~Mistlberger, \emph{{High precision determination of the
  gluon fusion Higgs boson cross-section at the LHC}},
  \href{https://dx.doi.org/10.1007/JHEP05(2016)058}{\emph{JHEP} {\bfseries 05}
  (2016) 058} [\href{https://arxiv.org/abs/1602.00695}{{\ttfamily
  arXiv:1602.00695}}].
%%CITATION = ARXIV:1602.00695;%%

\bibitem{Larin:1991fx}
S.A.~Larin, F.V.~Tkachov and J.A.M.~Vermaseren, \emph{{The
  $\mathcal{O}(\alpha^3)$ QCD correction to the lowest moment of the
  longitudinal structure function in deep inelastic electron-nucleon
  scattering}},
  \href{https://dx.doi.org/10.1016/0370-2693(91)91023-O}{\emph{Phys. Lett.}
  {\bfseries B272} (1991) 121}.
%%CITATION = PHLTA,B272,121;%%

\bibitem{Larin:1993vu}
S.A.~Larin, T.~van~Ritbergen and J.A.M.~Vermaseren, \emph{{The
  next-next-to-leading QCD approximation for non-singlet moments of deep
  inelastic structure functions}},
  \href{https://dx.doi.org/10.1016/0550-3213(94)90268-2}{\emph{Nucl. Phys.}
  {\bfseries B427} (1994) 41}.
%%CITATION = NUPHA,B427,41;%%

\bibitem{Larin:1996wd}
S.A.~Larin, P.~Nogueira, T.~van~Ritbergen and J.A.M.~Vermaseren, \emph{{The
  3-loop QCD calculation of the moments of deep inelastic structure
  functions}},
  \href{https://dx.doi.org/10.1016/S0550-3213(97)80038-7}{\emph{Nucl. Phys.}
  {\bfseries B492} (1997) 338}
  [\href{https://arxiv.org/abs/hep-ph/9605317}{{\ttfamily hep-ph/9605317}}].
%%CITATION = HEP-PH/9605317;%%

\bibitem{Retey:2000nq}
A.~R\'etey and J.A.M.~Vermaseren, \emph{{Some higher moments of deep inelastic
  structure functions at next-to-next-to-leading order of perturbative QCD}},
  \href{https://dx.doi.org/10.1016/S0550-3213(01)00149-3}{\emph{Nucl. Phys.}
  {\bfseries B604} (2001) 281}
  [\href{https://arxiv.org/abs/hep-ph/0007294}{{\ttfamily hep-ph/0007294}}].
%%CITATION = HEP-PH/0007294;%%

\bibitem{Blumlein:2004xt}
J.~Bl\"umlein and J.A.M.~Vermaseren, \emph{{The 16th moment of the non-singlet
  structure functions $F_2(x,Q^2)$ and $F_L(x,Q^2)$ to
  $\mathcal{O}(\alpha_s^3)$}},
  \href{https://dx.doi.org/10.1016/j.physletb.2004.11.059}{\emph{Phys. Lett.}
  {\bfseries B606} (2005) 130}
  [\href{https://arxiv.org/abs/hep-ph/0411111}{{\ttfamily hep-ph/0411111}}].
%%CITATION = HEP-PH/0411111;%%

\bibitem{Gorishnii:1989gt}
S.G.~Gorishnii, S.A.~Larin, L.R.~Surguladze and F.V.~Tkachov, \emph{{Mincer:
  Program for multiloop calculations in quantum field theory for the
  Schoonschip system}},
  \href{https://dx.doi.org/10.1016/0010-4655(89)90134-3}{\emph{Comput. Phys.
  Commun.} {\bfseries 55} (1989) 381}.
%%CITATION = CPHCB,55,381;%%

\bibitem{Larin:1991fz}
S.~Larin, F.~Tkachov and J.~Vermaseren, \emph{{The FORM version of MINCER}},
  preprint {\ttfamily NIKHEF-H-91-18}, 1991.
%%CITATION = NIKHEF-H-91-18 ETC.;%%

\bibitem{Chetyrkin:1981qh}
K.G.~Chetyrkin and F.V.~Tkachov, \emph{{Integration by parts: The algorithm to
  calculate $\beta$ functions in 4 loops}},
  \href{https://dx.doi.org/10.1016/0550-3213(81)90199-1}{\emph{Nucl. Phys.}
  {\bfseries B192} (1981) 159}.
%%CITATION = NUPHA,B192,159;%%

\bibitem{Moch:2014sna}
S.~Moch, J.A.M.~Vermaseren and A.~Vogt, \emph{{The three-loop splitting
  functions in QCD: The helicity-dependent case}},
  \href{https://dx.doi.org/10.1016/j.nuclphysb.2014.10.016}{\emph{Nucl. Phys.}
  {\bfseries B889} (2014) 351}
  [\href{https://arxiv.org/abs/1409.5131}{{\ttfamily arXiv:1409.5131}}].
%%CITATION = ARXIV:1409.5131;%%

\bibitem{Baikov:1996rk}
P.A.~Baikov, \emph{{Explicit solutions of the 3-loop vacuum integral recurrence
  relations}},
  \href{https://dx.doi.org/10.1016/0370-2693(96)00835-0}{\emph{Phys. Lett.}
  {\bfseries B385} (1996) 404}
  [\href{https://arxiv.org/abs/hep-ph/9603267}{{\ttfamily hep-ph/9603267}}].
%%CITATION = HEP-PH/9603267;%%

\bibitem{Baikov:1996iu}
P.A.~Baikov, \emph{{Explicit solutions of the multiloop integral recurrence
  relations and its application}},
  \href{https://dx.doi.org/10.1016/S0168-9002(97)00126-5}{\emph{Nucl. Instrum.
  Meth.} {\bfseries A389} (1997) 347}
  [\href{https://arxiv.org/abs/hep-ph/9611449}{{\ttfamily hep-ph/9611449}}].
%%CITATION = HEP-PH/9611449;%%

\bibitem{Baikov:2005nv}
P.A.~Baikov, \emph{{A practical criterion of irreducibility of multi-loop
  Feynman integrals}},
  \href{https://dx.doi.org/10.1016/j.physletb.2006.01.052}{\emph{Phys. Lett.}
  {\bfseries B634} (2006) 325}
  [\href{https://arxiv.org/abs/hep-ph/0507053}{{\ttfamily hep-ph/0507053}}].
%%CITATION = HEP-PH/0507053;%%

\bibitem{Laporta:2001dd}
S.~Laporta, \emph{{High-precision calculation of multi-loop Feynman integrals
  by difference equations}},
  \href{https://dx.doi.org/10.1142/S0217751X00002157}{\emph{Int. J. Mod. Phys.}
  {\bfseries A15} (2000) 5087}
  [\href{https://arxiv.org/abs/hep-ph/0102033}{{\ttfamily hep-ph/0102033}}].
%%CITATION = HEP-PH/0102033;%%

\bibitem{Anastasiou:2004vj}
C.~Anastasiou and A.~Lazopoulos, \emph{{Automatic integral reduction for higher
  order perturbative calculations}},
  \href{https://dx.doi.org/10.1088/1126-6708/2004/07/046}{\emph{JHEP}
  {\bfseries 07} (2004) 046}
  [\href{https://arxiv.org/abs/hep-ph/0404258}{{\ttfamily hep-ph/0404258}}].
%%CITATION = HEP-PH/0404258;%%

\bibitem{Smirnov:2008iw}
A.V.~Smirnov, \emph{{Algorithm FIRE -- Feynman Integral REduction}},
  \href{https://dx.doi.org/10.1088/1126-6708/2008/10/107}{\emph{JHEP}
  {\bfseries 10} (2008) 107} [\href{https://arxiv.org/abs/0807.3243}{{\ttfamily
  arXiv:0807.3243}}].
%%CITATION = ARXIV:0807.3243;%%

\bibitem{Smirnov:2014hma}
A.V.~Smirnov, \emph{{FIRE5: A C++ implementation of Feynman Integral
  REduction}},
  \href{https://dx.doi.org/10.1016/j.cpc.2014.11.024}{\emph{Comput. Phys.
  Commun.} {\bfseries 189} (2015) 182}
  [\href{https://arxiv.org/abs/1408.2372}{{\ttfamily arXiv:1408.2372}}].
%%CITATION = ARXIV:1408.2372;%%

\bibitem{Studerus:2009ye}
C.~Studerus, \emph{{Reduze -- Feynman integral reduction in C++}},
  \href{https://dx.doi.org/10.1016/j.cpc.2010.03.012}{\emph{Comput. Phys.
  Commun.} {\bfseries 181} (2010) 1293}
  [\href{https://arxiv.org/abs/0912.2546}{{\ttfamily arXiv:0912.2546}}].
%%CITATION = ARXIV:0912.2546;%%

\bibitem{vonManteuffel:2012np}
A.~von~Manteuffel and C.~Studerus, \emph{{Reduze 2 -- Distributed Feynman
  Integral Reduction}},  \href{https://arxiv.org/abs/1201.4330}{{\ttfamily
  arXiv:1201.4330}}.
%%CITATION = ARXIV:1201.4330;%%

\bibitem{Tkachov:1984xk}
F.V.~Tkachov, \emph{{An algorithm for calculating multiloop integrals}},
  \href{https://dx.doi.org/10.1007/BF01086253}{\emph{Theor. Math. Phys.}
  {\bfseries 56} (1983) 866}.
%%CITATION = TMPHA,56,866;%%

\bibitem{Moch:2004pa}
S.~Moch, J.A.M.~Vermaseren and A.~Vogt, \emph{{The three-loop splitting
  functions in QCD: the non-singlet case}},
  \href{https://dx.doi.org/10.1016/j.nuclphysb.2004.03.030}{\emph{Nucl. Phys.}
  {\bfseries B688} (2004) 101}
  [\href{https://arxiv.org/abs/hep-ph/0403192}{{\ttfamily hep-ph/0403192}}].
%%CITATION = HEP-PH/0403192;%%

\bibitem{Vogt:2004mw}
A.~Vogt, S.~Moch and J.A.M.~Vermaseren, \emph{{The three-loop splitting
  functions in QCD: the singlet case}},
  \href{https://dx.doi.org/10.1016/j.nuclphysb.2004.04.024}{\emph{Nucl. Phys.}
  {\bfseries B691} (2004) 129}
  [\href{https://arxiv.org/abs/hep-ph/0404111}{{\ttfamily hep-ph/0404111}}].
%%CITATION = HEP-PH/0404111;%%

\bibitem{Vermaseren:2005qc}
J.A.M.~Vermaseren, A.~Vogt and S.~Moch, \emph{{The third-order QCD corrections
  to deep-inelastic scattering by photon exchange}},
  \href{https://dx.doi.org/10.1016/j.nuclphysb.2005.06.020}{\emph{Nucl. Phys.}
  {\bfseries B724} (2005) 3}
  [\href{https://arxiv.org/abs/hep-ph/0504242}{{\ttfamily hep-ph/0504242}}].
%%CITATION = HEP-PH/0504242;%%

\bibitem{Lee:2012cn}
R.N.~Lee, \emph{{Presenting LiteRed: a tool for the Loop InTEgrals REDuction}},
   \href{https://arxiv.org/abs/1212.2685}{{\ttfamily arXiv:1212.2685}}.
%%CITATION = ARXIV:1212.2685;%%

\bibitem{Lee:2013mka}
R.N.~Lee, \emph{{LiteRed 1.4: a powerful tool for reduction of multiloop
  integrals}},
  \href{https://dx.doi.org/10.1088/1742-6596/523/1/012059}{\emph{J. Phys. Conf.
  Ser.} {\bfseries 523} (2014) 012059}
  [\href{https://arxiv.org/abs/1310.1145}{{\ttfamily arXiv:1310.1145}}].
%%CITATION = ARXIV:1310.1145;%%

\bibitem{Tentyukov:2007mu}
M.~Tentyukov and J.A.M.~Vermaseren, \emph{{The multithreaded version of FORM}},
  \href{https://dx.doi.org/10.1016/j.cpc.2010.04.009}{\emph{Comput. Phys.
  Commun.} {\bfseries 181} (2010) 1419}
  [\href{https://arxiv.org/abs/hep-ph/0702279}{{\ttfamily hep-ph/0702279}}].
%%CITATION = HEP-PH/0702279;%%

\bibitem{Kuipers:2012rf}
J.~Kuipers, T.~Ueda, J.A.M.~Vermaseren and J.~Vollinga, \emph{{FORM version
  4.0}}, \href{https://dx.doi.org/10.1016/j.cpc.2012.12.028}{\emph{Comput.
  Phys. Commun.} {\bfseries 184} (2013) 1453}
  [\href{https://arxiv.org/abs/1203.6543}{{\ttfamily arXiv:1203.6543}}].
%%CITATION = ARXIV:1203.6543;%%

\bibitem{Ruijl:2016pkm}
B.~Ruijl, T.~Ueda, J.A.M.~Vermaseren, J.~Davies and A.~Vogt, \emph{{First
  Forcer results on deep-inelastic scattering and related quantities}},
  {\emph{PoS} {\bfseries LL2016} (2016) 071}
  [\href{https://arxiv.org/abs/1605.08408}{{\ttfamily arXiv:1605.08408}}].
%%CITATION = ARXIV:1605.08408;%%

\bibitem{Davies:2016jie}
J.~Davies, A.~Vogt, B.~Ruijl, T.~Ueda and J.A.M.~Vermaseren, \emph{{Large-$n_f$
  contributions to the four-loop splitting functions in QCD}},
  \href{https://dx.doi.org/10.1016/j.nuclphysb.2016.12.012}{\emph{Nucl. Phys.}
  {\bfseries B915} (2017) 335}
  [\href{https://arxiv.org/abs/1610.07477}{{\ttfamily arXiv:1610.07477}}].
%%CITATION = ARXIV:1610.07477;%%

\bibitem{Ueda:2016sxw}
T.~Ueda, B.~Ruijl and J.A.M.~Vermaseren, \emph{{Calculating four-loop massless
  propagators with Forcer}},
  \href{https://dx.doi.org/10.1088/1742-6596/762/1/012060}{\emph{J. Phys. Conf.
  Ser.} {\bfseries 762} (2016) 012060}
  [\href{https://arxiv.org/abs/1604.08767}{{\ttfamily arXiv:1604.08767}}].
%%CITATION = ARXIV:1604.08767;%%

\bibitem{Ueda:2016yjm}
T.~Ueda, B.~Ruijl and J.A.M.~Vermaseren, \emph{{Forcer: a FORM program for
  4-loop massless propagators}}, {\emph{PoS} {\bfseries LL2016} (2016) 070}
  [\href{https://arxiv.org/abs/1607.07318}{{\ttfamily arXiv:1607.07318}}].
%%CITATION = ARXIV:1607.07318;%%

\bibitem{Ruijl:2017prop}
B.~Ruijl, T.~Ueda, J.A.M.~Vermaseren and A.~Vogt, \emph{{Four-loop QCD
  propagators and vertices with one vanishing external momentum}},
  \href{https://arxiv.org/abs/1703.08532}{{\ttfamily arXiv:1703.08532}}.
%%CITATION = ARXIV:1703.08532;%%

\bibitem{Herzog:2017ohr}
F.~Herzog, B.~Ruijl, T.~Ueda, J.A.M.~Vermaseren and A.~Vogt, \emph{{The
  five-loop beta function of Yang-Mills theory with fermions}},
  \href{https://dx.doi.org/10.1007/JHEP02(2017)090}{\emph{JHEP} {\bfseries 02}
  (2017) 090} [\href{https://arxiv.org/abs/1701.01404}{{\ttfamily
  arXiv:1701.01404}}].
%%CITATION = ARXIV:1701.01404;%%

\bibitem{Ruijl:2015aca}
B.~Ruijl, T.~Ueda and J.~Vermaseren, \emph{{The diamond rule for multi-loop
  Feynman diagrams}},
  \href{https://dx.doi.org/10.1016/j.physletb.2015.05.015}{\emph{Phys. Lett.}
  {\bfseries B746} (2015) 347}
  [\href{https://arxiv.org/abs/1504.08258}{{\ttfamily arXiv:1504.08258}}].
%%CITATION = ARXIV:1504.08258;%%

\bibitem{Browne2012}
C.~Browne, E.~Powley, D.~Whitehouse, S.~Lucas, P.~Cowling, P.~Rohlfshagen,
  S.~Tavener, D.~Perez, S.~Samothrakis and S.~Colton, \emph{{A Survey of
  {M}onte {C}arlo {T}ree {S}earch Methods}},
  \href{https://dx.doi.org/10.1109/TCIAIG.2012.2186810}{\emph{Computational
  Intelligence and AI in Games, IEEE Transactions on} {\bfseries 4} (2012) 1}.

\bibitem{Vermaseren:2004mc}
J.A.M.~Vermaseren, \emph{{The Rules of Physics}},
  \href{https://dx.doi.org/10.1016/j.nima.2004.07.093}{\emph{Nucl. Instrum.
  Meth.} {\bfseries A534} (2004) 232}.
%%CITATION = NUIMA,A534,232;%%

\bibitem{Baikov:2010hf}
P.A.~Baikov and K.G.~Chetyrkin, \emph{{Four loop massless propagators: An
  algebraic evaluation of all master integrals}},
  \href{https://dx.doi.org/10.1016/j.nuclphysb.2010.05.004}{\emph{Nucl. Phys.}
  {\bfseries B837} (2010) 186}
  [\href{https://arxiv.org/abs/1004.1153}{{\ttfamily arXiv:1004.1153}}].
%%CITATION = ARXIV:1004.1153;%%

\bibitem{Lee:2011jt}
R.N.~Lee, A.V.~Smirnov and V.A.~Smirnov, \emph{{Master integrals for four-loop
  massless propagators up to weight twelve}},
  \href{https://dx.doi.org/10.1016/j.nuclphysb.2011.11.005}{\emph{Nucl. Phys.}
  {\bfseries B856} (2012) 95}
  [\href{https://arxiv.org/abs/1108.0732}{{\ttfamily arXiv:1108.0732}}].
%%CITATION = ARXIV:1108.0732;%%

\bibitem{igraph}
G.~Csardi and T.~Nepusz, \emph{The igraph software package for complex network
  research}, {\emph{InterJournal} {\bfseries Complex Systems} (2006) 1695},
  \url{http://igraph.org}.

\bibitem{DBLP:journals/corr/abs-1102-1523}
S.~van~der~Walt, S.C.~Colbert and G.~Varoquaux, \emph{{The NumPy Array: A
  Structure for Efficient Numerical Computation}},
  \href{https://dx.doi.org/10.1109/MCSE.2011.37}{\emph{Computing in Science
  {\&} Engineering} {\bfseries 13} (2011) 22}
  [\href{https://arxiv.org/abs/1102.1523}{{\ttfamily arXiv:1102.1523}}],
  \url{http://www.numpy.org}.

\bibitem{sympy}
A.~Meurer et~al., \emph{{SymPy}: symbolic computing in {P}ython},
  \href{https://dx.doi.org/10.7717/peerj-cs.103}{\emph{PeeJ Computer Science}
  (2017) 3:e103}, \url{http://www.sympy.org}.

\bibitem{QGRAF}
P.~Nogueira, \emph{{Automatic Feynman Graph Generation}},
  \href{https://dx.doi.org/10.1006/jcph.1993.1074}{\emph{J. Comput. Phys.}
  {\bfseries 105} (1993) 279}.
%%CITATION = JCTPA,105,279;%%

\bibitem{vanRitbergen:1998pn}
T.~van~Ritbergen, A.N.~Schellekens and J.A.M.~Vermaseren, \emph{{Group theory
  factors for Feynman diagrams}},
  \href{https://dx.doi.org/10.1142/S0217751X99000038}{\emph{Int. J. Mod. Phys.}
  {\bfseries A14} (1999) 41}
  [\href{https://arxiv.org/abs/hep-ph/9802376}{{\ttfamily hep-ph/9802376}}].
%%CITATION = HEP-PH/9802376;%%

\bibitem{vanRitbergen:1997va}
T.~van~Ritbergen, J.A.M.~Vermaseren and S.A.~Larin, \emph{{The four-loop
  $\beta$-function in quantum chromodynamics}},
  \href{https://dx.doi.org/10.1016/S0370-2693(97)00370-5}{\emph{Phys. Lett.}
  {\bfseries B400} (1997) 379}
  [\href{https://arxiv.org/abs/hep-ph/9701390}{{\ttfamily hep-ph/9701390}}].
%%CITATION = HEP-PH/9701390;%%

\bibitem{Czakon:2004bu}
M.~Czakon, \emph{{The four-loop QCD $\beta$-function and anomalous
  dimensions}},
  \href{https://dx.doi.org/10.1016/j.nuclphysb.2005.01.012}{\emph{Nucl. Phys.}
  {\bfseries B710} (2005) 485}
  [\href{https://arxiv.org/abs/hep-ph/0411261}{{\ttfamily hep-ph/0411261}}].
%%CITATION = HEP-PH/0411261;%%

\bibitem{Moch2017}
S.~Moch, B.~Ruijl, T.~Ueda, J.A.M.~Vermaseren and A.~Vogt, \emph{Four-loop
  non-singlet splitting functions in the planar limit and beyond},  to appear.

\bibitem{RuijlZurich}
B.~Ruijl, \emph{{Towards five loop calculations in QCD}},  seminar in Zurich,
  December 2016,
  \url{http://www.physik.uzh.ch/en/seminars/ttpseminar/HS2016.html}.

\bibitem{Lee:2016ixa}
J.~Henn, A.V.~Smirnov, V.A.~Smirnov, M.~Steinhauser and R.N.~Lee,
  \emph{{Four-loop photon quark form factor and cusp anomalous dimension in the
  large-$N_c$ limit of QCD}},
  \href{https://arxiv.org/abs/1612.04389}{{\ttfamily arXiv:1612.04389}}.
%%CITATION = ARXIV:1612.04389;%%

\bibitem{Baikov:2016tgj}
P.A.~Baikov, K.G.~Chetyrkin and J.H.~K\"uhn, \emph{{Five-Loop Running of the
  QCD Coupling Constant}},
  \href{https://dx.doi.org/10.1103/PhysRevLett.118.082002}{\emph{Phys. Rev.
  Lett.} {\bfseries 118} (2017) 082002}
  [\href{https://arxiv.org/abs/1606.08659}{{\ttfamily arXiv:1606.08659}}].
%%CITATION = ARXIV:1606.08659;%%

\bibitem{Herzog:2017bjx}
F.~Herzog and B.~Ruijl, \emph{{The R*-operation for Feynman graphs with generic
  numerators}},  \href{https://arxiv.org/abs/1703.03776}{{\ttfamily
  arXiv:1703.03776}}.
%%CITATION = ARXIV:1703.03776;%%

\bibitem{Abbott80}
L.F.~Abbott, \emph{{The Background Field Method Beyond One Loop}},
  \href{https://dx.doi.org/10.1016/0550-3213(81)90371-0}{\emph{Nucl. Phys.}
  {\bfseries B185} (1981) 189}.
%%CITATION = NUPHA,B185,189;%%

\bibitem{AbbottGS83}
L.F.~Abbott, M.T.~Grisaru and R.K.~Schaefer, \emph{{The Background Field Method
  and the S Matrix}},
  \href{https://dx.doi.org/10.1016/0550-3213(83)90337-1}{\emph{Nucl. Phys.}
  {\bfseries B229} (1983) 372}.
%%CITATION = NUPHA,B229,372;%%

\bibitem{Tkachov:1981wb}
F.V.~Tkachov, \emph{{A theorem on analytical calculability of 4-loop
  renormalization group functions}},
  \href{https://dx.doi.org/10.1016/0370-2693(81)90288-4}{\emph{Phys. Lett.}
  {\bfseries B100} (1981) 65}.
%%CITATION = PHLTA,B100,65;%%

\bibitem{Vermaseren2014}
J.A.M.~Vermaseren, \emph{Status of {FORM}},  talk presented in Durbach,
  September 2014.

\end{thebibliography}\endgroup

%--#] Bibliography : 

\end{document}